# Online Prediction for Streaming Observational Data

## Bertrand Clarke [1]  Aleena Chanda[2]

[1] *Department of Statistics, University of Nebraska-Lincoln , e-mail:* bclarke3@unl.edu

[2] *Department of Statistics, University of Nebraska-Lincoln, , e-mail:* achanda2@huskers.unl.edu

**Abstract:**

The automated collection of streaming observational data has become standard and defies most traditional analytic techniques. It is not just that models are hard to identify, there may not be any model that can be safely and usefully assumed. Indeed, frequently it is only predictions that can be made and assessed. Problems for this kind of data are often called M-Open and have motivated new approaches and philosophies.

This paper will review some of the most successful recent predictive methods for the M-Open problem class. Techniques include predictors using expert advice such as the Shtarkov solution, Bayesian nonparametrics such as Gaussian process priors, hash function based predictors such as the **Count-Min** sketch, and conformal prediction. Throughout, the properties of the predictors are presented and compared from a principled standpoint.

**MSC2020 subject classifications:** Primary 62L99, 62C99; secondary 62M20.

## Contents

## 1. Introduction

The automated collection of large scale observational data sets has become commonplace. Often this goes by the name of webscraping, prediction along a string, or, as we call it here, streaming data – leaving it understood that we mean data less structured than time series.

For present purposes, we think of the paradigm streaming data set as high volume, continually and rapidly generated, and flowing endlessly i.e., as long as required with no meaningful beginning or end. The data is observational; we are not trying to intervene in any way. We have a data generator (DG) that simply runs. Our task is to process the stream of data outcome by outcome. That is, our processing is a stream, too. Since there is no meaningful stopping point we do not use batch processing. In particular, since the output for time $n$ must be generated before the outcome for time $n + 1$ is received, our analysis is in real time. Hence, our analytic procedures are time sensitive and incremental in the data. We do not have time to redo an analysis that we want to disavow. Another constraint is that we have limited data storage so while we can accumulate some data or some summaries of data we must discard most of it. Taken together, our analysis must be 'one pass' – we look at the new data point and our accumulated data summary, and then compute our output for the next time step in one running-time bounded procedure.

Moreover, we also assume there is not enough structure in the DG to allow traditional modeling. In particular, there is nothing stable enough about the data stream for modeling to be effective. For instance, if there are occasional jumps in the data they are too variable to be modeled effectively. Otherwise put, whatever the application and terminology, the data are too complex to fit into a traditional probabilistic modeling framework. Thus, modeling for these data is of limited effectiveness compared to the traditional data types in Statistics and we seek a form of modeling that imposes much weaker constraints on the data. This is logical because when data has fewer and weaker properties that we are accustomed to seeing, we should use techniques that make weaker assumptions and are less prescriptive.

This opens the question of whether probabilistic modeling can be effective at all. In part this is a philosophical question that we ignore here. Nevertheless, we will see that some sorts of probabilistic modeling are more effective than others suggesting that the role of probability theory is substantial but of a different character than in conventional statistical analyses.

Perhaps surprisingly, there are many data sources like this. A few are: sensor data, financial data, weather data, real-time online purchases, the internet of things, and much agricultural data. Indeed, despite the profusion of techniques practitioners have for these settings, the development of general analytic techniques for streaming data seems to have lagged their demand. That is, technological progress has pressed the world of data collection, management, and analysis to leap ahead of what methodologists have developed to date.

Admitting that the kind of streaming data we want to analyze is so non-representative of the data types we have traditionally tried to model means that for the most part we are thrown back on prediction and the properties of predictors such as robustness as our main analytic framework. For streaming decisions, we may also have an assessment of the costs of the decisions we have made, but we do not treat this case here. Our decisions will only be about what value we think the next outcome will take i.e., the $n$-th stage point prediction for the outcome at the $n + 1$ time step.

We start to structure our thinking about sequential prediction by recalling the division of problems into three classes based on the relative positions of the data generator (DG) and the mathematical structures we have proposed to understand it. These originate in [7] and were called $\mathcal{M}$-closed, -completed, and -open. The idea is that we have a DG and what we will generically call a collection of 'models', or better for present purposes, 'predictors'. Here we update this tripartite partition of problem classes as follows.

In the simplest case, called $\mathcal{M}$-closed, one of the models is true in the sense that it provides a complete and correct description of the DG. That is, the only unknown aspect of the problem is which 'model' in our class is true. It is understood that at least in an asymptotic sense, the true model gives the (unique) optimal predictor; see [42] Theorem 2 for an instance of this. Other inferences proceed from a chosen model, taking variability into account.

One step up in complexity is the $\mathcal{M}$-complete class of problems. In this problem class we assume there is a class of predictors and a DG but we do not assume any of the predictors are optimal. Since every valid scientific model must give a unique predictor, models can be included in the predictor class of an $\mathcal{M}$-complete problem. Thus, from the modeling standpoint, the true model which is assumed to exist, and would be available via its predictor, is not in the class. It is understood that despite this, the resulting bias is not usually the issue. Instead, the intuition is that the true model and/or optimal predictor is inaccessible for some reason. Perhaps it is just too complicated for us even to conceptualize or perhaps it is inaccessible because it grows in complexity as data accumulate and can never be obtained. The setting usually assumed (if impossible to verify) is that any model class that might be theoretically credible can only be approximated under restricted circumstances. So, the model or predictor class used is an approximation, hopefully a good one, at least in the sense that model bias is smaller than other sources of error.

Because there is an optimal predictor, it can be transformed into a model, at least approximately. That is, it is possible to propose a sequence of models whose predictors converge to the optimal predictor. Consequently, expectations and convergences still exist and can still be used.

This line of reasoning leads to the concept of a belief model. It is easy, and not too far wrong, to think of a belief model as a partially validated approximation to the true model because invocation of a belief model is a pragmatic solution to the real problem of model identification. Thus, features of the predictor or model class are sometimes assessed, perhaps for robustness of inferences, tractability of analysis, or communication of results, in view of a 'belief model' whose status is subjective in that different analysts may have different belief models – each in principle striving to approximate the true model. See [49] and [37] for discussion on this process.

In practice, however, the dictum 'all models are wrong but some are useful' has limited utility because we usually do have ever weaker ways to assess bias as the complexity of the DG increases. That is: All models are guilty of bias until proven innocent. Nevertheless, we can compare the performance of predictors with reference to each other and ignore the status of any belief or true model.

A conceptual example might be: Consider a true model so complex that we can only describe it by listing some of its properties and some of its possible covariates. We might formulate a class of non-linear belief models that had fewer covariates and some complex error structure. Then, we might consider an $\mathcal{M}$-complete model class consisting of linear models with a simplified error structure that we thought was applicable under some restricted circumstances.

The most complex class of problems is called $\mathcal{M}$-open. In this case, we do not have a belief model and we are left to assess the elements of our model class based on the data we have at hand and nothing more. In this context, usually predictors and their properties are all that we have to examine. An easy way to think of this is that there is no model that can be usefully and safely identified, i.e., there is no true model so expectations, convergences, and evaluations based on anything more than the data are infeasible.

As a generality, prediction in the $\mathcal{M}$-open problem class should satisfy the prequential principle. An elementary form of the prequential principle is that 'evaluations of predictors should be disjoint from their construction', see [20]. This criterion arose from the desire to avoid model-based evaluations on the grounds that using a model to generate a predictor and then to evaluate its predictive performance was a sort of double use of the model that should be avoided in parallel to avoiding the double use of the data. We impose this criterion here without further discussion. Henceforth, we also assume our data come from an $\mathcal{M}$-Open DG.

More precisely, we can imagine that in $\mathcal{M}$-open settings there is a DG that produces $\{y_1, \ldots, y_n, \ldots\}$ but that the outcomes do not arise from a stochastic process. Recall the Kolmogorov sufficient conditions for a collection of random variables to form a stochastic process from, say, [40], p. 11. The two key assumptions are that a collection of random variables satisfies the permutation condition in which the permutation of random variables is well-behaved and the marginalization condition which ensures joint distributions yield the correct marginal distributions. While it is not clear that failing to satisfy these two conditions is enough to make a DG $\mathcal{M}$-open, for present purposes we assume that $\mathcal{M}$-open problems have DG's outside the class of stochastic processes.

It is an open question, and perhaps not answerable, what the typical case of $\mathcal{M}$-open data looks like. Indeed, there cannot be any clear pattern including one of noise. After all, a data set that looks like it came from a $N(\mu, 1)$ gives an $\mathcal{M}$-closed problem. On the other hand, a protein synthesis problem involving numerous biochemical inputs, catalysts, and other cellular conditions is very likely to be $\mathcal{M}$-complete since we may be convinced there is a model that can be approximated under various circumstances even if we cannot write it down. A possible example of an $\mathcal{M}$-open data set might be taking the world's great literature, converting it to a string of letters or characters, and then trying to produce i.e., predict, a new piece of great literature. Even in this example predictions can be made, possibly even nontrivially, but there is likely no model or even optimal predictor that could usefully predict the next great novel.

The reader wishing to try to develop some intuition for what $\mathcal{M}$-open data looks like can skip ahead to Sec. 6 where we present example data sets from DG's that we regard as plausibly $\mathcal{M}$-open. Indeed, since the term $\mathcal{M}$-open is not precisely defined two people may reasonably disagree about whether a data is complex enough to be genuinely $\mathcal{M}$-open or is actually $\mathcal{M}$-complete. We expect that over time, the class of $\mathcal{M}$-open problems will be divided into smaller classes based on other properties relevant to prediction. The data sets we use here are not as large as we typically mean by the term streaming data, however, they suffice for evaluating predictive procedures.

We comment that with $\mathcal{M}$-open data, interpretations will never be as strong as when modeling is feasible. Critics will see this as a deficiency. However, if there is little stable enough to interpret, interpretations are suspect at best. The best response to this criticism is that a good predictor can be examined in context to see what it implies about the DG, e.g., what explanatory variables may be involved, what the typical variability is etc. In short, statements that stop well short of announcing a model. This might best be seen proposing a *class* of good predictors whose common properties might be apparent and linked to the DG. To this end, in our computations below, we examine robustness as well as predictive error.

Sequential prediction has a long history in statistics even if most of it does not relate to streaming data. Perhaps the earliest prediction oriented treatment of statistics is [2]. It offered the first unified framework for prediction in $\mathcal{M}$-closed problems. Prior to this book, predictive thinking was scattered in special cases across various fields and applications – regression and early econometrics for instance. Nearly 20 years later, [22] began to persuade people that prediction should be the central organizing principle of Statistics in place of estimation, testing, and modeling more generally. Predictive inference has always sought to focus on observables and uncertainty about the DG more than classical inference such as estimation and testing at the (high) cost of assuming well-defined models. Concurrently with the development of these ideas, [20] provided the prequential principle as a contrast with the likelihood principle and the idea of revising models regularly as data accumulate. Separately, [19] noted that bias was undetectable from within the Bayes paradigm necessitating some form of empirical approach.

More recently the celebrated textbook [10] provided a treatment of sequential prediction that was nearly encyclopedic up to the early 00's in particular for what we now call 'predictions with expert advice' and other standard predictors e.g., exponentially weighted average forecasts. This text has numerous specific examples of techniques with (often pointwise) performance bounds, e.g., on regret that we will see in Sec. 2. The results are often only for one predictive round at a time although some results are sequential. Some results are for the $\mathcal{M}$-open case, but many are for $\mathcal{M}$-complete or -closed; these cases are often not clearly distinguished. One of the points of this book is that the range of point predictors is enormous. So, a key question always is how well a given predictor behaves, especially as data accumulate.

For ease of exposition, we limit our attention to streaming data where the vectors $y = (y_1, \ldots, y_n, \ldots)$ have $y_i \in (m, M)$ for all $i$ for positive $m, M$ with $M > m$. Importantly, we make no necessary assumptions about any distributional properties of the $y_i$'s. In practice, there are other response variables and covariates and this is important but we do not address it here.

At this time of writing, there are only four established methods to form one-step-ahead point predictors that can be said to be specifically for streaming data. All four will be discussed here roughly in historical order. A fifth technique – recommender systems – that can be used for streaming data, will not be treated here since it is primarily for discrete data and, while it gives outputs that can be taken as predictions, there is no obvious analog to predictive error for ranked recommendations. The point of this paper is to have one place where the four main techniques are described systematically, recognized as addressing the same problem – one-step-ahead prediction for streaming data, and usefully contrasted.

The first technique is the Shtarkov solution. It rests on a finding a predictor the achieves the minimax regret under a log criterion – the regret is the difference between the best possible prediction that could have been made and the prediction that was actually made. This predictor not used very much, often for computational reasons, but is conceptually important as a framework for thinking about sequential $\mathcal{M}$-open prediction. It emerged from the information theory literature in the late 80's and has both a Bayesian and a frequentist version, with the Bayesian version generally being better behaved. In this technique, the prior is put over a class of experts, not DG's, whose predictions are combined. Bayesians prefer to generate whole distributions for observables rather than just point predictors. However, it is unclear what this means for streaming $\mathcal{M}$-open data where there is no true distribution.

The second technique is a stochastic process approach. Even though we rule out stochastic processes as a model for the DG, stochastic processes, if used well, and properly interpreted, can give valid point predictors. The main one developed here is the Gaussian process, or more exactly, since this class is mainly Bayesian, a Gaussian process prior on a function space. When using this class we have to be careful to ensure posterior convergence does not put us back in an $\mathcal{M}$-complete or -closed case where there is a true model. Other Bayes methods can also be used with the appropriate caveats.

The third class of techniques originates from computer science and rests on randomly generated 'hash' functions. Hash functions are functions that are not one-to-one – they reduce their domains – and so can be used for data compression. The name probably originates from its normal English usage – hash functions metaphorically chop and mix their domains to give a sort of uniformity on their range. In practice we have to choose several, perhaps many hash functions, but by allowing a small, controlled amount, of error we can achieve good prediction and high computational efficiency.

For the sake of completeness, we include a short section on conformal prediction. Performance assessment of conformal predictors is normally in an $\mathcal{M}$-complete or -open context but their empiricism makes them applicable in $\mathcal{M}$-open sequential prediction problems as well.

We will also present examples, results, and computational comparisons of the first four methods. The goal of our computational work is to produce examples of when one method outperforms the others.

The rest of this paper proceeds as follows. In Sec. 2 we present the minimax regret approach, focusing on the Shtarkov solution and predictor. In Sec. 3, we adapt Bayesian thinking to $\mathcal{M}$-open DG's and present some basic techniques using Gaussian process priors and Dirichlet process priors. In Sec. 4, we present the concept of hash functions as a technique to achieve good prediction, running time, and data storage properties simultaneously. Here we will explicitly use techniques, including imposing a 'one-pass' requirement, to ensure that our predictors scale up to high volume data. In a short Sec. 5 we briefly discuss conformal prediction. We conclude with computational results in Sec. 6 and more general discussion of predictive methodologies in Sec. 7. Many of the routine proofs are relegated to Appendices A, B, C, and D.

## 2. Shtarkov

The Shtarkov solution, see [47], is a density that can be used for prediction under a minimax regret optimality criterion. The criterion assumes that there are multiple experts each giving predictions for a future outcome and that a forecaster may use these to form the actual prediction that will be announced. The forecaster's goal is only to match the predictive performance of the best expert. In practice, the 'experts' are represented by their candidate models that represent their honest views of what the next outcome will be.

Often this is phrased as a game of $n$ rounds between a Forecaster and Nature, overseen by an MC. At the start of the first round, the MC calls for each expert to announce a predictor for the first outcome $y_1$, assumed univariate. The experts are indexed by $\theta$ and are represented by predictive densities say $p(y|\theta)$, for $\theta \in \Theta$ where $\Theta$ is the set of experts. For ease of exposition, we assume $\Theta$ is a connected open set in $d$-dimensional real space and that $p(y|\theta)$ is continuous and bounded on $\Theta$ in both its arguments.

Next, the Forecaster uses these $p(y_1|\theta)$'s to form a predictive density $q(y_1)$ that will be used to generate the actual predictions for the actual outcome $y_1$. Seeing all this, Nature generates the actual outcome $y_1$ by any mechanism at all – probabilistic or not. Thus, the prediction task is $\mathcal{M}$-Open. The MC compares the prediction and the outcome by computing $q(y_1)$ and instructing the Forecaster to pay Nature $\ln(1/q(y_1))$, concluding the round; when $\ln(1/q(y_1))$ is negative it is the amount Nature pays the Forecaster. We assume that each player wants to make as much money (or lose as little money) as possible.

If $n$ rounds of the game are played sequentially, then Forecaster's predictor $q(y_i)$ for outcome $y_i$, $i = 2, \ldots, n$ may use the earlier observed $y_i$'s. Also, the experts may use the earlier data to form the predictors they announce. The Forecaster only uses the earlier data by way of the experts' predictors.

This game uses ln as if it were a loss function. In fact, we are using ln as a scoring rule: it measures how representative an outcome is of a density. Values of $y$ with higher values of $q(y)$ are more representative (of $q$) than values of $y$ with lower values of $q(y)$. There are many scoring rules. They have different properties and forecasts are sensitive to which one is chosen.

A physical interpretation comes from information theory. For discrete outcomes, the values $\ln(1/q(y))$ are approximately the Shannon code lengths when $q$ is the source distribution and therefore it is a measure of complexity – higher complexity $y$'s correspond to longer code words and lower probabilities. When the base of the logarithm is two, code length is a cost because it is the number of bits that must be used in a binary representation of the outcome. An extensive discussion of the ln scoring rule, the role of experts, and sequential prediction more generally is in [14].

The question is how the Forecaster should choose $q$ and make use of the experts' predictors. The Forecaster may decide that, rather than form predictions independently based on no information, it may be best to 'follow the best expert'. That is, consistently choose the predictor from the expert $\theta$ who gives the best predictions. This is formalized in the concept of regret: Regret is the loss the forecaster incurs beyond the loss of the best performing expert. The best performing expert depends on the data and may vary from round to round.

The difference between the Forecaster's prediction and an expert's density under the ln scoring rule is

$$\mathsf{Regret}(q, \theta, y) = \log \frac{1}{q(y)} - \log \frac{1}{p(y|\theta)} = \log \frac{p(y|\theta)}{q(y)}. \tag{1}$$

The maximal regret is

$$\mathsf{R_{max}}(q) = \sup_y \sup_\theta \log \frac{p(y|\theta)}{q(y)}. \tag{2}$$

The optimal $q(\cdot)$ minimizes (2) and is given by the Shtarkov solution

$$q_{\mathsf{opt,F}}(y) = \arg_q \; \left[ \min_q \sup_y \sup_\theta \ln \frac{p(y|\theta)}{q(y)} \right] = \frac{p(y|\hat{\theta})}{\int p(y|\hat{\theta}) dy} \tag{3}$$

where $\hat{\theta} = \theta(\hat{y})$ is the maximal likelihood estimate (MLE), provided the integral exists. So, Nature can maximize the cost to the Forecaster by choosing $y = \arg \max_y \; \ln 1/(q_{\mathsf{opt,F}}(y))$ when the maximum exists, provided Nature knows the the Forecaster will use $q_{\mathsf{opt,F}}$.

The expression in bracekts in (3) is the minimax regret, so $q_{\mathsf{opt,F}}$ is the minimax predictor that we call the Shtarkov solution. A proof of (3) is in [47], see also [4], exp. (5). If the Forecaster has access to experts weighted by $w(\theta)$ the optimum $q_{\mathsf{opt,B}}$ is the Bayes Shtarkov solution

$$q_{\mathsf{opt,B}}(y) = \frac{w(\tilde{\theta}(y^n)) p(y^n | \tilde{\theta}(y^n))}{\sum_{y^n} w(\tilde{\theta}(y^n)) p(y^n | \tilde{\theta}(y^n))}, \tag{4}$$

where $\tilde{\theta}$ is the maximum posterior likelihood estimator (MPLE), i.e., the posterior mode. In $\mathcal{M}$-open problems the status of $w$ as a prior is unclear but it can be regarded simply as a pre-data preference for some experts over others, perhaps in the sense of reliability. In this view, the experts are essentially regarded as analogous to actions in a decision theory problem rather than as distributions. Without further comment, we take $w$ to be continuous. For each $n$, the Shtarkov solution gives a density often called the normalized maximum likelihood (NML). In the Bayes case these are normalized maximum posterior likelihoods (NMPL). For convenience, we write $q_{\mathsf{opt}}$ when a statement applies to both $q_{\mathsf{opt,F}}$ and $q_{\mathsf{opt,B}}$.

Recall, the Kolmogorov Extension Theorem identifies two conditions that are sufficient for a set of distributions to form a stochastic process; see [40], p.

11. One of these is a marginalization condition and it is reasonable to conjecture that $q_{opt}$ does not satisfy it. Thus, it is unlikely that either the NML or NMPL is a stochastic process in general. However, [4] and [17] argue that the NML, and by extension the NMPL, are close to the mixture of distributions $m(y) = \int w(\theta) p_\theta(y) d\theta$ in the sense that both asymptotically achieve the minimax regret. That is, the sequence of densities the NML or the NMPL represents is asymptotically to a stochastic process. Note that $q_{opt}$ is finite $n$ optimal while $m_n$ isn't even though it is asymptotically optimal.

A continual question is how to choose the experts since in practice we rarely have a bank of experts who will announce densities. When experts are available, they usually only give their predictions, not predictors for general use. In fact, the experts here are simply a parametric family and we are free to choose it however we wish. Regarding the $p(\cdot|\theta)$'s as announced by experts is simply an interpretation to ensure that our method makes sense in an $\mathcal{M}$-open context; we don't have to assume a parametric family is true to get useful results.

The Shtarkov solution can be used to give predictions. First, because it is a density we can form highest density prediction regions given a desired level of predictive cofidence $1 - \alpha$. However, since we don't assume the underlying DG is a probability we don't really believe the spread of $q_{opt}$. So, we only use $q_{opt}$ to give point predictors. Often, the mode of the Shtarkov solution is used as a point predictor because $q_{opt}$ can be very highly skewed (when it is not symmetric and unimodal); see [31] and the examples below.

Here, we use the maximum of

$$q_{opt,\mathsf{F}}(y^n, y_{n+1}) \quad \text{or} \quad q_{opt,\mathsf{B}}(y^n, y_{n+1})$$

over $y_{n+1}$ holding $y^n$ fixed, and write the predictions as $\hat{y}_{n+1,\mathsf{F}}(y^n)$ or $\hat{y}_{n+1,\mathsf{B}}(y^n)$. We call these the frequentist and Bayes Shtarkov predictors, respectively. This would be equivalent to finding the maximum of

$$q_{\mathsf{Sht}}(y_{n+1}|y^n) = \frac{q_{opt}(y^{n+1})}{q_{opt}(y^n)}$$

provided that the conditional density is well-defined. Since we must assume that the normalizing constants in $q_{opt}(y^{n+1})$ and $q_{opt}(y^{n+1})$ exist, it is equivalent to find

$$\arg\max_{y_{n+1}} p(y^n, y_{n+1}|\theta(y^{\hat{n+1}})).$$

The next subsection looks at sufficient conditions for the NML and NPML to exist. Then we turn to some illustrative examples. Outside well behaved parametric families of experts few Shtarkov predictors can be given in closed form. However, as a generality, they can be given numerically, e.g., in the binomial case below. We conclude this section with a brief discussion of these and other issues in Shtarkov predictors.

### 2.1. *Existence of* $q_{opt,F}$

The main impediment to using the NML or NMPL as a density is ensuring it exists. Since results ensuring the existence of MLE's and MPLE's under mild conditions are well-known, the task here is to find conditions under which the denominators in (3) and (4) are finite. Similar problems can arise if other optimal predictors are used, see [14]. In fact, the denominator of the NML or NMPL can be infinite making the solution undefined. For instance, the NML constant for the Exponential($\lambda$) is infinite because it is the integral of

$$p(y^n|\hat{\lambda}) = \frac{1}{\bar{y}}e^{-n} \tag{5}$$

where $\hat{\lambda} = 1/\bar{y}$. However, it is easy to see that the NML of the Binomial($N, p$) is finite because its support is a finite set.

In some cases, an asymptotic result from [43] can be used. Effectively it gives

$$q_{opt,F}(y^n) = \frac{d}{2}\ln\frac{n}{2\pi} + C + o(1) \tag{6}$$

and identifies $C$; using essentially the same proof gives an analogous result for $q_{opt,B}$. One limitation of this approach is that it can be difficult to determine whether the hypotheses are satisfied. Another limitation is that $n$ must be large so finding the exact value of the constant may be difficult if that is desired. Moreover, the hypotheses that lead to expressions like (6) are too strong: we don't need a nice asymptotic expression; we only want existence or, at most, useful finite sample approximations.

Some authors try to avoid the nonexistence of the NML and/or NMPL or the ineffectiveness of asymptotics by focusing on the Bayesian mixture distribution as an approximation for finite samples, see [5] and [17].

Here, we denote the normalizing constants in the NML and NMPL by

$$D_{n,F} = \int p(y^n|\hat{\theta})dy^n \quad \text{and} \quad D_{n,B} = \int w(\tilde{\theta})p(y^n|\tilde{\theta})dy^n.$$

There are two types of hypotheses that ensure the two forms of $D_n$ exist as bounded, strictly positive real numbers. The first gives a result stronger than is actually needed: it gives a rate for the normalization constant as well as existence. Even though its hypotheses are mild, they are hard to verify. The second type of hypotheses are easy to check but are stronger than required.

The first type of hypotheses are based on [43], see also [31]. With some informality, we have the following.

**Theorem 2.1.** *Assume the following:*

1. *Let $I_n(\theta)$ be the $n^{th}$ stage Fisher Information and suppose there is an $I(\theta)$ so that*

$$I_n(\theta) = -\frac{1}{n}E\left[\frac{\partial^2 \log p(y^n|\theta)}{\partial\theta_i\partial\theta_j}\right] \to I(\theta),$$

*as $n \to \infty$, and suppose $\exists\, c_1, c_2$ so that $\forall \theta \in \Theta$, $0 < c_1 \le |I(\theta)| \le c_2 < \infty$.*

2. *The elements of $I(\theta)$ are continuous on $\Theta$.*
3.

$$\int_\Theta \sqrt{I(\theta)} \mathrm{d}\theta < \infty.$$

4. *The posterior mode $\tilde\theta$ and the MLE $\hat\theta$ satisfy a uniform central limit theorem. That is, for $\tilde\theta$ we have*

$$\xi \;=\; \sqrt{n}(\tilde\theta(y^n) - \theta) \xrightarrow{\mathcal{L}} N(0, I^{-1}(\theta))$$

*uniformly for $\theta \in \Theta$ and similarly for $\hat\theta$.*

5. *There is a positive definite matrix $C_0$ so that*

$$I(y^n, \tilde\theta) \;=\; \left( -\frac{1}{n} \left\{ \frac{\partial^2 \log p(y^n|\theta)}{\partial\theta_i\partial\theta_j} \right\}_{\theta=\tilde\theta} \right)_{i,j=1,\cdots,k} < C_0 < \infty$$

*assuming $\tilde\theta \in \Theta$, and similarly for $\hat\theta$. In addition, the family*

$$I_{ij}(y^n, \theta(\xi)) \;=\; -\frac{1}{n} \frac{\partial^2 \log p(y^n|\theta(\xi))}{\partial\theta_i\partial\theta_j},$$

*where $\theta(\xi) = \tilde\theta + \xi/\sqrt{n}$ is equicontinuous at $\xi = 0$ for $n \geq 1$, $1 \leq i$, $j \leq k$ and similarly for $\hat\theta$.*

*Then, $D_{n,\mathsf{F}}$ and $D_{n,\mathsf{B}}$ exist and hence so do $q_{\mathsf{opt,F}}(\cdot)$ and $q_{\mathsf{opt,B}}(\cdot)$, respectively.*

*Proof.* This follows from an examination of the proofs in [43] and [31]. The convergences assumed in the hypotheses necessitate a restriction to sequences $y^n$ for which $\hat\theta$ and $\tilde\theta$ are in the parameter space $\Theta$. □

The second type of result has a much simpler argument, partially explaining why its hypotheses are stronger albeit easier to check.

**Theorem 2.2.** *Assume the hypotheses:*

1. *The parameter space is bounded, convex, and the interior of its closure.*
2. *There is an $N$ so that for $n > N$ the likelihood function, resp. the joint density for the parameter and data, is continuous and strictly convex in $\theta$.*
3. *The densities $\{p(y|\theta)\}$ have common bounded support in $y$ as $\theta$ varies.*
4. *The densities $\{p(y|\theta)\}$ are continuous as real valued functions of two real vector valued arguments $y$ and $\theta$.*
5. *The prior density $w(\cdot)$ is positive and continuous on the parameter space.*

*Then, $D_{n,\mathsf{F}}$ and $D_{n,\mathsf{B}}$ exist and hence so do $q_{\mathsf{opt,F}}(\cdot)$ and $q_{\mathsf{opt,B}}(\cdot)$, respectively.*

**Remark 1:** Although the densities and random variables are assumed bounded, we conjecture that this can be relaxed. In particular, we think Condition 3 can be improved by using the hypotheses of Theorem 2.1 in [36].

*Proof.* This is a straightforward exercise in real analysis under the stated hypotheses, see Appendix A. □

The examples in the next subsection – Normal, Binomial, Exponential, and Gamma – probably satisfy the hypotheses of Theorem 2.1 but this is hard to verify without easy hypotheses for its Conditions 4 and 5 (and for the criterion on sets that was omitted from the statement for simplicity). On the other hand, it is easy to see that the binomial satisfies the hypotheses of Theorem 2.2 and the other examples satisfy the hypotheses for compact subsets of the parameter space and truncations of the random variables.

## 2.2. Examples

In this subsection we start by presenting the Bayes and Frequentist Shtarkov point predictors for three normal families of experts. They correspond to the cases i) $\mu$ unknown, $\sigma$ known, ii) $\mu$ known, $\sigma$ unknown, and iii) $\mu$ and $\sigma$ unknown. The experts are not models; they are only predictors and do not have any necessary physical correlates in terms of the DG.

After giving these six normal case predictors, we turn to the binomial – frequentist and Bayesian. Predictors in this case cannot be worked out in closed form. So, we approximate them computationally. For breadth, we also present exponential and Gamma family examples. Some of these can be worked out in closed form; they give intuitively reasonable answers.

### 2.2.1. Normal Distribution

The simplest case is the frequentist normal mean problem with known variance. For data $y^n$, write the MLE as $\hat{\mu} = \bar{y} = \bar{y}_n$ for the sample mean from $n$ observations. The frequentist Shtarkov solution for mean $\mu$ and variance $\sigma^2$ is

$$p(y^n | \hat{\mu}_n, \sigma^2) = \left( \frac{1}{\sigma^2 2\pi} \right)^{\frac{n}{2}} e^{-\frac{1}{2\sigma^2} \sum_{i=1}^{n} (y_i - \bar{y}_n)^2}, \tag{7}$$

normalized to integrate to one over $y^n$. To get a predictor for $n+1$, write

$$p(y^{n+1} | \hat{\mu}_{n+1}, \sigma^2) = \left( \frac{1}{\sigma^2 2\pi} \right)^{\frac{n+1}{2}} e^{-\frac{1}{2\sigma^2} \sum_{i=1}^{n+1} (y_i - \bar{y}_{n+1})^2} \tag{8}$$

and set

$$\bar{y}_{n+1} = \frac{n\bar{y}_n + y_{n+1}}{n+1}. \tag{9}$$

Using (9) in (8), re-arranging, and differentiating (twice) with respect to $y_{n+1}$, gives the MLE

$$\hat{y}_{n+1} = \bar{y}_n \tag{10}$$

that maximizes (8). For details, see Appendix B.1.

In this very simple case, we can actually identify the normalizing constant for (7). Starting with

$$\int \frac{1}{(2\pi\sigma^2)^{\frac{n}{2}}} e^{-\frac{1}{2\sigma^2}\sum_{i=1}^n (y_i - \bar{y})^2} dy^n, \tag{11}$$

consider the transformation $u = Hy$ where $H$ equals

$$\begin{pmatrix} \frac{1}{\sqrt{n}} & \frac{1}{\sqrt{n}} & \frac{1}{\sqrt{n}} & \cdots & \frac{1}{\sqrt{n}} \\ \frac{1}{\sqrt{2}} & \frac{-1}{\sqrt{2}} & 0 & \cdots & 0 \\ \frac{1}{\sqrt{6}} & \frac{1}{\sqrt{6}} & \frac{-2}{\sqrt{6}} & \cdots & 0 \\ \cdots & \cdots & \cdots & \cdots & \cdots \\ \frac{1}{\sqrt{n(n-1)}} & \frac{1}{\sqrt{n(n-1)}} & \frac{1}{\sqrt{n(n-1)}} & \cdots & \frac{-(n-1)}{\sqrt{n(n-1)}} \end{pmatrix}. \tag{12}$$

This gives $u_2, u_3, \cdots u_n$ are IID $N(0, \sigma^2)$ and

$$\sum_{i=2}^n u_i^2 = \sum_{i=1}^n (y_i - \bar{y})^2. \tag{13}$$

Using (13) in (11) we find that

$$\int \frac{1}{(2\pi\sigma^2)^{\frac{n}{2}}} e^{-\frac{1}{2\sigma^2}\sum_{i=1}^n (y_i - \hat{\mu}_{MLE})^2} dy^n$$

$$= \int \frac{1}{(2\pi\sigma^2)^{\frac{n}{2}}} e^{-\frac{1}{2\sigma^2}\sum_{i=2}^n u_i^2} du^{n-1} \tag{14}$$

$$= \frac{(2\pi\sigma^2)^{\frac{n-1}{2}}}{(2\pi\sigma^2)^{\frac{n}{2}}} \int_{\mathbf{R^{n-1}}} \frac{1}{(2\pi\sigma^2)^{\frac{n-1}{2}}} e^{-\frac{1}{2\sigma^2}\sum_{i=2}^n u_i^2} du^{n-1}$$

$$= \frac{1}{\sqrt{2\pi\sigma^2}}. \tag{15}$$

We recognize (15) as the standard deviation for a single normal outcome.

The next simplest case is the derivation of the Shtarkov predictor for the normal case where $\sigma^2$ is unknown but $\mu$ is known. The MLE is $\hat{\sigma}_n^2 = \frac{1}{n}\sum_{i=1}^n (y_i - \mu)^2$. So,

$$p(y^n|\mu, \hat{\sigma}_n^2) = \frac{n^{\frac{n}{2}} e^{-\frac{n}{2}}}{(2\pi)^{\frac{n}{2}} \sum_{i=1}^n (y_i - \mu)^2}.$$

Using this for $n+1$ gives

$$p(y^{n+1}|\mu, \hat{\sigma}_{n+1}^2) \propto \frac{1}{\sum_{i=1}^n (y_i - \mu)^2 + (y_{n+1} - \mu)^2}.$$

It is now easy to see that the minimum occurs at $\hat{y}_{n+1} = \mu$.

The third frequentist normal case allows both $\mu$ and $\sigma^2$ to be unknown. We have the MLE's $\hat{\mu}_n = \bar{y}_n$ and $\hat{\sigma}_n^2 = \frac{1}{n} \sum_{i=1}^n (y_i - \bar{y}_n)^2$. So,

$$
\begin{aligned}
p(y^n | \hat{\mu}_n, \hat{\sigma}_n^2) &= \frac{n^{\frac{n}{2}} e^{-\frac{n}{2}}}{(2\pi)^{\frac{n}{2}} \left[ \sum_{i=1}^n (y_i - \bar{y}_n)^2 \right]^{\frac{n}{2}}} \\
&\propto \frac{1}{\left[ \sum_{i=1}^n (y_i - \bar{y}_n)^2 \right]^{\frac{n}{2}}}.
\end{aligned}
$$

Writing this for $n+1$, taking logarithms and using (9), we get

$$
\begin{aligned}
\ln p(y^{n+1} | \hat{\mu}_{n+1}, \hat{\sigma}_{n+1}^2) &\propto -\frac{n+1}{2} \ln \sum_{i=1}^{n+1} (y_i - \bar{y}_{n+1})^2 \\
&= -\frac{n+1}{2} \ln \left[ \sum_{i=1}^n y_i^2 + y_{n+1}^2 - (n+1) \left( \frac{n\bar{y}_n + y_{n+1}}{n+1} \right)^2 \right]
\end{aligned}
$$

$$(16)$$

Differentiating with respect to $y_{n+1}$, setting the derivative equal to zero, and re-arranging gives

$$
\hat{y}_{n+1} - \frac{1}{n+1} (n\bar{y}_n + \hat{y}_{n+1}) = 0
$$

Solving this gives $\hat{y}_{n+1} = \bar{y}_n$. This maximizes the NML as shown in Appendix B.1 and so is the Shtarkov predictor.

Next we turn to the Bayes Shtarkov predictors for the normal case using conjugate priors. We start with the simplest case in which $\mu$ is unknown but $\sigma^2$ is known. If we choose the prior

$$
p(\mu | \mu_0, \sigma_0^2) = \left( \frac{1}{\sigma_0^2 2\pi} \right)^{\frac{1}{2}} e^{-\frac{1}{2\sigma_0^2} (\mu - \mu_0)^2}. \tag{17}
$$

for $\mu$, the posterior is available in closed form and the maximum posterior likelihood estimate for $\mu$ is

$$
\hat{\mu}_{MPLE} = \frac{1}{\tau_n} \left( \frac{\mu_0}{\sigma_0^2} + \frac{n\bar{y}_n}{\sigma^2} \right)
$$

where $\tau_n = \frac{1}{\sigma_0^2} + \frac{n}{\sigma^2}$. Writing the maximized joint likelihood and prior as

$$
p(y^n | \hat{\mu}_n, \sigma^2, \mu_0, \sigma_0^2) = p(y^n | \hat{\mu}_n, \sigma^2) \times p(\hat{\mu}_n | \mu_0, \sigma_0^2),
$$

we see that for $n+1$ we get

$$
\begin{aligned}
p(y^{n+1}|\hat{\mu}_{n+1}, \sigma^2, \mu_0, \sigma_0^2) &= \left(\frac{1}{\sigma^2 2\pi}\right)^{\frac{n+1}{2}} \left(\frac{1}{\sigma_0^2 2\pi}\right)^{\frac{1}{2}} \\
&\times e^{-\frac{1}{2\sigma^2} \sum_{i=1}^{n+1} \left[y_i - \frac{1}{\tau_{n+1}} \left(\frac{\mu_0}{\sigma_0^2} + \frac{(n+1)\bar{y}_{n+1}}{\sigma^2}\right)\right]^2} \\
&\times e^{-\frac{1}{2\sigma_0^2} \left[\frac{1}{\tau_{n+1}}\left(\frac{\mu_0}{\sigma_0^2} + \frac{(n+1)\bar{y}_{n+1}}{\sigma^2}\right) - \mu_0\right]^2}.
\end{aligned}
\tag{18}
$$

Taking logs, using (9) in (18) and re-arranging gives an expression that can be readily differentiated with respect to $y_{n+1}$. It turns out that the maximum is achieved at

$$
\hat{y}_{n+1} = \hat{\mu}_{MPLE},
$$

see Appendix B.1 for details.

Turning to the case where $\sigma$ is unknown and $\mu$ is known, we see it is similar to the corresponding frequentist case. The Inv-Gamma$(\alpha, \beta)$ is

$$
p(\sigma^2|\alpha, \beta) = \frac{\beta^\alpha}{\Gamma(\alpha)} \left(\frac{1}{\sigma^2}\right)^{\alpha+1} e^{-\frac{\beta}{\sigma^2}}.
\tag{19}
$$

So,

$$
\begin{aligned}
p(y^n|\mu, \sigma^2) \times p(\sigma^2|\alpha, \beta) &= \left(\frac{1}{\sigma^2 2\pi}\right)^{\frac{n}{2}} e^{-\frac{1}{2\sigma^2} \sum_{i=1}^n (y_i - \mu)^2} \\
&\times \frac{\beta^\alpha}{\Gamma(\alpha)} \left(\frac{1}{\sigma^2}\right)^{\alpha+1} e^{-\frac{\beta}{\sigma^2}}. \\
&\propto \left(\frac{1}{\sigma^2 2\pi}\right)^{\alpha + \frac{n}{2} + 1} e^{-\frac{1}{\sigma^2} \left[\beta + \frac{1}{2} \sum_{i=1}^n (y_i - \mu)^2\right]}.
\end{aligned}
\tag{20}
$$

Hence, $\sigma^2|\mu, \alpha, \beta, y^n \sim$ Inv-Gamma$(\alpha + \frac{n}{2}, \beta + \frac{1}{2}\sum_{i=1}^n (y_i - \mu)^2)$ giving

$$
\hat{\sigma}_{MPLE}^2 = \frac{\beta + \frac{1}{2}\sum_{i=1}^n (y_i - \mu)^2}{\alpha + \frac{n}{2} + 1}.
\tag{21}
$$

Letting $p(y^n|\hat{\sigma}^2, \mu, \alpha, \beta) = p(y^n|\mu, \hat{\sigma}^2) \times p(\hat{\sigma}^2|\alpha, \beta)$ we find

$$
\begin{aligned}
p(y^n|\hat{\sigma}^2, \mu, \alpha, \beta) &\propto \left[\frac{\alpha + \frac{n}{2} + 1}{\beta + \frac{1}{2}\sum_{i=1}^n (y_i - \mu)^2}\right]^{\alpha + \frac{n}{2} + 1} \\
&\times e^{-\frac{\alpha + \frac{n}{2} + 1}{\beta + \frac{1}{2}\sum_{i=1}^n (y_i - \mu)^2} \left[\beta + \frac{1}{2}\sum_{i=1}^n (y_i - \mu)^2\right]}.
\end{aligned}
\tag{22}
$$

Some routine calculus arguments give that

$$
\hat{y}_{n+1} = \mu
$$

is the maximum, see Appendix B.1.

Extending these to the Bayes Shtarkov normal case where both $\mu$ and $\sigma^2$ are unknown, we use the standard priors

$$p(\mu|\mu_0, \sigma_0^2) = \left(\frac{1}{\sigma_0^2 2\pi}\right)^{\frac{1}{2}} e^{-\frac{1}{2\sigma_0^2}(\mu-\mu_0)^2} \quad \text{and} \tag{23}$$

$$p(\sigma^2|\alpha, \beta) = \frac{\beta^\alpha}{\Gamma(\alpha)}\left(\frac{1}{\sigma^2}\right)^{\alpha+1} e^{-\frac{\beta}{\sigma^2}}. \tag{24}$$

So, the joint likelihood and prior is

$$p(y^n|\mu, \sigma^2) \times p(\mu|\mu_0, \sigma_0^2) \times p(\sigma^2|\alpha, \beta)$$

$$\propto \left(\frac{1}{\sigma^2}\right)^{\frac{n}{2}+\alpha+1}\left(\frac{1}{\sigma^2}\right)^{\frac{1}{2}} e^{-\frac{1}{\sigma^2}[\beta+\frac{1}{2}\sum_{i=1}^{n}(y_i-\mu)^2+\frac{1}{2}(\frac{\mu-\mu_0}{\sigma_0})^2]}, \tag{25}$$

and it is straightforward to see that

$$\mu|\sigma^2, \sigma_0^2, \mu_0 \sim \mathcal{N}\left(\frac{n\bar{y}+\frac{\mu_0}{\sigma_0^2}}{n+\frac{1}{\sigma_0^2}}, \frac{\sigma^2\sigma_0^2}{n\sigma_0^2+1}\right) \quad \text{and}$$

$$\sigma^2|\alpha, \beta \sim \mathcal{IG}\left(\alpha+\frac{n}{2}, \beta+\frac{1}{2}\left\{\sum_{i=1}^{n}y_i^2+\frac{\mu_0^2}{\sigma_0^2}-\frac{(n\bar{y}+\frac{\mu_0}{\sigma_0^2})^2}{n+\frac{1}{\sigma_0^2}}\right\}\right).$$

Thus, we have

$$\hat{\mu}_{MPLE} = \frac{n\bar{y}+\frac{\mu_0}{\sigma_0^2}}{n+\frac{1}{\sigma_0^2}} \tag{26}$$

$$\hat{\sigma}_{MPLE}^2 = \frac{\beta+\frac{1}{2}\left\{\sum_{i=1}^{n}y_i^2+\frac{\mu_0^2}{\sigma_0^2}-\frac{(n\bar{y}+\frac{\mu_0}{\sigma_0^2})^2}{n+\frac{1}{\sigma_0^2}}\right\}}{\alpha+\frac{n}{2}+1}. \tag{27}$$

Now, some calculus arguments, see Appendix B.1, lead to

$$p(y^{n+1}|\hat{\mu}_{n+1,MPLE}, \hat{\sigma}^2{}_{n+1,MPLE}) \times p(\hat{\mu}_{n+1,MPLE}|\mu_0, \sigma_0^2)p(\hat{\sigma}_{n+1,MPLE}^2|\alpha, \beta)$$

being maximized for given $y^n$ at

$$\hat{y}_{n+1} = \frac{n\bar{y}_n+\frac{\mu_0}{\sigma_0^2}}{n+\frac{1}{\sigma_0^2}}.$$

### 2.2.2. *Binomial Distribution*

As a second example, we look at the binomial. Let $Y \sim \text{Bin}(N, \theta)$ have probability mass function denoted

$$p(y|\theta) = \binom{N}{y}\theta^y(1-\theta)^{N-y} \tag{28}$$

for $0 \le \theta \le 1$ and $y = 0, 1, 2, \cdots, N$. For the frequentist case, recall that for $n$ samples $y_1, y_2, \cdots, y_n$, the MLE of $\theta$ is

$$\hat{\theta}_{MLE} = \frac{\sum_{i=1}^n y_i}{Nn} = \frac{\bar{y}_n}{N}. \tag{29}$$

We have that for fixed $N$, $\binom{N}{y}\theta^y(1-\theta)^{N-y}$ is bounded as a function of $y$ and $\theta$. Thus, $\sum p(y^n|\hat{\theta}) = \sum_{y_i=0}^N \prod_{i=1}^n \binom{N}{y}\hat{\theta}^y(1-\hat{\theta})^{N-y}$ is bounded so the frequentist Shtarkov solution exists by Theorem 2.2. Moreover, for any well-behaved prior, a similar observation gives that the Bayes Shtarkov solution exists, too.

To find the frequentist Shtarkov predictor for a binomial class of experts, note that the log-likelihood function of $\theta|y_1, y_2, \cdots, y_n$ is

$$p(y^n|\theta) \;=\; \left\{\prod_{i=1}^n \binom{N}{y_i}\right\}\theta^{\sum_{i=1}^n y_i}(1-\theta)^{Nn-\sum_{i=1}^n y_i}. \tag{30}$$

So, replacing $\theta$ in (30) by its MLE gives

$$\begin{aligned}
p(y^n|\hat{\theta}_{MLE}) &= \left\{\prod_{i=1}^n \binom{N}{y_i}\right\}\left(\frac{\bar{y}_n}{N}\right)^{\sum_{i=1}^n y_i}\left(1 - \frac{\bar{y}_n}{N}\right)^{Nn-\sum_{i=1}^n y_i} \\
&= \left\{\prod_{i=1}^n \binom{N}{y_i}\right\}\left(\frac{\bar{y}_n}{N}\right)^{\sum_{i=1}^n y_i}\left(\frac{N-\bar{y}_n}{N}\right)^{Nn-\sum_{i=1}^n y_i}.
\end{aligned} \tag{31}$$

Taking logarithms on both sides of (31) and re-arranging gives that

$$\begin{aligned}
\ln p(y^{n+1}|\hat{\theta}_{MLE}) &\propto \ln\binom{N}{y_{n+1}} + (n\bar{y}_n + y_{n+1})\ln(n\bar{y}_n + y_{n+1}) \\
&\quad + \{N(n+1) - (n\bar{y}_n + y_{n+1})\}\ln\{N(n+1) - (n\bar{y}_n + y_{n+1})\},
\end{aligned} \tag{32}$$

see Appendix B.2. We cannot optimise (32) to get a closed-form expression for $y_{n+1}$. Also, $y_{n+1}$ is discrete, so it is not reasonable to differentiate (32) w.r.t. $y_{n+1}$. Hence, we adopt a computational approach.

Fig. 1 shows a plot of the log-likelihood in (32) for the case $N = 30$ and $n = 10$. The range of $y_{n+1}$ is $\{0, 1, \ldots, 30\}$. The actual range of $\bar{y}_{n+1}$ is the same, but we used .5 to 30 in increments of .5 to get a smoother plot, omitting $\bar{y} = 0$ because $\ln(n\bar{y}_n + y_{n+1})$ in (32) is undefined for $y_{n+1} = \bar{y}_n = 0$.

For fixed $\bar{y}$, the curve in the surface of Fig. 1 is concave so we can maximize $\hat{y}_{n+1}$ uniquely over the range of $\bar{y}_n$. It is seen that for $\bar{y}$ near zero, the maximum is achieved by $y_{n+1} = 0$ and for $\bar{y} = 30$, the maximum is achieved by $y_{n+1} = 30$. The optimal values of $y_{n+1}$ increase with $\bar{y}$.

Surprisingly, as $n$ increases the maximum does not become stronger. For instance, Fig. 2 shows the plots of the loglikelihood in (32) for for $n = 10$ and 25 using red and green dots respectively. The red dots are the values in the line $\bar{y} = 10$ in the surface plotted in Fig. 1. The two dotted curves in Fig. 2 are nearly identical; that is, the curvature does not increase with $n$.

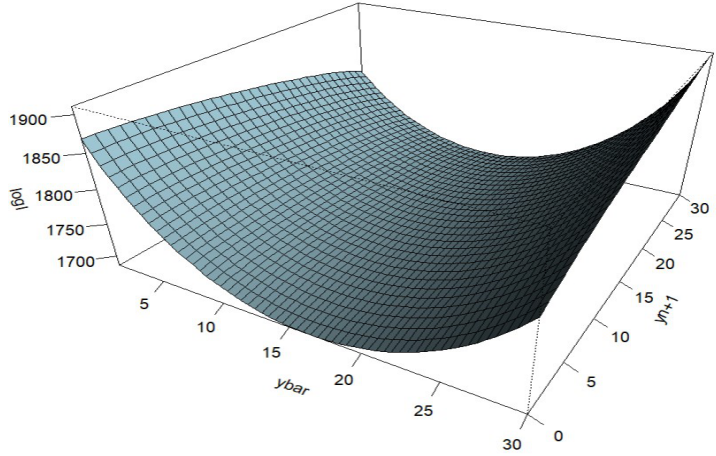

*Fig 1. Perspective plot of (32) as a function of $\bar{y}$ and $y_{n+1}$. The optimal $\hat{y}_{n+1}$ for a given $\bar{y}$ is the maximum on the surface across the line $\bar{y}$. Note that as $\bar{y}$ moves from zero to 30, the location of the maximum moves from zero to 30 as well.*

The Bayes Shtarkov predictor for binomial experts is similar. Let us use the conjugate prior for $\theta$ which is given by

$$w(\theta|\alpha,\beta) \;\;=\;\; \frac{1}{Beta(\alpha,\beta)}\theta^{\alpha-1}(1-\theta)^{\beta-1}; \alpha>0, \beta>0, \tag{33}$$

where the Beta function is $Beta(\alpha,\beta)=\frac{\Gamma(\alpha)\Gamma(\beta)}{\Gamma(\alpha+\beta)}$ in terms of the Gamma function. Then, from Appendix B.2, we see the analog to (32) is (34):

$$\ln[p(y^{n+1}|\hat{\theta}_{MPLE})w(\hat{\theta}_{MPLE},\alpha,\beta)] \tag{34}$$

$$\propto \ln\binom{N}{y_{n+1}} + (n\bar{y}_n + y_{n+1} + \alpha - 1)\ln(n\bar{y}_n + y_{n+1} + \alpha - 1)$$

$$+(N(n+1)+\beta-(n\bar{y}_n+y_{n+1})-1)\ln(\beta+N(n+1)-n\bar{y}_n-y_{n+1}-1).$$

As in the frequentist binomial case, there is no closed form solution for $\hat{y}_{n+1}$. As before, we set $N=30$ and $n=10$ and used values of $\bar{y}$ ranging from .5 to 30 in increments of .5. For each value of $\bar{y}$ we plotted (34) for $y_{n+1}$ ranging from zero to 30 in steps of size 1. The resulting surface is plotted in Fig. 3. It is seen that for each $\bar{y}$ the height of the joint likelihood is concave and there is a unique maximal value of $y_{n+1}$ that we take as $\hat{y}_{n+1}$.

We plotted the Bayes analog of Fig. 2 and it showed the same feature, namely the curvature does not increase with $n$. So, again, we do not get a stronger maximum as sample size increases.

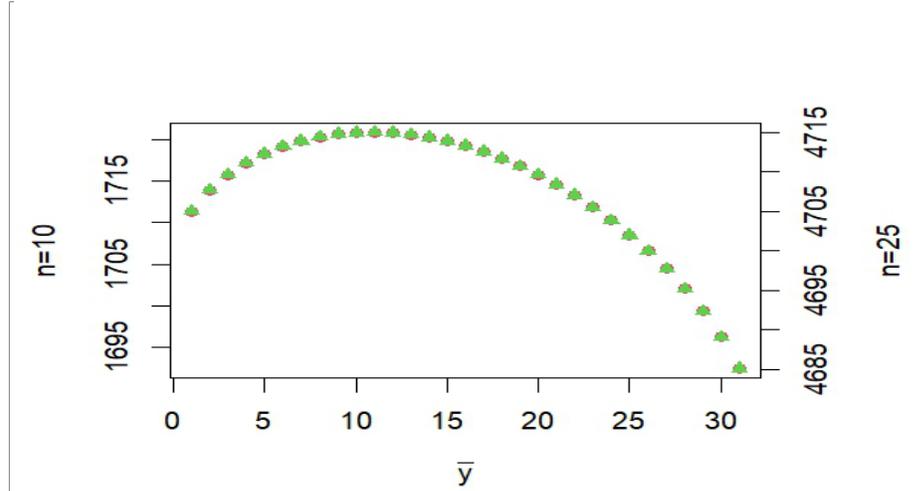

FIG 2. *For $\bar{y}_n = 10$ and $n = 10, 25$ we plot the loglikelihood for $y_{n+1}$. The maximum at about 11 indicates optimal value for $\hat{y}_{n+1}$. The red dots are for $n = 10$ and the green dots are for $n = 25$. The are equal to within the resolution of the image file.*

### 2.2.3. Exponential Distribution

This is a surprisingly simple example that gives an intuitive result that is – counter-intuitively – not useful. For $n$ outcomes we have the density

$$p(y^n|\theta) = \theta^n e^{-\theta \sum_{i=1}^n y_i}. \tag{35}$$

The MLE is $\hat{\theta}_{MLE} = 1/\bar{y}_n$ and the maximized likelihood is

$$p(y^n|\hat{\theta}_{MLE}) = \left(\frac{1}{\bar{y}_n}\right)^n e^{-\frac{1}{\bar{y}_n} n \bar{y}_n} = \left(\frac{1}{\bar{y}_n}\right)^n e^{-n}. \tag{36}$$

Now, for $n$ samples $y_1, y_2, \cdots, y_n$ and $y_{n+1}$, we can use (9) and rewrite (36) as:

$$p(y^{n+1}|\hat{\theta}_{MLE}) = \left(\frac{1}{\frac{n\bar{y}_n}{n+1} + \frac{y_{n+1}}{n+1}}\right)^{n+1} e^{-n+1}. \tag{37}$$

It is now easy to see that maximizing over $y_{n+1}$ for fixed $\bar{y}$ gives $\hat{y}_{n+1} = 0$. That is, since the exponential has a mode at zero and most of the probability piles up there it is optimal to predict zero all the time.

The Bayes case is similar. Let us use the conjugate prior

$$w(\theta|\alpha, \beta) = \frac{\beta^\alpha}{\Gamma(\alpha)} \theta^{\alpha-1} e^{-\beta\theta}. \tag{38}$$

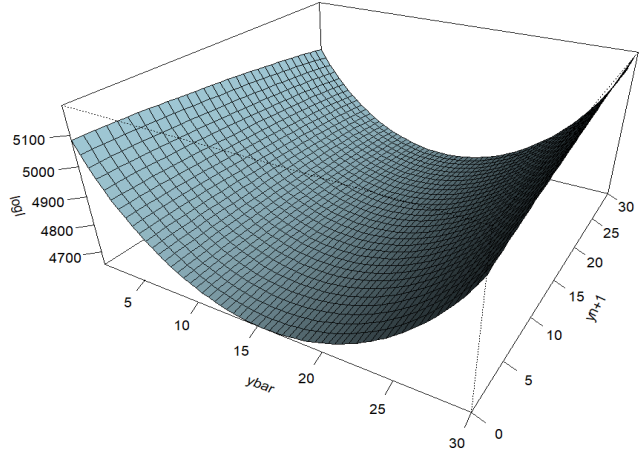

Fɪɢ 3. *Perspective plot of $\bar{y}$ and $y_{n+1}$ for $N = 30$ and $n = 10$. The optimal $\hat{y}_{n+1}$ for a given $\bar{y}$ is the maximum on the surface across the line $\bar{y}$. Note that as $\bar{y}$ moves from zero to 30 the location of the maximum moves from zero to 30 as well.*

Then,

$$p(y^n|\theta)w(\theta) \propto \theta^{n+\alpha-1}e^{-\sum_{i=1}^{n}y_i+\beta} \tag{39}$$

and the posterior distribution of $\theta|y^n, \alpha, \beta$ is Gamma with parameters $(n + \alpha, \sum_{i=1}^{n}y_i + \beta)$. Hence,

$$\hat{\theta}_{MPLE} = \frac{n+\alpha-1}{\sum_{i=1}^{n}y_i+\beta}. \tag{40}$$

Now,

$$
\begin{aligned}
p(y^n|\hat{\theta}_{MPLE})w(\hat{\theta}_{MPLE}) &= \frac{\beta^\alpha}{\Gamma(\alpha)}\left(\frac{n+\alpha-1}{\sum_{i=1}^{n}y_i+\beta}\right)^{n+\alpha-1} \\
&\quad \times e^{-(\sum_{i=1}^{n}y_i+\beta)\frac{n+\alpha-1}{\sum_{i=1}^{n}y_i+\beta}} \\
&= \frac{\beta^\alpha}{\Gamma(\alpha)}(n+\alpha-1)^{n+\alpha-1}e^{-(n+\alpha-1)} \\
&\quad \times \left(\frac{1}{\sum_{i=1}^{n}y_i+\beta}\right)^{n+\alpha-1}. 
\end{aligned}
\tag{41}
$$

Hence,

$$p(y^{n+1}|\hat{\theta}_{MPLE})w(\hat{\theta}_{MPLE}) \propto \left(\frac{1}{\sum_{i=1}^{n}y_i+y_{n+1}+\beta}\right)^{n+\alpha}. \tag{42}$$

Again, (42) is maximised at $y_{n+1} = 0$ making the predictor $\hat{y}_{n+1} \equiv 0$. This is independent of the choice of conjugate prior but may be the typical case.

### 2.2.4. Gamma Distributions

The $Gamma(\alpha, \theta)$ is a generalization of the exponential family given by

$$p_\alpha(y^n|\theta) \quad = \quad \frac{\theta^{n\alpha}}{\{\Gamma(\alpha)^n\}}\left\{\prod_{i=1}^n y_i^{\alpha-1}\right\}e^{-\theta\sum_{i=1}^n y_i}. \tag{43}$$

to predict $Y_{n+1}$. For any fixed $\alpha$, we can find the Shtarkov predictor $\hat{y}_{n+1}$ optimizing over $\theta$ in the frequentist case. The impediment to optimizing over $\alpha$ as well is that differentiation brings in the digamma functions that are difficult to use. The Bayes case has the same problem and working with the conjugate prior for the joint parameter $(\alpha, \theta)$ is also difficult. So, here, we treat $\alpha$ only as an index not as a parameter.

In the frequentist case, taking logarithms on both sides of (43) gives

$$\ln p_\alpha(y^n|\theta) \quad = \quad n\alpha\ln\theta - \theta\sum_{i=1}^n y_i + \ln\frac{\{\prod_{i=1}^n y_i^{\alpha-1}\}}{\Gamma(\alpha)^n}. \tag{44}$$

The MLE is $\hat{\theta}_{MLE} = \alpha/\bar{y}_n$, Using this for $n+1$ outcomes gives

$$\begin{aligned}
p_\alpha(y^{n+1}|\hat{\theta}_{MLE}) \quad &= \quad \left(\frac{\alpha}{\bar{y}_{n+1}}\right)^{(n+1)\alpha}e^{-\frac{\alpha}{\bar{y}_{n+1}}(n+1)\bar{y}_{n+1}}\left\{\prod_{i=1}^n y_i^{\alpha-1}\right\}y_{n+1}^{\alpha-1} \\
&\propto \quad \frac{y_{n+1}^{\alpha-1}}{(n\bar{y}_n + y_{n+1})^{(n+1)\alpha}}. 
\end{aligned} \tag{45}$$

This leads to

$$\hat{y}_{n+1} \quad = \quad \frac{n(\alpha-1)\bar{y}_n}{n\alpha+1} \tag{46}$$

as our Shtarkov predictor, see Appendix B.3. Note that for $\alpha = 1$ we recover the Shtarkov predictor for the exponential family.

Turning to the Bayes case, let us choose the prior for $\theta$ to be

$$w(\theta|\alpha_0, \beta_0) \quad = \quad \frac{\beta_0^{\alpha_0}}{\Gamma(\alpha_0)}\theta^{\alpha_0-1}e^{-\beta_0\theta}. \tag{47}$$

Then the joint prior and likelihood is

$$\begin{aligned}
p(y^n|\theta, \alpha) \times w(\theta|\alpha_0, \beta_0) \quad &= \quad \frac{\beta_0^{\alpha_0}}{\Gamma(\alpha_0)}\prod_{i=1}^n y_i^{\alpha-1}\frac{1}{\Gamma(\alpha)^n}\theta^{n\alpha+\alpha_0-1}e^{-\theta(n\bar{y}_n+\beta_0)} \\
&\propto \quad \theta^{n\alpha+\alpha_0-1}e^{-\theta(n\bar{y}_n+\beta_0)}. 
\end{aligned} \tag{48}$$

So, the posterior distribution of $\theta|y^n, \alpha, \alpha_0, \beta_0$ is a Gamma distribution with parameters $(n\alpha + \alpha_0, n\bar{y}_n + \beta_0)$. Hence,

$$\hat{\theta}_{MPLE} \quad = \quad \frac{n\alpha + \alpha_0 - 1}{n\bar{y}_n + \beta_0}. \tag{49}$$

Now, differentiating the log joint prior and likelihood, setting equal to zero and re-arranging gives

$$\hat{y}_{n+1} = \frac{(\alpha - 1)(\beta_0 + n\bar{y}_n)}{n\alpha + \alpha_0}. \tag{50}$$

Now, (i) for $\alpha > 1$, (48) is maximized by (50); (ii) for $\alpha = 1$, (48) is maximized by $\hat{y}_{n+1} = 0$; and (iii) for $\alpha < 1$ (48) is maximized by $\hat{y}_{n+1} = 0$ provided $n$ is large enough; see Appendix B.3.

### *2.3. Other Cases*

The examples worked out here for the binomial generalize to the multinomial. Indeed, a lot of work has been done in this case because the computing becomes much more difficult especially outside the case of conjugate priors. For details and further work, see [10] Chap. 3, [44] [29], and [31]. In particular, [5] focuses on general computational solutions.

The Shtarkov solution i.e., normalized maximum likelihood, comes up frequently because it defines a code length that can be used for many statistical purposes including model selection, see [24] and [48], as well as prediction, see [30] amongst others.

## 3. Bayesian Methods

Philosophically, Bayesian methods fit well with streaming data problems because Bayesians condition on past data and have a concept of updating upon receipt of more data. Moreover, as a matter of practice, they treat collected data as deterministic, i.e., as if they no longer have any stochastic properties. That is, the variability in the data is 'transmuted' into variability of an estimand or a future value $Y_{n+1}$ that follows a conditional distribution given the data. So, strictly speaking, Bayesians only need a likelihood to form posterior quantities. It is desirable that the likelihood come from a model but this is not necessary, as will be explained shortly.

By contrast, frequentist inferences rest on the sampling distribution which is derived from a probability model for the DG. So, the frequentist regards data as outcomes of a random variable that can in principle be repeatedly sampled. In $\mathcal{M}$-open problems there is no probability model and hence no sampling distribution so it is unclear how frequentist methods can be applied at all.

Since Bayesians do not use a sampling distribution as a description for post-experimental variability, they do not have to put a distribution on the data. This is consistent with assuming the data are $\mathcal{M}$-open.

The Bayesian tendency to conceptualize received data as fixed rests on the interpretation of conditional probability. Specifically, [15] presents results such as posterior normality in a way that shows how the convergences needed can be regarded as nonprobabilistic. That is, to characterize the behavior of posterior quantities, it is enough if the data sets form a well-defined deterministic sequence rather than having any distributional properties. In essence, this means that Bayesian inference can be regarded as simply an input-output relation: deterministic data in, probabilistic inferences out. There is no contradiction in regarding the posterior distribution as simply a set function on the parameter space indexed by the past data regarded as deterministic inputs.

Quoting [15]: 'our asymptotic results are nonprobabilistic in nature; that is, the data sets $\mathcal{D}_n$ are not necessarily related across $n$, but simply form a well-defined deterministic sequence.' In short, much like our treatment of $\mathcal{M}$-open data in Sec. 2, Bayesian inferences only depend on having a string of data. No distributional properties of the sequence are required to characterize the behavior of posterior quantities. [15] also states that convergences in this deterministic sense can be improved to probabilistic convergences if the assumptions are taken as probabilistic as well.

As a final philosophical point, one could regard frequentist procedures as an input-output relation from a string of data to a sampling distribution. The problem then is that this is just not consistent with the frequentist view that the likelihood is actually a probabilistic model under repeated sampling. There is no frequentist analog to [15].

To use this 'deterministic' property of Bayesian methods in a streaming context, ideally we would like to avoid convergence of, say, the posterior predictive to a limit under stochastic assumptions, IID being the simplest. The reason is that if the posterior predictive converges to a specific density e.g., $p(y_{n+1}|y^n) \to p(y_{n+1}|\theta_0)$ in some mode under some set of regularity conditions as $n$ increases, it's as if we are saying that, even apart from the mode, the $\mathcal{M}$-open data eventually have a distribution and hence are not truly $\mathcal{M}$-Open. More to the point, no theorem states that convergence of the posterior predictive to a limiting distribution means that the data comes from that distribution. So, this is simply an intuitively reasonable desideratum. Unfortunately, the failure of a convergence such as a posterior predictive density does not imply a sequence of data is necessarily $\mathcal{M}$-open.

Given the foregoing, there are at least three ways to think about Bayesian prediction in streaming $\mathcal{M}$-Open problems that respect the lack of data distributions and possible concerns about convergences.

First, and easiest, is not to worry about it. Just let the data do it: if the data really are $\mathcal{M}$-open then there is no optimal predictor or 'true' distribution although one may be better than another. So, in principle, the posterior predictive, for instance, needn't converge anywhere and if it does, the convergence is irrelevant. As noted above, even if the posterior predictive did converge to a limit in some sense, this does not imply the data were generated by the limit. Relying on the data is a tidy approach the problem, if perhaps unsatisfying.

Second, for streaming data where there is no replication, there is no loss in thinking of each $Y_i \sim p_i(y_i)$ for a collection of $p_i$'s that are completely unrelated to each other. If the $p_i$'s were chosen probabilistically, e.g., independent and tied together with a prior, the problem would become $\mathcal{M}$-complete: there is a true model but it is essentially inaccessible to us.

Since this sort of 'model' is unimplementable, it may be worthwhile to interpret it by a more effective approximation. One possibility, suggested by [35], is to regard each $Y_i$ as having a random bias. That is, regard each $Y_i$ as being of the form $Y_i + A_i$ where the $A_i$'s form a sequence of independent random bias terms and we only observe the sum $y_i + a_i$. This is still $\mathcal{M}$-open because the data are. It will only be our analytic techniques that makes use of the random bias. The hope is that the flexibility added to the predictor by the bias will mimic the behavior of $\mathcal{M}$-open data thereby reducing the error. Otherwise put, reducing the inferential power of the predictor by including the random bias may 'fool' the predictor into being a better approximation of an optimal predictor for data that do not have a distribution.

Third, as an alternative to using a random bias, we can prevent the posterior distribution from concentrating at a limit point, if desired, by ensuring its variance variance does not go to zero. That is, with the interpretation from [15] for, say, the posterior convergence in mind, if the posterior itself converges to a nontrivial distribution, or simply does not converge at all, then the posterior predictive distribution will not concentrate at a member of the models used to form it. Indeed, the posterior, and hence posterior predictive, may converge somewhere but it will not have the same meaning as in $\mathcal{M}$-complete or -closed cases. This alternative is difficult to formulate in practice even though it may be the most realistic. Hence, here we only think about Bayes predictors in the $\mathcal{M}$-open context in the first two ways.

The remaining problems with Bayesian techniques are mainly computational. Standard Bayes theory advocates conditioning on all the data even though there is evidence this can be suboptimal predictively. Nevertheless, we follow this here and note further that as data accumulate, in some schemes like Gaussian process priors (GPP's), it is impossible to reduce the data to a small set of sufficient or nearly sufficient statistics and the number of parameters grows with $n$. Moreover, since running time is important, we will impose a 'one-pass' constraint effectively limiting the amount of data we use to form a posterior predictive helping to ensure it does not converge. Details on this are in Sec. 6.

In this section we will give three nonparametric Bayes predictors. One comes from using zero mean GPP's for function estimation; the second is similar but the GPP's has a mean given by a random bias term; and the third simply uses a Dirichlet process prior on distributions. All of these are relatively familiar but now they are being used in an $\mathcal{M}$-open context.

### 3.1. Gaussian Process Priors

We start with our two GPP based predictors. The posterior predictive distribution for a new function value under a mean zero GPP is well-known and we simply quote it here. In the case that the GPP has a random additive bias we state what the posterior predictive density is and identify one way to obtain values for the hyperparameters to specify the predictor fully.

### 3.1.1. No Bias

Given a stream of data $Y_i$, for $i = 1, \ldots, n$, the idea is to assume $Y_i = f_i + \epsilon_i$, $i = 1, \cdots, n$ where the $\epsilon_i$'s are IID $N(0, \sigma^2)$. Now, the $i^{th}$ data point $y_i$ is an outcome of $Y_i$ and we can equip $f = (f_1, f_2, \cdots, f_n)^T$ with a Gaussian process prior. That is,

$$f \sim \mathcal{N}(a, \sigma^2 K_{11}), \tag{51}$$

where $a = (a_1, a_2, \cdots, a_n)^T$ and $K_{11} = \left( \left( k_{ij} \right) \right)$ for $i, j = 1, \cdots, n$. In this subsection, we assume the means $a_i = 0$ for all $i$. So, letting $\epsilon = (\epsilon_1, \ldots, \epsilon_n)^T$, the joint distribution of $Y = (Y_1, Y_2, \cdots, Y_n)^T$ and $Y_{n+1}$ is

$$
\begin{aligned}
\begin{pmatrix} Y \\ \vdots \\ Y_{n+1} \end{pmatrix} &= \begin{pmatrix} f \\ \vdots \\ f_{n+1} \end{pmatrix} + \begin{pmatrix} \epsilon \\ \vdots \\ \epsilon_{n+1} \end{pmatrix} \\
&\sim \mathcal{N} \left[ \begin{pmatrix} 0 \\ \vdots \\ 0 \end{pmatrix}, \sigma^2 \begin{pmatrix} K_{11} + I & \vdots & K_{12} \\ \cdots & \vdots & \cdots \\ K_{21} & \vdots & K_{22} + 1 \end{pmatrix} \right]
\end{aligned} \tag{52}
$$

where $K_{12} = (k_{1,n+1}, k_{2,n+1}, \cdots, k_{n,n})^T$, $K_{21} = K_{12}^T$, and $K_{22} = k_{n+1,n+1}$. Compactly, we write (52) as

$$Y^{n+1} \sim \mathcal{N}(0^{n+1}, \sigma^2(I + K)_{n+1 \times n+1}). \tag{53}$$

It is well known from normal theory that the predictive distribution of $Y_{n+1}$ given $Y^n$ is the conditional

$$Y_{n+1} | Y^n \quad \sim \quad \mathcal{N}(\mu^*, \Sigma^*),$$

where

$$\mu^* = \sigma^2 K_{12} \{\sigma^2(K_{11} + I)\}^{-1} y = K_{12} \{(K_{11} + I)\}^{-1} y \tag{54}$$

and

$$
\begin{aligned}
\Sigma^* &= \sigma^2(K_{22} + 1) - K_{21} \{\sigma^2(K_{11} + I)\}^{-1} \sigma^2 K_{12} \\
&= \sigma^2(K_{22} + 1) - K_{21}(K_{11} + I)^{-1} K_{12}.
\end{aligned} \tag{55}
$$

In the zero bias case of all $a_i = 0$, the optimal point predictor (under squared error loss for instance) is simply the conditional mean $\mu^*$ in (54).

To complete the specification, it remains to estimate $\sigma^2$. From (53), using $n$ in place of $n+1$ gives $Y^n \sim \mathcal{N}(0, \sigma^2(I + K)_{n \times n})$. Hence, $(I + K)^{\frac{1}{2}} Y^n \sim \mathcal{N}(0, \sigma^2 I)$. Define $Y' = (I + K)^{\frac{1}{2}} Y^n$. Since the sample variance is an unbiased estimate of population variance, we estimate $\sigma^2$ by $S'_2 = \frac{1}{n-1} \sum_{i=1}^{n} (y'_i - \bar{y'})^2$.

### 3.1.2. *IID Random Additive Bias*

The suggestion for the technique in this subsection comes from [35]. From (52), we see that

$$Y = Y^n \sim \mathcal{N}(a, \sigma^2(I_{n \times n} + K_{n \times n})).$$

So, the likelihood is given by

$$
\begin{aligned}
\mathcal{L}_1(a, \sigma^2 | y) &= \mathcal{N}(a, \sigma^2(I_{n \times n} + K_{n \times n}))(y) \\
&= \frac{e^{-\frac{1}{2\sigma^2}(y-a)'(I_{n \times n} + K_{n \times n})^{-1}(y-a)}}{(2\pi)^{\frac{n}{2}}(\sigma^2)^{\frac{n}{2}}|I_{n \times n} + K_{n \times n}|^{\frac{1}{2}}}.
\end{aligned}
\tag{56}
$$

It remains to choose priors for $a$ and $\sigma^2$. First, we equip $a = (a_1, \ldots, a_n)^T$ with the distribution

$$a \sim \mathcal{N}(\gamma \mathbf{1}_n, \sigma^2 \delta^2 I_{n \times n}), \tag{57}$$

where $\mathbf{1}_n$ denotes the vector of ones of length $n$. We treat $\gamma$ as a hyperparameter representing the mean of the IID random bias in the $Y_i$'s and use an empirical Bayes approach to obtain a serviceable value for it. Also, as discussed below, we treat $\delta^2 > 0$ as a deterministic parameter chosen for convenience. To complete the specification of the model we equip $\sigma^2$ with an inverse-Gamma distribution

$$\sigma^2 \sim \mathcal{IG}(\alpha, \beta). \tag{58}$$

Below, we will also use an empirical Bayes approach to find serviceable values for $\alpha$ and $\beta$. Thus, explicitly, our joint prior is

$$w(a, \sigma^2 | \alpha, \beta, \gamma, \delta) = \mathcal{N}(\gamma 1, \sigma^2 \delta^2 I_{n \times n}) \ \mathcal{IG}(\alpha, \beta)$$

$$= \frac{e^{-\frac{1}{2\sigma^2}(a - \gamma 1)'(\delta^2 I_{n \times n})^{-1}(a - \gamma 1)}}{(2\pi)^{\frac{n}{2}}(\sigma^2 \delta^2)^{\frac{n}{2}}} \frac{\beta^\alpha}{\Gamma(\alpha)} \times \left(\frac{1}{\sigma^2}\right)^{\alpha+1} e^{-\frac{\beta}{\sigma^2}}. \tag{59}$$

The following theorem gives a predictive distribution for $Y_{n+1}$.

**Theorem 3.1.** *The posterior predictive distribution of a future observation $Y_{n+1}$ given the past observations $y^n$ is*

$$m(y_{n+1} | y^n) = St_{2\alpha+n}\left(A_1, \frac{\beta^*}{\frac{2\alpha+n}{2}}\right)(y_{n+1}). \tag{60}$$

*In* (60), *$St_v(\theta, \Sigma)$ denotes the Student's t distribution with $v$ degrees of freedom, location parameter $\theta$ and scale parameter $\Sigma$.*

*The form of $\beta^*$ is given by $\beta^* = \beta + A_2$ and the form of $A_1$ is given by $A_1 = \frac{g_2 - (y^n)^T g_1^n}{g_1}$ where, $g_1^n$, $g_1$, $g_2$, and $A_2$ can be explicitly written as functions of $y^n$, $\gamma$, $\delta$, and the $(n+1) \times (n+1)$ variance matrix $K$.*

For a proof of this result, see [13].

It is immediate from (60) that $A_1$ is the appropriate point predictor for $Y_{n+1}$. Indeed, the main effect of the random bias on point prediction is to make the location dependent on the data in a nonlinear way, cf. (54). Moreover, even though it may seem inconsistent to $Y_{n+1}$ when we assume it doesn't have one, the posterior predictive is only a statement of our belief for the location – not a statement about the actual behavior of $Y_{n+1}$.

To estimate $\alpha$, $\beta$, $\gamma$, and $\delta$ we first start with $\gamma$ and adopt a maximum likelihood approach. We first write the joint likelihood of $y^n$ and $a^n$ given $\gamma$, $\delta^2$ and $\sigma^2$. Then, we integrate out the $a^n$ and write the result as product of a function of $\gamma$ and a function of the other parameters that can be ignored. More explicitly, $\mathcal{L}(y^n, a^n | \sigma^2, \delta^2, \gamma)$ equals

$$\frac{1}{(2\pi)^{\frac{n}{2}} (\sigma^2)^{\frac{n}{2}} |I_{n \times n} + K_{n \times n}|^{\frac{1}{2}}} e^{-\frac{1}{2\sigma^2} (y^n - a^n)' (I_{n \times n} + K_{n \times n})^{-1} (y^n - a^n)}$$

$$\times \frac{1}{(2\pi)^{\frac{n}{2}} (\sigma^2 \delta^2)^{\frac{n}{2}}} e^{-\frac{1}{2\sigma^2} (a^n - \gamma 1^n)' (\delta^2 I_{n \times n})^{-1} (a^n - \gamma 1^n)}.$$

Integrating out over $a^n$ and re-arranging gives

$$\begin{aligned}
\mathcal{L}_2(y^n | \gamma, \delta^2, \sigma^2) &= h(\gamma) \frac{|V_{n \times n}|^{\frac{1}{2}}}{(2\pi \sigma^2 \delta^2)^{\frac{n}{2}} |(I + K)_{n \times n}|^{\frac{1}{2}}} \\
&\quad \times e^{-\frac{1}{2\sigma^2} \left[ y'^n \left\{ (I+K)_{n \times n}^{-1} + (I+K)_{n \times n}^{-1} V_{n \times n} (I+K)_{n \times n}^{-1} \right\} y^n \right]} \quad (61)
\end{aligned}$$

where

$$h(\gamma) = e^{-\frac{1}{2\sigma^2} \left[ -2\gamma y'^n \frac{(I+K)_{n \times n}^{-1} V_{n \times n}}{\delta^2} 1 + \gamma^2 1' \left( \frac{I}{\delta^2} - \frac{V_{n \times n}}{\delta^4} \right) 1^n \right]}$$

and the other factor on the right side of (61) can be discarded. Taking logarithms on both sides of $h$ gives

$$\ln h(\gamma) = -\frac{1}{2\sigma^2} \left[ -2\gamma y'^n \frac{(I+K)_{n \times n}^{-1} V_{n \times n}}{\delta^2} 1^n + \frac{\gamma^2}{\delta^2} 1'^n \left( I_{n \times n} - \frac{V_{n \times n}}{\delta^2} \right) 1^n \right]. \quad (62)$$

Differentiating (62) with respect to $\gamma$, and equating it to zero leads to

$$\hat{\gamma} = \frac{y'^n (I+K)_{n \times n}^{-1} V_{n \times n} 1^n}{1'^n \left( I_{n \times n} - \frac{V_{n \times n}}{\delta^2} \right) 1^n}.$$

A second derivative argument gives that $\hat{\gamma}$ is typically a local minimum, at least for small $\delta > 0$; see [13].

Next, we find effective values for $\alpha$ and $\beta$. We do this for the sake of completeness because $\alpha$ and $\beta$ are only required to identify the posterior predictive from Theorem 3.1; they are not needed to identify the point predictor $A_1$ for $Y_{n+1}$. Our procedure rests on a method of moments argument.

Recall from (56) that $Y \sim \mathcal{N}(a, \sigma^2 (I + K)_{n \times n})$. Hence, $(I + K)^{\frac{1}{2}} Y \sim \mathcal{N}(a, \sigma^2 I)$. So, if we define $Y' = (I + K)^{\frac{1}{2}} Y$ we can set $S_2' = \frac{1}{n-1} \sum_{i=1}^{n} (y_i' - \overline{y'})^2$. We use the first two moments of $S_2'$ to find values $\hat{\alpha}$ and $\hat{\beta}$ for given $\delta > 0$. [1] In fact, we have

$$\sigma^2 = E \frac{S_2'}{1 + \delta^2}, \tag{63}$$

see (140) in Appendix C.1; the argument giving (63) is different from the argument for $S_2'$ at the end of Subsec. 3.1.1. So, for given $\delta > 0$, we use

$$\hat{\sigma}^2 = \frac{S_2'}{1 + \delta^2}$$

in our first moment condition based on $S_2'$, because it is unbiased.

Since we have two hyperparameters $\alpha$ and $\beta$, we need the second moment of $\hat{\sigma}^2$ as well. From (147), in Appendix C.1 we have that

$$Var(\hat{\sigma}^2) = \frac{2\sigma^2}{(n-1)^2} \left[ \sigma^2 + \frac{2n\gamma^2}{1 + \delta^2} \right]. \tag{64}$$

So, we can approximate (64) as

$$\widehat{Var}(\hat{\sigma}^2) = \frac{2\hat{\sigma}^2}{(n-1)^2} \left[ \hat{\sigma}^2 + \frac{2n\gamma^2}{1 + \delta^2} \right].$$

To finish this specification, recall (58). For the inverse gamma we have

$$\hat{\sigma}^2 \approx E(\sigma^2) = \frac{\beta}{\alpha - 1}$$

$$\widehat{Var}(\hat{\sigma}^2) \approx Var(\sigma^2) = \frac{\beta^2}{(\alpha - 1)^2 (\alpha - 2)}.$$

From these equations we can solve solve for $\alpha$ and $\beta$ to find

$$\hat{\alpha} \approx \frac{E^2(\hat{\sigma}^2)}{Var(\hat{\sigma}^2)} + 2 \tag{65}$$

$$\hat{\beta} \approx E(\hat{\sigma}^2)(\hat{\alpha} - 1). \tag{66}$$

Now, by plugging in estimates for the expectations we get values for $\hat{\alpha}$ and $\hat{\beta}$.

Finding a good value for $\delta$ is more problematic. One way to find a value would be by maximum likelihood: find a version of $\mathcal{L}$ analogous to (61), but with only one factor being a function of $\delta^2$. That is, form the likelihood $\mathcal{L}_3(y|\gamma, \delta^2, \sigma^2)$ by integrating out $a$ from the product of (56) and (57) and maximizing it over $\delta$.

---

[1] For contrast, at the end of Subsec. 3.1.1, we could estimate $\sigma$ by using $\hat{\sigma}^2 = S_2'$ because $\sigma^2 = E S_2'$. This does not hold here because of the random bias.

Unfortunately, we cannot simply differentiate $g(\delta^2)$, set the derivative to zero, and solve. The resulting equations are just too complicated to be useful.

In principle, one could do a grid search over $\delta^2$ in an interval $\mathbf{I} \subset \mathbb{R}^+$ to maximize $\mathcal{L}_3(y|\gamma, \delta^2, \sigma^2)$. However, in computational work not shown here, we found that the optimal $\delta \in \mathbf{I}$ was almost always the left hand end point, even as $\mathbf{I}$ moved closer and closer to 0. Hence, pragmatically, in our computations here, we simply chose $\delta$ to be small. Note that if $\delta = 0$, the prior on $a$ is degenerate and in fact $\gamma = 0$ as well. This means that the mean and variance of our bias $a$ is zero i.e., there is no bias. Practical choices of $\delta$ in our examples were around .1 so that variability in the prior would not overwhelm the data. To compensate, we did robustness analyses to verify the stability of our results using values of $\delta$ as large as 1.

### 3.1.3. INID Random Additive Bias

For comparison with the earlier two examples using GPP's, consider the case where the distribution of $a^n$ is independent but not identical. i.e.,

$$a^n \sim \mathcal{N}(\gamma^n, \sigma^2 \delta^2 I_{n \times n}), \tag{67}$$

where $\gamma^n = (\gamma_1, \ldots, \gamma_n)^T \in \mathbb{R}^n$. This means that at the $n+1$ stage when we want to predict $Y_{n+1}$, we have an extra location parameter $\gamma_{n+1}$ in addition to the first $n$ parameters in $\gamma^n$. Theorem 3.1 can be formally extended to this case. We have the following.

**Theorem 3.2.** *The posterior predictive distribution of the future observation $Y_{n+1}$ given the past observations $y^n$ is*

$$m(y_{n+1}|y^n) = St_{2\alpha+n}\left(A_1^*, \frac{\beta^{**}}{\frac{2\alpha+n}{2}}\right)(y_{n+1}). \tag{68}$$

*In* (68), *the form of $\beta^{**}$ is given by $\beta^{**} = \beta + A_2^*$ and the form of $A_1^*$ is given by $A_1^* = \frac{g_2^* - y'^n g_1^{*n}}{g_1^*}$ where $g_1^{n*}$, $g_1^*$, $g_2^*$, and $A_2^*$ can be explicitly written as functions of $y^n$, $\gamma^{n+1}$, $\delta$, and the $(n+1) \times (n+1)$ variance matrix $K$.*

*Proof.* A proof of this result follows by replacing $\gamma$ in the proof of Theorem 3.1 with $\gamma^n$ and doing a line-by-line verification. A sketch of the proof can be found in Appendix C.2. □

It is problematic to use Theorem 3.2 in practice. To see the impediments, consider finding values of the parameters.

Suppose we seek a value for $\gamma^n$, the first $n$ biases. If we use the analog of (61) from the IID case for the present INID case, we get a more general form of the function $h$. Taking logarithms on both sides of $h$ gives an analog of (62):

$$\ln h(\gamma^n) = -\frac{1}{2\sigma^2}\left[-\frac{2}{\delta^2}y'^n(I+K)_{n \times n}^{-1}V_{n \times n}\gamma^n + \gamma'^n\left(\frac{I_{n \times n}}{\delta^2} - \frac{V_{n \times n}}{\delta^4}\right)\gamma^n\right].$$

Differentiating this w.r.t. $\gamma^n$ and setting the derivative to zero gives

$$\hat{\gamma}^n = \left(I_{n \times n} - \frac{V_{n \times n}}{\delta^2}\right)^{-1} (I + K)_{n \times n}^{-1} V_{n \times n} y^n. \tag{69}$$

A second derivative argument, essentially the same as before, gives that $\hat{\gamma}^n$ is again a maximum.

However, if we replace $n$ by $n+1$ in (69) we see that we need $y_{n+1}$ to estimate $\gamma_{n+1}$ or a value of $\gamma_{n+1}$ to predict $y_{n+1}$ and both are unknown. Hence, the non-identical biases make it impossible to predict $y_{n+1}$. One way around this is to assign a value to $\gamma_{n+1}$ by some other technique; we have only explored this in one case namely taking $\gamma_{n+1} = \bar{\gamma}_n$.

A few further comments about the form of $A_1^*$ and $\beta^{**}$. First, $A_1^*$ involves $\gamma^{n+1}$ only because of $g_2^*$; $g_1^{n*}$ and $g_1^*$ do not involve $\gamma^{n+1}$. Second, only $g_2^*$ and $A_2^*$ involve $\gamma^{n+1}$. Third, similar to Subsubsec. 3.1.2, we can use a method of moments argument to find $\hat{\alpha}$ and $\hat{\beta}$. The details are given in Appendix C.3. Thus, even though we have a form for the predictive distribution of $Y_{n+1}|y^n$, in practice we are not able to use it without extra information.

### 3.2. *Dirichlet*

Suppose a discrete prior $G$ is distributed according to a Dirichlet Process (DPP), and write $G \sim DP(\alpha, G_0)$ where $\alpha$ is the mass concentration parameter (that we take to be one) and $G_0$ is the base measure with $\mathbb{E}(F) = G_0$. Then, by construction, we have the following standard results; see [23].

If the sample space $\mathbb{R}$ is partitioned into $A_1, A_2, \cdots, A_k$, the random vector of probabilities $G(A_1), G(A_2), \cdots, G(A_k)$ has a Dirichlet distribution, i.e.,

$$p(G(A_1), G(A_2), \cdots, G(A_k)) \sim Dirichlet(\alpha(A_1), \alpha(A_2), \cdots, \alpha(A_k)),$$

where $\alpha(\mathbb{R}) = M$, which we take here to be one.

By conjugacy, the posterior distribution of

$$G(A_1), G(A_2), \cdots, G(A_k)|Y_1, Y_2, \cdots, Y_n$$

is also Dirichlet but with parameters $\alpha(A_j) + n_j$ where

$$n_j = \sum_{i=1}^{n} I(Y_i \in A_j); j = 1, 2, \ldots, k.$$

If $Y_j^{'}; j = 1, 2, \cdots, k$ are the distinct observations in $\{Y_i; i = 1, 2, \cdots, n\}$, the posterior predictive distribution of $Y_{n+1}|Y_1, Y_2, \cdots, Y_n$ is

$$Y_{n+1}|Y_1, Y_2, \cdots, Y_n = \begin{cases} \delta_{Y_j^{'}}, \text{with probability} \\ \frac{n_j}{M+n}; j = 1, 2, \cdots, \text{k; and} \\ F_0, \text{with probability } \frac{M}{M+n} \end{cases}.$$

Now, our DPP predictor is

$$\hat{Y}_{n+1} \;=\; \sum_{j=1}^{k} y'_j \frac{n_j}{M+n} + \frac{M}{M+n}\mathsf{median}(F_0). \tag{70}$$

## 4. A Computer Science View of Data Streams

Computer science has developed its own terminology and techniques for streaming data. The algorithms are designed for rapidly arriving, massive data that cannot be stored as is. Hence, both running time and storage bounds have to be met. Here is one representative example. It is estimated that there are several hundred million routers in the world at this time of writing. Each sends and receives IP packets; contemporary routers handle a few million IP packets per second. This amounts to over a terabyte of data per day per router. To be useful, statistical analysis of the traffic through routers has to be done in real time or nearly real time and scale to the volume of data.

The generic setting for analyzing streaming data presumes discrete data arriving at discrete time points, that, as far as possible, should only be looked at once. Each data point – called a 'token' – is assumed to be in a set of the form $[U] = \{1, \ldots, U\}$ for some $U \in \mathbb{N}$ where $U$ is very large; $U$ is sometimes called the universe. Thus, following standard computer science practice, we have a stream $\sigma = (y_1, \ldots, y_n, \ldots)$ simply regarded as a string of tokens without any stochastic properties. The overall goal is to extract whatever information we can from $\sigma$ by identifying whatever regularities it has. As a practical point, we would also like to enforce a storage bound i.e., an upper bound on random access working memory, of the form $\ln n + \ln U$ bits. Note that no relationship between the sizes of $U$ and $n$ can be assumed.

Despite regarding the data as non-stochastic, computer scientists often use probability modeling to design procedures for analyzing data streams. Sometimes probability is used to analyze the performance of a method, i.e., a method is strictly empirical and probabilities are used to assess its performance under various scenarios. This is much like what the statistician faces in fixed effects linear regression: The least squares criterion can be used to define an estimator for the regression parameters but its performance depends on the error structure assumed. For instance, assuming independent and identical measurement errors $\epsilon_i$'s on the $y_i$'s gives different inferences from assuming a serial dependence structure. In these cases, computer scientists, like statisticians, are thinking of $\mathcal{M}$-open data and considering how to impose probabilistic assumptions that make the problem $\mathcal{M}$–complete or -open. We will return to this sort of empiricism in Sec. 5.

Instead of assuming a probability structure on the data, probability is used to form the predictor or other object used for inference. That is, the computer scientist may randomly choose elements, such as hash functions defined below, from a set of elements and use the selection to form the object that will be used for inference. The techniques here mimic techniques from data compression in information theory that use random coding algorithms, usually to achieve some sort of data compression. This latter form of modeling – intended for $\mathcal{M}$-open problems – is the focus of this section. The material we present here is derived substantially from the expositions [11], [38], [39], and [12], amongst others.

While no treatment of the topic can be exhaustive, many commonly occurring problems in streaming data have been well-studied in computer science leading to a plethora of procedures. These problems have good solutions i.e., achieve satisfactory bounds on storage and running times, even if they are still open to development. Here are five amongst many:

1. Estimating the number of distinct elements in $\sigma$;
2. Estimating the probabilities of the distinct elements in $\sigma$;
3. Identifying the most commonly occurring elements in $\sigma$;
4. Estimating moments and quantiles of $\sigma$;
5. Finding a representative sample for $\sigma$.

Getting useful answers to the first two problems is the bare minimum needed for prediction. Problem 3) is an extension of problem 2). Problem 4) is a way to find predictors such as a moments or median values and a good solution to problem 5) is the natural way to permit the streaming usage of any pre-specified predictor: simply evaluate it on the representative subset, update the subset as more data is received, and predict again using the updated representative subset.

In this section, statistical terms like estimate, moments, and quantiles are understood to be in a streaming sense, not in the sense of having a fixed finite dimensional sample space. For instance, a streaming median is not defined in the usual way – any $\lambda$ satisfying $P(Y < \lambda) = P(Y > \lambda)$ or by some sort of stochastic process generalization of this – because it is allowed to evolve with the stream; see [12], Sec. 11, for an alternative treatment from a computer science perspective. Often, the computer science approach to streaming data has a Bayesian feel because, as will be seen, it is standard to put a probabilistic structure not on the data but on components of the objects used for inference. This is not Bayesian in any orthodox sense but is analogous to using a prior in hierarchical Bayes.

The rest of this section proceeds as follows. First, we give some of the formalities needed for the computer science treatment of streaming data. After that, we explain one well-known procedure, the $\mathsf{Count-Min}$ sketch, and give a numerical example since the thinking behind it will be unfamiliar to most statisticians. The output of the $\mathsf{Count-Min}$ sketch generates a DF that can be used for prediction. Finally, we develop a statistical view of the $\mathsf{Count-Min}$ sketch to see that it meshes well with established statistical treatments.

### *4.1. Some Formalities*

There are four important pre-requisite concepts to introduce.

First, we need the concept of a data structure: this is a formal way to organize, store, and present the data we have accumulated so we can compute with it.

Some of this amounts to making the data 'analysis ready', but this also includes ensuring the way the data are presented can be readily updated sequentially to generate a desired output.

Second, the typical algorithm for streaming data has a well defined data structure that goes through three steps. It begins with an initialization done before beginning to process $\sigma$, followed by a processing stage that iterates with each $y_i$, i.e., each time a token is received. Finally, there is an output that answers our question about the stream, whether it is a prediction, a decision, or response to some other 'query'. These algorithms are called streaming if they iteratively process a data stream $\sigma$ to provide their output in 'one pass' subject to a memory and running time constraint. Here, one pass means that the algorithm looks at each token in sequence once and never goes back. It updates its output from the data structure from time step to time step by summarizing the received data in a way that prevents the summary from growing too fast in size. There are multi-pass algorithms but they are less desirable.

Third, we need the concept of a sketch. Roughly, in computer science, a sketch is an algorithm that compresses the data stream so that some function of the stream can be effectively computed. More formally, a sketch is a data structure, say $DS$, that can be regarded as a function from each $\sigma_n = (y_1, \ldots, y_n)$ to an output at time $n$, say $y_n^*$, i.e., $DS : \sigma_n \to y_n^*$, with a 'concatenation' property. That is, if the concatenation of two streams $\sigma$ and $\tau$ is denoted $\sigma \circ \tau$, there is a space efficient algorithm, say COMB that can combine them so that

$$\mathsf{COMB}\left((DS(\sigma), DS(\tau)\right) = DS(\sigma \circ \tau).$$

That is, there is an efficient way to derive the sketch for the concatenated stream from the individual streams. Essentially, the $DS$ organizes the data in some way that lets the output of two streams be merged easily. There are many important sketches described in the references; more recently see, for example, [3] who developed sketches for classification in a Bayesian context.

Fourth, probabilistic hash functions are, arguably, the key mathematical quantities that make one pass sketches with storage and running time bounds effective for statistical tasks with streaming data. By definition, a hash function is any function that assigns fixed length values to its argument. In practice, a hash function typically maps a finite set of objects onto a smaller set of objects. As a result, hash functions are not one-to-one. On the other hand, they provide data compression: the lack of one-to-one-ness means that information in the range is strictly less than the information in the domain. The idea is that the information loss is not very big or not very important. Moreover, typically, sketches use many probabilistically chosen hash functions and this controls the information loss. The effectiveness of this form of data compression requires $n$ be very high compared to other inputs to the sketch such as $U$, $V$, the number of distinct objects in the stream (which must also be large), etc. These are the scenarios in which the benefits of data compression outweigh the cost of using multiple hash functions.

While there are good sketches for all five tasks listed in the first part of this section, we only exemplify them here by one technique for obtaining a streaming DF in the continuous case.

To begin, fix sets $[U]$ and $[V]$ with $U > V$ and let

$$\mathcal{H} \subseteq \{h : [U] \to [V]\}.$$

The class $\mathcal{H}$ is called a hash family and the elements of $\mathcal{H}$ are the hash functions. Clearly $\#(\mathcal{H}) \leq U^V$ and it would take at most $V \log U$ bits to encode all of $\mathcal{H}$. Let $W$ be a probability distribution on $\mathcal{H}$ with the property that $\forall u, u' \ \forall v, v'$: $u \neq u'$ implies

$$W(\{H(u) = v\} \cap \{H(u') = v'\}) = \frac{1}{V^2}, \tag{71}$$

where $H$ is the random variable with outcomes $h \in \mathcal{H}$. Formally, expression (71) is called (strongly) 2-universal because two conditions on $H$ are imposed [2]. Points at which a hash function $h$ is not one-to-one give 'collisions': a collision occurs for $h$ when there are $u, u' \in [U]$ so that $h(u) = h(u')$. Under (71), this happens with probability $W(H(u) = H(u')) = 1/V$ and controls the number of times we lose information about a value in $[U]$ by assigning the same $v$ to two different $u$'s. The smaller $V$ is, the more collisions will occur but the greater the data compression will be.

The point of using the functions in $\mathcal{H}$ is to give up one-to-oneness so the functions can be coded with less storage by allowing for a smaller $V$. To get around the resulting collisions, the sketches will use multiple hash functions and combine them. It is understood (but not examined here) that these hash-function based techniques are used in settings where the storage for multiple hash functions will be less than the storage for the single correct function.

Here, we only address some aspects of the second problem on the list at the start of this section, namely, estimating probabilities of distinct elements in a stream, for the sake of illustrating the main ideas. Essentially, we give a one pass sketch based on probabilistically chosen hash functions. The surprising point is we can extend this sketch to continuous data, obtain an estimated empirical distribution function, and use it to predict the next outcome.

### *4.2. Mechanics of the* Count-Min *Sketch*

In its most basic form, the **Count-Min** sketch is an effective way to estimate the frequencies of events in a stream when the number of distinct events is very large. Normalizing the estimated frequencies gives estimates of 'probabilities' without having to give a precise definition of probability in $\mathcal{M}$-open contexts. The normalized estimated frequencies can be used to form an *estimate* of the empirical DF (EDF), when the actual EDF would be ineffective to use, for instance when the range of the data is too large. **Count-Min** sketches are actually a class of procedures that give a provably good approximation to exact counts or frequencies of values under a storage bound.

---
[2]Strictly speaking, 2-universal only requires '$\leq 1/V^2$' in (71). However, here we are assuming that $\mathcal{H}$ contains all possible $h$'s so we get equality.

Aside from being practical in some settings, using the estimated EDF (EEDF) is reasonable from a principled standpoint. In $\mathcal{M}$-open problems, there is no defined population, let alone a DF. Indeed, for $\mathcal{M}$-open problems, neither the EDF nor the EEDF need converge to a limit so by using the EEDF we are tracking the EDF along the stream.

To see how this sketch makes this possible, it is worth going through its details and then giving a numerical example. The reason is that the use of sketches and data structures (together) is quite different from how statisticians use probability modeling for data even though the two approaches play analogous roles.

First, consider the naive approach of simply constructing an EDF. Start with a data stream $\sigma = (y_1, y_2, \ldots)$ assuming values in a finite set $[U]$, see [18]. Let $a(i) = (a_1(i), \ldots, a_U(i))$ be the number of occurrences of each $u = 1, \ldots, U$ up to time $n$. That is, for each $u \in [U]$ let

$$a_u(n) = \mathbf{card}\left(\{y_i|\ i \leq n \text{ and } y_i = u\}\right).$$

We update the vector $a(n)$ to $a(n+1)$ upon receipt of $y_{n+1}$ by incrementing its $u$-th coordinate by one. That is, for each $u$

$$a_u(n+1) = \begin{cases} a_u(n) + 1 & \text{if } y_{n+1} = u \\ a_u(n) & \text{if } y_{n+1} \neq u. \end{cases} \tag{72}$$

Obviously, $a(n)/n$ is a probability vector on $[U]$ at time $n$. So, $a(n)/n$ gives an EDF $\hat{F}_n$ on $[U]$.

Even though $\hat{F}_n$ is not an EDF or estimated DF for any random variable, it can be used to generate a point predictor for $y_{n+1}$: simply find the mean, median, or other location estimator it defines and choose the value of $[U]$ closest to it. This can be extended to continuous streams and to give PI's.

There are at least three problems with this approach if $U$ is very large. First, it can be inefficient if we insist on only using one pass procedures. Second, we want to control the storage. Third, we want to control the error. By using the data compression properties of hash functions we can resolve all three problems.

To begin the description of the Count-Min sketch, we assume $U$ is given and that $\epsilon > 0$ and $\delta > 0$ have been chosen; in Subsec. 4.3 they will be used to characterize bounds on the error of the procedure. For the moment, we set $V = \lceil 2/\epsilon \rceil$ and choose $d = \lceil \log(1/\delta) \rceil$ hash functions $h_1, \ldots, h_d$ independently at random from $\mathcal{H} = \{h : [U] \to [V]\}$ using a probability that satisfies (71).

Given these choices, we define a data structure for the Count part of the Count-Min sketch. For each $i$, we form a $d \times V$ matrix

$$C(i) = ((c_{jv}(i)))_{j=1,\ldots,d; v=1,\ldots,V}$$

that is initialized at zero for $i = 0$. For $i \geq 1$, we update $C(i-1)$ by setting each $c_{jv}(i) = \mathsf{Count}(j, h_j(y_i))$. The function Count updates here as $a_u$ does in (72) – but using $V$ in place of $U$ thereby allowing collisions. That is, the $(j, v)$ element of $C(i)$ is the the number of times the $j$-th hash function has assumed the value $v \in [V]$ on the elements of the sequence $y_1, \ldots, y_i$. Now we have a sequence of matrices $C(0)$, $C(1)$, and so on.

We sequentially apply the Min part of the Count-Min sketch to the $C(i)$'s. Specifically, given a count matrix $C(i)$, we choose the minimum entry over the $d$ elements in each of the $v$ columns. That is, for each $i$ and $v$ we find

$$m_v(i) = \min C(i)[v] = \min_{j=1}^{d} c_{jv}(i) = \min_{j=1}^{d} \mathsf{Count}(j, v), \tag{73}$$

the minimum of the $v$-th column $C(i)[v]$ of $C(i)$ where $v$ ranges over the values of $h_j(y_i)$. If we normalize this by writing $f_v(i) = m_v(i)/i$ we get a discrete probability on the elements of $\sigma$. When we use this sort of relative frequency in Subsec. 4.3, we write the corresponding EEDF simply as $\hat{F}$.

The output of the Count-Min sketch is the estimated frequencies of the tokens in $\sigma$ and these values have the nice properties we want. Specifically, they are readily computed iteratively i.e., in one pass, they are combinable in the sense that if we have two streams we can add their two data structures as matrices. Less obviously, we can also control the error and the storage required. Separately, the EEDF from the estimated frequencies has nice statistical properties if we treat the problem as $\mathcal{M}$-closed or $\mathcal{M}$-complete. That is, if we compare the EEDF $\hat{F}$ (that we have) to the EDF $\hat{F}_n$ (that we only have conceptually) we can show it has the usual convergence properties we would expect.

To get a sense for what how this class of sketches is actually computed, consider a toy example. Let $\sigma$ be stream formed from four values $\{A, B, C, D\}$ and suppose the first 10 elements are $[A, B, C, A, A, C, D, B, D, A]$. Choose $\delta = e^{-5}$ and $\epsilon = 2/3$. Then, $d = 5$ and $V = 3$ so we want five randomly chosen pairwise independent hash functions each taking $\{A, B, C, D\}$ to $\{1, 2, 3\}$. Suppose these are given by the columns in the table:

| | $h_1$ | $h_2$ | $h_3$ | $h_4$ | $h_5$ |
|---|---|---|---|---|---|
| $A$ | 1 | 1 | 1 | 1 | 3 |
| $B$ | 2 | 2 | 1 | 3 | 2 |
| $C$ | 3 | 2 | 2 | 2 | 1 |
| $D$ | 3 | 3 | 2 | 2 | 1 |

The data structure of the Count-Min sketch is a $5 \times 3$ array where the $j^{th}$ row corresponds to the $j^{th}$ hash function and the columns correspond to the three possible values a hash function can assign to the elements of the stream. Initially all entries are 0. For the Count part, each element of the stream is passed through all of the hash functions. When an item $y_i$ appears in the stream the count of the cell corresponding to $(j, h_j(y_i))$ increases by 1 for each $j$. Starting with $y_1 = A$, $h_1(A) = 1$. So the count of the cell $(h_1,1)$ increases and the frequency changes from 0 to 1. Since $h_2(A) = 1$ also, the count of the cell $(h_2,1)$ increases by one and its frequency changes from 0 to 1. Again, $h_3(A) = 1$. So the frequency of $(h_3,1)$ changes from 0 to 1. Likewise, $h_4(A) = 1$. So, the cell corresponding to $(h_4,1)$ changes from 0 to 1. Finally for this iteration, $h_5(A) = 3$ and hence the cell frequency of $(h_5,3)$ gets updated from 0 to 1. So the $5 \times 3$ table of zeros is updated to the following table.

|       | 1   | 2 | 3   |
|-------|-----|---|-----|
| $h_1$ | 1   | 0 | 0   |
| $h_2$ | 1   | 0 | 0   |
| $h_3$ | 1   | 0 | 0   |
| $h_4$ | 1   | 0 | 0   |
| $h_5$ | 0   | 0 | 1   |

The second element of the stream is $y_2 = B$. Like the first element $A$ we shall pass this element of the stream through all the hash functions. We have, $h_1(B) = 2, h_2(B) = 2, h_3(B) = 1, h_4(B) = 3, h_5(B) = 2$. So, the frequencies of the cells corresponding to $(h_1, 2)$, $(h_2, 2)$, $(h_3, 1)$, $(h_4, 3)$, and $(h_5, 2)$ are incremented by one. This is the second time we are incrementing the cell corresponding to $(h_3, 1)$ so it moves from 1 to 2. The newly updated table is the following.

|       | 1 | 2   | 3   |
|-------|---|-----|-----|
| $h_1$ | 1 | 1   | 0   |
| $h_2$ | 1 | 1   | 0   |
| $h_3$ | 2 | 0   | 0   |
| $h_4$ | 1 | 0   | 1   |
| $h_5$ | 0 | 1   | 1   |

We use the same procedure for the remaining 8 elements of the stream. The sum across rows, i.e., evaluations of each hash function, is 10, the length of the stream. The final table coming out of Count is the following.

|       | 1 | 2 | 3 |
|-------|---|---|---|
| $h_1$ | 4 | 2 | 4 |
| $h_2$ | 4 | 4 | 2 |
| $h_3$ | 6 | 4 | 0 |
| $h_4$ | 4 | 4 | 2 |
| $h_5$ | 4 | 2 | 4 |

Next, we apply the Min stage. That is, we obtain an estimate for the frequency of each element in the stream from the last table. In this stage, we take the minimum of the counts over $j$ i.e., corresponding to the cells $(j, h_j(y_i))$ as $j$ ranges over $1, 2, \cdots, d$. Thus, for $n = 10$, we see from the final table that for $A$, Count$(h_1, 1) = 4$, Count$(h_2, 1) = 4$, Count$(h_3, 1) = 6$, Count$(h_4, 1) = 4$, and Count$(h_5, 3) = 4$. (to be clear, we see that, in the last value for instance, that the '4' is from cell $(5,3)$ because $h_5(A) = 3$.) Now, the estimated frequency of $A$ is $\min(4, 4, 6, 4, 4) = 4$.

Similarly, for $B$, we have Count$(h_1, 2) = 2$, Count$(h_2, 2) = 4$, Count$(h_3, 1) = 6$, Count$(h_4, 3) = 2$, and Count$(h_5, 2) = 2$. So the estimated frequency of $B$ is $\min(2, 4, 6, 2, 2) = 2$. Doing the same for $C$ and $D$ gives the output of the Count − Min sketch. Here is the comparison of the actual frequencies and the estimated frequencies for the distinct elements in the finite stream:

| $\sigma$ | Actual Frequency | Estimated Frequency |
|---|---|---|
| $A$ | 4 | 4 |
| $B$ | 2 | 2 |
| $C$ | 2 | 4 |
| $D$ | 2 | 2 |

In this table, we got exact equality for for $A$, $B$, and $D$ but the value for $C$ is a little bit higher. In general, the estimated frequency for each element is greater than or equal to the actual frequency because of collisions in the hash functions, i.e., when $h_j(y) = h_j(y')$ for some $y \neq y'$. For streams with many more possible elements the agreement will typically not be nearly as good unless $n$ is very large. Obviously, in practice we would go through the Count part followed by the Min part for each $i$ and use the resulting probability to issue predictions or other decisions for the next time step.

### *4.3. Statistical aspects of the* Count-Min *Sketch*

Recall the setting of Subsec. 4.2, i.e., we have a stream $\sigma = (y_1, y_2, \ldots)$ from $[U]$ and are given $\epsilon, \delta > 0$. So, we have $d = \lceil \ln(1/\delta) \rceil$ randomly chosen 2-universal hash functions $h : [U] \to [V]$ where $V = \lceil 2/\epsilon \rceil$ leading to an EEDF $\hat{F}$ for the EDF $\hat{F}_n$ for a discrete stream. We start by extending from a discrete stream to a continuous stream and then state some properties of the extension. These properties will be from an $\mathcal{M}$-closed or -complete standpoint because otherwise it is unclear how to give formal properties. This means we are assuming the methods are good for $\mathcal{M}$-open cases because they perform as they should in $\mathcal{M}$-closed and -complete settings.

#### *4.3.1. Extension to continuous streams*

To discuss the statistical properties of the output from the Count $-$ Min sketch, we convert from discrete outcomes to streams of continuous outcomes. It is straightforward to do this and continuous data is convenient for many analytic purposes, e.g., taking limits is easier than trying to relate specific rates of increasing $[U]$, $[V]$, and $d$ to the behavior of $\hat{F}$.

Consider a stream $(y_1, y_2, \ldots, y_n, \ldots)$ that consists of real numbers from the range $(m, M)$ where we take $m = 0$ and $M \in \mathbb{R}$. Fix $K \in \mathbb{N}$ assumed large; this will play the role of $U$ but be under our control. Then, partition $(0, M]$ into $K$ intervals each of length $M/K$. Denote the $k^{th}$ interval by $I_k = I_{Kk}$, for $k = 1, 2, \cdots, K$. That is,

$$I_k = I_{Kk} = \left[ (k-1)\frac{M}{K}, k\frac{M}{K} \right). \tag{74}$$

Our goal will be to use the Count $-$ Min sketch on the discretized $[0, M]$ to produce an EEDF that we can use to predict $y_{n+1}$ after seeing $(y_1, \cdots, y_n)$.

Let
$$a_k = a_{Kk}(n) = \#\{y_i \in I_k \mid i = 1, \cdots, n\}$$

so that $a_k(n)$ is the frequency of tokens in the stream $(y_1, \cdots, y_n)$ that fall in $I_k$. Let $d = d_K$ and randomly choose hash functions $h_1, \cdots, h_{d_K}$ where, for $j = 1, \ldots, d_K$, each $d_j : [K] \to [w_K]$ for some $w_K$ that plays the role of $V$. (Thus, we are choosing $V \approx w_K$, where $K$ plays the role of $U$.) For storage bounds, we want $w_K \ll K$. More generally, we want $w_K$ and $d_K$ small as functions of $K$ but will have to allow them to increase slowly with $K$ while $K$ itself increases.

We extend the $h_j$'s to the continuous domain $[0, M]$ simply by defining

$$\tilde{h}_j : [0, M) \longrightarrow \{1, 2, \cdots, w_K\}$$

where $\tilde{h}_j(s) = h_j(k)$ for $s \in I_k$. In terms of the tokens, this means for $i \leq n$ and $y_i \in I_k$ we have

$$\tilde{h}_j(y_i) = h_j(k).$$

Following the $\mathsf{Count-Min}$ procedure, we define an estimate of $a_k$ (frequency of the $k^{th}$ interval) at time $n$. For the $j^{th}$ hash function $h_j$, an interval $k$ and time $n$, we set

$$\hat{a}_{jk} = \mathsf{Count}_n(j, h_j(k)) = \# \left( \{i \leq n \mid \tilde{h}_j(y_i) = h_j(k)\} \right)$$

so that the estimate $\hat{a}_k$ of $a_k$ becomes

$$\hat{a}_k(n) = \min_j \ \hat{a}_{jk}(n) \geq 0. \tag{75}$$

Now, the estimated EDF (EEDF) generated by the $\mathsf{Count-Min}$ sketch is

$$\hat{F}(x) = \sum_{k \leq x} \frac{\hat{a}_k(n)}{n} \tag{76}$$

and the actual EDF is

$$\hat{F}_n(x) = \sum_{k \leq x} \frac{a_k(n)}{n}. \tag{77}$$

The EEDF is only an estimate of the EDF because the $\mathsf{Count-Min}$ sketch only gives an estimate of the frequencies. The reason is that the EEDF is based on hash functions so that it will satisfy a storage bound that we will shortly state.

To finish the present line of reasoning, we use (76) to define point predictions. (The EEDF also gives interval predictions, but it is unclear what the interval means in $\mathcal{M}$-open settings.) In our computed results we use two predictors:

1. Weighted mean: our prediction is the weighted mean of the midpoints of the intervals $I_k$; $u = 1, 2, \cdots, K$ defined in (74), where the weights are $\hat{a}_k$ as defined in (75). Formally, denote the mid-point of interval $I_k$ by $m_k$. Then,

$$\hat{y}_{n+1} = \hat{y}_{K, n+1} = \sum_{k=1}^{K} m_k \frac{\hat{a}_k(n)}{n} \tag{78}$$

2. Weighted median: our prediction is the weighted median of the $m_k$'s, with weights $w_k = \hat{a}_k / \sum_{k=1}^{K} \hat{a}_k$, defined as the average of $m_{q-1}$ and $m_q$, where $m_q$ satisfies

$$\sum_{i=1}^{q-1} w_i \leq \frac{1}{2} \text{ and } \sum_{i=q+1}^{K} w_i \leq \frac{1}{2}. \tag{79}$$

### 4.3.2. Desirable properties of the extension

Here we state four results for use with streaming $\mathcal{M}$-open data: an error bound, a storage bound for the error bound, a convergence result for $\hat{F}$ to the EEDF, and a convergence result for $\hat{F}$ to $F$, when it exists. The mode of convergence to the EEDF is defined by the probability on the hash functions not by any probability associated with the stream, cf. the discussion in [13]. The mode of convergence of $\hat{F}$ to $F$ includes the probability associated with $F$.

Our first result is that $\hat{a}_k$ is a good estimate of the frequency of an interval $I_k$, essentially a consistency result for fixed $K$. Let $||a||_1 = \sum_{k=1}^{K} a_k(n)$ be the sum over $k$ of the number of elements in $y$ up to time $n$ that land in $I_K$, where $K$ and $n$ are suppressed in the notation $||a||_1$. We have the following; it is similar to the guarantee for the $\mathsf{Count - Min}$ sketch, see [39] and [13].

**Proposition 4.1.** *Let $W$ correspond to the probability in* (71)*. Then, $\forall \epsilon > 0$ and $\forall \delta > 0$, $\exists\ N$ such that $\forall d_K > N$, we have*

$$W(\forall j = 1, \cdots, d_K; \hat{a}_{jk}(n) \leq a_k(n) + \epsilon ||a||_1) \leq \delta.$$

**Remark**: Here, $||a||_1 = n$ because we are looking at data streams in the cash register model of streaming data i.e., items only accumulate. It is immediate from Prop. 4.1 that we get $\hat{a}_k(n) \leq a_k(n) + \epsilon ||a||_1$ from the minimum in (75). Also, by construction we get $a_k(n) \leq \hat{a}_k(n)$. So, we have upper and lower bounds.

Next, we address the storage requirement for the procedure used in Prop. 4.1. Heuristically, observe that the storage is upper bounded by the number of hash functions $\log(1/\delta)$ multiplied by the number of values each hash function can take, namely $e/\epsilon$ giving $\mathcal{O}((1/\epsilon)\log(1/\delta))$. Adapting the proof in [39] to our present setting, we see that storage of the order $\mathcal{O}(1/\epsilon)$ will suffice, see [13].

**Theorem 4.1.** *Let $\eta > 0$. Assume the storage available is $\Omega(1/\epsilon)$[3]. Then, under the probability in* (71)*, we have that*

$$P(\hat{a}_{jk} \leq a_k + \epsilon ||a||_1) \leq \eta.$$

We extend Prop. 4.1 by letting $K, d_K, n \to \infty$ at appropriate rates to get a consistency result for the EEDF. That is, our EEDF converges to an EDF based on the streaming data that is not necessarily the true DF since it needn't exist. We have the following.

---

[3]$\Omega$-notation gives a lower bound in contrast to big-$\mathcal{O}$ notation that gives an upper bound.

**Theorem 4.2.** *Let $x \in (0, M]$. Then, pointwise in $x$,*

$$\hat{F}(x) - \hat{F}_n(x) \overset{W}{\to} 0 \ \text{as} \ d_K, K, \text{and} \ n \longrightarrow \infty.$$

Unsurprisingly, if the stream comes from an $\mathcal{M}$-closed or -complete source $F$, then the EDF converges to $F$ (by the usual law of large numbers) and so does the EEDF $\hat{F}$. We state this as the following.

**Corollary 4.1.** *If there exists an $F$ such that the $Y_i$'s are independently and identically distributed according to $F$, then, under the hypotheses of Theorem 4.2, we have that the $\mathsf{Count - Min}$ sketch generated estimate $\hat{F}$ of $F$ is consistent for $F$, that is*

$$\hat{F} \longrightarrow F,$$

*in the joint mode of convergence defined by the $W$ used in Theorem 4.2 and the DF the $Y_i$'s follow.*

Like the EDF, the EEDF tracks the location of the data. Recall that a $100(1 - \alpha)\%$ PI is given by $(\hat{F}_n^{-1}(\alpha/2), \hat{F}_n^{-1}(1 - \alpha/2))$. As $n$ increases, the EEDF follows the data i.e., the relationship between $y_n$ and $y_{n+1}$ may be different from the relationship between $y_{100n}$ and $y_{100n+1}$. The location of $\hat{F}$ moves to track where the preponderance of data is. There are relatively standard methods, see [14], to force more recent data to be weighted more and these techniques can be combined with the EEDF if desired.

To conclude this section, we make a few observations about how $\hat{F}$ can be expected to behave. First, it is possible to prove a Glivenko-Cantelli Theorem for the convergence of $\hat{F}$ to $\hat{F}_n$ and again observe the reduction when the $Y_i$'s follow an $F$. We cite [16] for the standard form and proof of such theorems; see also [46]; cf. [13]. We want results about $\hat{F}$ because it is fully empirical. These results can likely be generalized to many dependent data settings. Likewise, we suggest versions of other major theorems for the EDF such as Donsker's theorem and the Kiefer-Dvoretzky-Wolfowitz theorem can also be established for $\hat{F}$.

## 5. Conformal prediction

The central premise of conformal prediction (CP) is that future data looks like past data, i.e., it 'conforms'. Leaving aside the philosophical question as to what that means in $\mathcal{M}$-Open settings, the practical implication is that we find data dependent quantities and associate a prediction interval to them. That is, we treat the data received as a set of real numbers and form expressions for a future value without reference to any probability structure.

A caricature of conformal methods would be the following. Take the mean and SD of $n$ observations and announce $\bar{y} \pm c_\alpha \hat{\sigma}$ as a $1 - \alpha 100\%$ PI provided we had some way to interpret the $\alpha$ e.g., perhaps using a percentile from an EDF based on the accumulated data, but not necessarily saying that the data followed any distribution. Then, make distributional assumptions about the data and investigate the behavior of the predictor under those assumptions. See [32] for an extensive analysis of conformal predictors under stochastic assumptions.

This idea is far from new: in standard linear regression we obtain estimators as a result of an optimization under squared error and only derive inferential properties of them after specifying an error structure. Indeed, in the conventional setting, the metric structure (squared error) is assigned independently of the error structure (normality) and there are multiple other viable choices.

It is easy to see that conformal prediction is applicable to $\mathcal{M}$-open problems because, like many computer science techniques, it is purely empirical and does not necessitate assumptions to be stated explicitly. For instance, it is easy to see that the derivation of RVM's is a deterministic optimization. After deriving the predictor, we assume an error structure with which to prove consistency and other properties of RVM's. Shtarkov predictors have similar properties.

Accordingly, in this short section we only present the empirical aspects of conformal prediction, ignoring the accumulated analyses of these methods in $\mathcal{M}$-complete and -closed settings. Our discussion owes much to the 'computer science view' as expressed in [45], [6], and [50].

### 5.1. *No explanatory variables*

CP is based on the concept of a conformity measure. There are many choices, but each of them provides an analog of concept of confidence for PI's. This is empirical and does not invoke an error structure. A nonconformity measure assesses how close a potential future value $y_{n+1}$ is to the accumulated data $y_1, \ldots, y_n$. Formally, conformity measures are non-negative functions $C : \mathbb{R}^n \times \mathbb{R} \to \mathbb{R}$ assumed symmetric in their first argument. They formalize the idea of distance between a set of outcomes of size $n$ and a future outcome. It is tempting to think of these sets as $y^n$ and $y_{n+1}$ respectively, but the sense of conformity used here is more subtle and has a feel of cross-validation.

To understand the intuition, fix a data vector $y^{n+1}$ where $y^n$ is the data we have and $y_{n+1}$ is a candidate future value. Let $i \in \{1, \ldots, n+1\}$ and remove $y_i$ from $y^{n+1}$. Write the $y_i$-deleted $y^{n+1}$ as $y^{n+1 \setminus i} = (y_1, \ldots, \hat{y}_i, \ldots, y_{n+1})$, where the hat indicates the deletion.

We use $C$ to compare $y_i$ with $\hat{y}^{n+1 \setminus i}$. So, write

$$C_i(y_{n+1}) = C(\hat{y}^{n+1 \setminus i}, y_i).$$

Given that we have chosen $C$ properly, $C_i(y_{n+1})$ is large when $\hat{y}^{n+1 \setminus i}$ 'conforms' with $y_i$ (because of $y_{n+1}$), intuitively when $y_i$ is close from the 'middle' of the values in $y^{n+1 \setminus i}$. Now, the proportion of times, out of $n+1$, that $C_i(y_{n+1}) \leq C_{n+1}(y_{n+1})$ measures how similar $y_{n+1}$ is to $y^n$. We use $n+1$ because for $i = n+1$ the inequality holds trivially $- C_{n+1}(y_{n+1}) \leq C_{n+1}(y_{n+1})$ by definition. The proportion of times the reverse inequality holds measures how similar $y_{n+1}$ is to $y^n$. Thus, if this proportion is high enough we want to put the candidate value $y_{n+1}$ into our PI. The reasoning is that the higher this proportion is, the more $y_{n+1}$ conforms to $y^n$, as a set.

Let $\alpha > 0$ be the level of a conformal PI for $y^{n+1}$ based on $y^n$. We define

$$PI(\alpha, y^n) = \left\{ y_{n+1} \mid \frac{\#\left(\{i | C_i(y_{n+1}) \geq C_{n+1}(y_{n+1})\}\right)}{n+1} \geq \alpha \right\}. \qquad (80)$$

That is, if a value $y_{n+1}$ gives a high enough proportion of large enough $C_i$'s, it is put in $PI(\alpha, y^n)$. Clearly, for $\alpha_1 < \alpha_2$ we have $PI(\alpha_2, y^n) \subset PI(\alpha_1, y^n)$.

It remains to propose suitable conformity measures or, equivalently, nonconformity measures. Perhaps the easiest is the 'distance to average' (DTA) nonconformity: $C(y^n, y_{n+1}) = |\bar{y} - y_{n+1}|$. The closer the potential future value $y_{n+1}$ is to the mean, the more it conforms to the existing data. While intuitively reasonable, this choice of $C$ is has a computational limitation: it requires computing $n + 1$ means at stage $n$. To avoid this, it is common to use $C(y^n, y_{n+1}) = |\bar{y}_{n+1} - y_{n+1}|$. We see that

$$C(y^n, y_{n+1}) = |\frac{n\bar{y} + y_{n+1}}{n+1} - y_{n+1}| = \frac{n}{n+1}|\bar{y} - y_{n+1}|$$

so the two forms of the DTA are equivalent as $n$ increases.

As a generality, we can use density estimators as nonconformities. For instance, [33] proposes that we set $C(y^n, y_{n+1}) = \hat{p}(y_{n+1}; y^n)$ where $\hat{p}$ is a density estimator for the true density of the stream (which may not exist) using $y^n$. Establishing results that ensure consistency of $\hat{p}$ identifies regularity conditions on $C$ ensuring it reduces properly in an $\mathcal{M}$-closed problem. Focusing on kernel methods, [33] give results on bandwidth selection and studied the actual prediction regions. It is possible to get prediction regions that have as many disjoint intervals as there are modes in the true density; it is not clear how to overcome this possible problem. This line of inquiry is continued in [32] which re-interprets conformity in terms of probability of mis-coverage and focuses on high-dimensional prediction problems that are often ignored.

Another relatively easy choice of conformity measure is Bayesian, namely the posterior predictive density as used in [8]. The simplest is of course based on the normal. Suppose the $Y_i$'s are IID $N(\theta, \sigma^2)$ with a normal prior assigned to $\theta$, $\theta \sim N(\mu, \tau^2\sigma^2)$ and an inverse Gamma assigned to $\sigma$, i.e., $1/\sigma^2 \sim \mathsf{Gamma}(a/2, b/2)$ for some $a, b > 0$. Then, for given $\mu$ and $\tau$, set $C(y^n, y_{n+1}) = p(y_{n+1}|y^n)$. This choice reflects conformity in the sense of a new data point being representative of the predictive density. In Sec. 6, we use their package for Bayesian CP; there is also a CP package for the techniques in [32].

Conditions to ensure that the PI is an actual interval are in [8] and [25] gives conditions to ensure the region is efficient, or Bayes optimal, in the sense that the prediction region is, asymptotically, as small as possible given the level, assuming the model is true. These amount to regularity conditions to ensure the method reduces to what is should be in $\mathcal{M}$-closed problems. For models other than what [8] study, it is possible that PI's are union of disjoint intervals.

### 5.2. *Explanatory Variables Present*

To see how conformal prediction extends to regression with explanatory variables consider simple linear regression. For data $\mathcal{D}_n = \{(x, y_1), \ldots, (x_n, y_n)\}$ suppose we write $y = \hat{a} + \hat{b}x$. Then, for a possible predicted value $\hat{y}_{n+1}$ at $x_{n+1}$ write the nonconformity

$$C(\mathcal{D}_n, (x_{n+1}, y_{n+1})) = |y_{n+1} - \hat{y}_{n+1}| = |y_{n+1} - \hat{a} - \hat{b}x_{n+1}|.$$

It is seen that the nonconformity uses the predictor class as an input.

Now, the procedure from Subsec. 5.1 can be generalized directly. To wit: Given $\mathcal{D}_n$ and $x_{n+1}$, consider the candidate future outcome $y_{n+1}$ and write

$$C_i(y_{n+1}) = C(\mathcal{D}_{n+1} \setminus \{(x_i, y_i)\}, \{(x_i, y_i)\}).$$

We write the analog of (80) again putting values of $y_{n+1}$ into it for $x_{n+1}$ when a high enough proportion of the $C_i$'s are larger than $C_{n+1}(y_{n+1})$. The procedure is the same for any other class of predictors for $Y$ from $X$.

It is seen that the conformity measure assesses the degree to which a candidate new data point conforms to the earlier data and that its conformity depends on the predictor class. Thus the nonconformity of a new value with the older values will differ if a different regression technique is used. For example, the prediction will depend on whether a linear model, an RVM, or a random forest is used. This is typical for predictive techniques but it is unclear how reasonable it is for $\mathcal{M}$-open data.

Indeed, it is an open question whether $\mathcal{M}$-open data satisfies a variance-bias tradeoff. Likely it does, but not in any sense that can be readily formalized. Nevertheless, it is numerically reasonable to conjecture that if we have three predictor classes, small medium and large, the medium class may be better (have a higher conformity score) than the small or the large due to bias and variance, respectively. Moreover, if there are two predictor classes leading to two different conformity measures, it is possible to find the corresponding PI's for a given level $\alpha$ but the ordering of the two sets of $C_i$'s may not match and they may not be as comparable as desired. However, the performance of the PI's from two different conformities can be empirically compared in terms of coverage, for instance. These problems become more complex if two different nonconformity measures e.g., a Bayesian's posterior versus a kernel density estimate, are used as well as two classes of predictor families.

Work by [26] compares conformities based on random forests, neural nets, and nearest-neighbors methods and argues that random forests typically provide the most efficient prediction regions for certain fixed conformities. Recent work by [21] predictively compares Bayesian linear regression to ridge regression and argues that ridge regression with conformal prediction often works better.

## 6. Computational Comparisons

In this section we give computational results for 12 predictors and two data sets. Results from four more data sets are given in Appendix D. All six data sets are very complex and, we suggest, can be regarded as $\mathcal{M}$-open. We give cumulative predictive errors over the stream of data. We also perturb the data by adding independent $N(0, \tau^2)$ noise to each data point and recalculate the cumulative errors to assess the sensitivity of each method as well as its raw performance.

In the next subsection, we describe our settings formally. Then we present and discuss our computational findings and make some tentative recommendations. Details for the implementations of the methods are given in Appendix D.1.

### 6.1. Settings for the comparison

For each of the 12 predictors, we compute the cumulative $L^1$ error. That is, for each method and each data stream $(y_1, y_2, y_3, \ldots)$ we have a sequence of errors $|y_{n+1} - \hat{y}_{n+1}|$ where a given prediction $\hat{y}_i$ depends on $y_1, \ldots, y_n$ (and possibly a burn-in set $\mathcal{D}_b$) and we find the cumulative predictive error

$$CPE = CPE(n+1) = \frac{1}{n+1} \sum_{i=1}^{n+1} |y_i - \hat{y}_i|. \tag{81}$$

It is seen that

$$CPE(n+1) = \frac{1}{n+1} \left( nCPE(n) + |y_{n+1} - \hat{y}_{n+1}| \right)$$

so we can easily compute it recursively. Note that this assessment follows the prequential principle; see [20].

In addition to finding the cumulative error we assess the sensitivity of the methods to perturbations of the data. Following [34], we calculate a running variance of the CPE given by

$$\sigma_{RV}^2 = \frac{1}{n} \sum_{i=1}^{n} CPE_i^2 - \left( \frac{1}{n} \sum_{i=1}^{n} CPE_i \right)^2 \tag{82}$$

and then define a parameter $\tau$ ranging over $[0, \sigma_{RV}]$. Then we perturb our data by forming a new stream $y_i' = y_i + \eta_i$, where each $\eta_i \sim \mathcal{N}(0, \tau^2)$. The $\eta_i$'s are drawn independently and let us recalculate the cumulative risks as a function of $\tau$ for each of the 12 methods. In our graphs below we call these 'sensitivity' curves and denote them by $CPE(\tau)$. In practice, we approximate sensitivity curves on a finite grid in $[0, \sigma_{RV}]$ as shown in Subsec. 6.2.

Our 12 methods are listed in Table 1 together with the abbreviation we use in our graphs to follow and references for where the exact form of the predictor is given in this paper. As can be seen, the methods segregate into two classes.

First, some of them are effective for streaming data because the running time per prediction does not appreciably increase with stream length. This is the case for Sht, DPP, Med, and Mean.

Second, some predictors – the three GPP methods and conformal – have a running time per iteration that increases with stream length. Specifically, for all GPP methods the variance matrix increases in dimension and there is no accepted streaming method to estimate it. Also, for conformal prediction we used the fabContinuousPrediction package in **R** and we did not rewrite the code to make it one pass. Accordingly, we could not realistically use these methods as is. So, we used a 'representative set' in place of the full stream. Our representative set is based on streaming $K$-means but any streaming clustering procedure could be used. The strategy is to choose a large value of $K$ and use the cluster centers as a our representative set. The representative set has a fixed cardinality, $K$, but the set itself updates upon receipt of each new $y_n$. In addition, we used the representative set with Sht, DPP, Mean, and Med, to see if using a representative set made their predictions better or worse.

| Method; Class | Abbreviation | Reference |
|---|---|---|
| Shtarkov normal; Shtarkov | Sht | (10) |
| Shtarkov normal; Shtarkov | Sht_rep | |
| GPP, no random bias; Bayes | GPPnoRB (rep) | (54) |
| GPP, with random bias; Bayes | GPPRB (rep) | (60), Subsec. 3.1.2 |
| GPP, independent bias; Bayes | GPP_INID (rep) | (68), Subsec. 3.1.3, $\gamma_{n+1} = \bar{\gamma}_n$ |
| Dirichlet; Bayes | DPP | (70) |
| Dirichlet rep; Bayes | DPP_rep | |
| Weighted mean; Hash based | Mean | (78) |
| Weighted mean rep.; Hash based | Mean_rep | |
| Weighted median; Hash based | Med | (79) |
| Weighted median rep.; Hash based | Med_rep | |
| Conformal Bayes; conformal | Conf (rep) | Subsec. 5.1, cf. [8] |

*Table 1*

*List of the methods we computed. The only methods that we could run without using a representative subset (indicated by 'rep') were Sht, DPP, Mean, and Med. The Shtarkov normal predictor is the same whether 'σ' is known or not and was simply the mean for this case. GPP's and DPP's are the simplest nonparametric Bayes techniques. The hash based procedures are intrinsically one pass. Conformal is Bayes because of the choice of nonconformity. Blanks in the right hand column indicate repetitions from the previous line.*

### 6.2. Results

To compare empirical performances of the 12 methods, we ran them on six data sets. The sensitivity curves from two are presented here; the other four graphs are in Appendix D.2. We chose the two sets for this section because the median hash based methods give the best performance for the first while the two GPP based methods were best for the second. These two classes of methods seemed to be better than the other two classes of methods; see Subsec. 6.3.

The first data set, Walmart Sales, see [28], has $n = 200000$ and measurements for 6 variables in its columns. It contains data on the unit price of products in Walmart and their quantities sold over 2017-2020. We extracted the first 10000 rows of the data set for our computation where we multiplied "Quantity" by "UnitPrice" to get the total price and used it. In our graphs we used $\sigma_{RV} = 154$ and grid with with interval length 19. The sensitivity curves as a function of $\tau$ for the methods are plotted in Fig. 4. The values at $\tau = 0$ show the actual CPE for the unperturbed i.e., original, data.

As a generality, we prefer sensitivity curves that are low for $\tau$ at or near zero and rise slowly as $\tau$ increases. The tradeoff between the value at zero and the rate of increase is hard to assess.

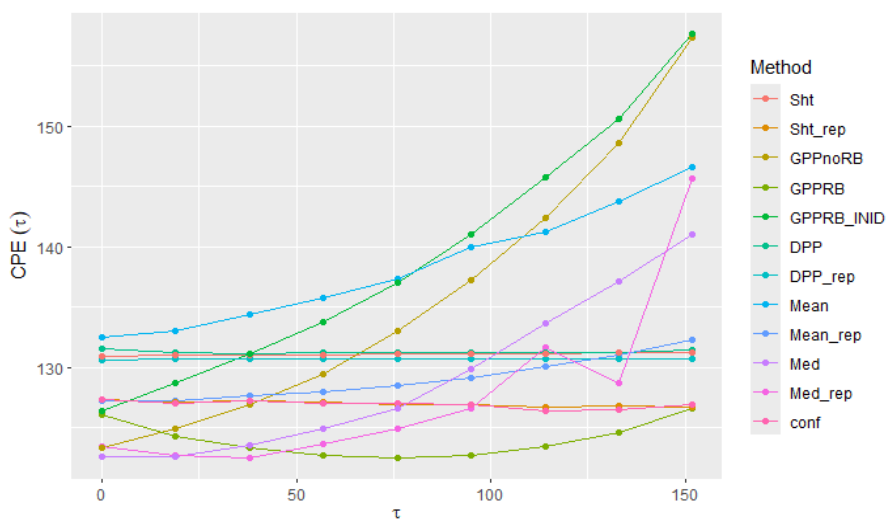

FIG 4. *Sensitivity curves for the* Walmart Sales *data, i.e., cumulative predictive error as a function of the perturbation $CPE(\tau)$. The two median methods perform best and the sensitivity curves for five methods flatline. The other methods give intermediate performance.*

The curves in Fig. 4 show that the medians (representative and one pass) have the lowest error for $\tau = 0$ and are generally rising to the right at a moderate rate compared to the other curves. The mean curves (representative and one-pass) rise as desired but are higher indicating worse CPE. The three GPP sensitivity curves either have higher initial error or rise too fast. Notably, the GPPRB curve decreases slowly with $\tau$ before rising slowly; this indicates that the method performs better when the data has small perturbation than when it has zero perturbation. We regard this as bad but are unclear what it indicates. The curves for five methods – Sht, Sht_rep, Conformal, DPP, and DPP_rep – flatline indicating insensitivity to $\tau$, an undesirable property.

The second data set, Customer Shopping, see [27], has customer age, gender, product categories, quantity, price, etc. from 10 different shopping malls in Istanbul between 2021 and 2023. This data set contains 10 columns and like other data sets we have extracted the first 10000 rows from the 99457 rows of the data. For our computational purposes here we have used the column "price". For this data set we found $\sigma_{RV} = 890$ and used a grid with interval length 89 to assess sensitivity.

It is seen from Fig. 5 that GPPRB_INID and GPPnoRB performed best with a close competition from the median_rep. The median one pass method and mean_rep perform a little worse mainly because their sensitivity curves rise quickly as $\tau$ increases from zero. Again, the GPPnoRB decreases initially and then increases, both slowly. The mean curve does quite poorly. The same five methods as in Fig. 4 (Sht, Sht_rep, Conformal, DPP, and DPP_rep) flatline. That is, the best performing methods were almost the reverse of what we found in the other graph.

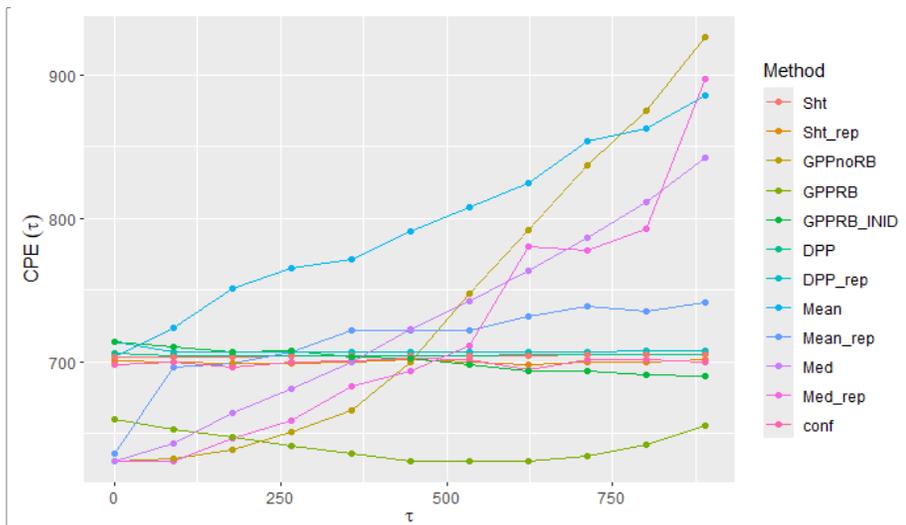

*Fig 5. Sensitivity curves for the* Customer Shopping *data in Istanbul. Two of the GPP methods (GPPnoRB and GPP_INID) performed best along with Med_rep The same five methods as in Fig. 4 flatlined. Other methods were intermediate.*

### 6.3. *Recommendations*

It is difficult to compare the performance of methods in any general sense because it is impossible at this time to produce a systematic comparison of the methods over the class of $\mathcal{M}$-open data sets. Indeed, it is unclear what important and well-defined subclasses of $\mathcal{M}$-open data sets would be.

Consequently, we simply looked at what we thought the top methods were for each data set and scored each method one point if it was a top performer and zero otherwise. This was purely heuristic and subjective. We ruled out methods that flatlined because we regarded them as insensitive to the data. We also ruled out methods that had sensitivity curves that rose too sharply or started at too high a value. In the 'strictly FWIW' category, we produced Table 2.

| method | Med | Med Rep | GPP no RB | GPP RB | GPP INID |
|--------|-----|---------|-----------|--------|----------|
| counts | 5   | 2       | 2         | 2      | 2        |

Table 2

*Informal counts of which methods perform best. Only methods with a positive score are shown; all other methods had count zero. Honorable mentions include Mean for the data set* Real Estate *but this would not change the informal conclusions.*

It is seen that the only two classes of methods that were ever best in our examples were median hash function based methods and GPP's of one sort or another. This does not mean that all other methods – DPP, conformal, Shtarkov – should be discarded because we only used the simplest conformal and Shtarkov methods and DPP's are arguably the simplest of nonparametric Bayes procedures. Moreover, we see that as a generality, using a representative set did not obviously impair performance of the methods. Although it is unclear what the representative set represents, and other choices of representative sets are possible, it seems using such a set can speed computation with essentially no cost in performance. Recalling that $K$-means effectively restricts clusters to convex sets this is surprising. On the other hand, we have only computed a few one-dimensional examples so the convexity restriction may be trivial.

## 7. Discussion

The 12 methods we computed represent four classes of predictive techniques: Shtarkov, Bayes, hash function based, and conformal that we have reviewed in detail. Our computations suggest that Bayes methods (GPP's in particular) and methods based on hash functions are likely to perform overall better than the other two classes of methods for $\mathcal{M}$-open data sets.

There are at least two caveats with this summary. First, the concept of $\mathcal{M}$-open is very broad and encompasses many data sets that are unrelated. Indeed, we conjecture that as study of this class of data proceeds recognizable subsets of the class of $\mathcal{M}$-open data will be defined and have different properties. Second, the sensitivity curves of one of the GPP methods (GPPnoRB) sometimes decreased and then increased indicating that it gave better predictions for small perturbations than for zero perturbations or large perturbations and it remains unclear what this means. If we were modeling, we would conjecture that the model was wrong but that a better model could be found if some aspect like bias could be corrected. That way we would ensure that as for an improved model we would get low error initially and gradually increasing error as the perturbations increased.

Probability modeling is used in all four conceptual classes of techniques. However, because we have focused on $\mathcal{M}$-open data sets we cannot use probability modeling in the usual way. We have never posited a distribution for the data; we have only used probability to construct the predictors. Summarizing:

1. Shtarkov predictors use probabilistic experts but the DG is not assumed to be probabilistic.
2. Bayesian techniques are reinterpreted as in [15], so we do not have a probability distribution on the data.
3. Hash based methods have a probability only on the choice of hash functions in the predictor; this is like a Bayesian's prior.
4. Conformal methods are fully empirical and only bring in probability as a way to assess performance. Here, the nonconformity is derived from a posterior density but this is not a distribution on the data.

A clear implication of the material we have presented here is that probability based methods remain fundamental to prediction for $\mathcal{M}$-open data. However, the way probability modeling should be done changes: we do not use probability modeling for the DG, we only use it for the construction of the predictor. This is consistent with the Bayesian view on how to use pre-experimental information.

No survey such as this can be truly comprehensive and, arguably, the biggest conceptual omission is recommender systems. These are not strictly speaking predictors because they give ranks rather than predictions but the highest ranked object could be taken as a prediction. Recommender systems are mainly if not exclusively for discrete data and it is unclear how to adapt them to continuous streams. There is relatively little theoretical development of recommender systems, however, they typically use statistical quantities such as correlations and conditional probabilities (in the form of association rules) without assuming the usual error structure in statistical contexts. The interested reader is referred to the now-classic texts [Aggarwal,2016] and [41].

We conclude with a restatement of our main findings: We cannot assume a distribution on data when it is from an $\mathcal{M}$-open DG. Nevertheless, we can form and assess predictors. These predictors can be assessed for stability as well as low predictive error. In principle, good predictors can be examined to make statements about the DG. In preliminary computations, it appears that Bayesian techniques and hash function based procedures (from computer science) are generally most effective in terms of low predictive error, relatively high stability, and manageable computational burden. Probability modeling remains the best approach to predictor formation but is used to form the predictor not to model the DG.

## Appendix A: Proofs from Sec. 2.1

Proof of Theorem 2.2:

First we observe that the MLE, resp. MPLE, exists and is unique under the first two hypotheses. Hence it is enough to show that the denominators $D_{n,\mathsf{F}}$ and $D_{n,\mathsf{B}}$ are finite and bounded away from zero for fixed $n$.

We begin with $D_{n,\mathsf{F}}$. First, we see $\int_{y^n} p(y^n|\hat{\theta})dy^n < \infty$: Under hypotheses 1, 3, and 4 there is an $R > 0$ so that $p(y|\theta) < R$ and, $|y| < R$. Now,

$$\int_{y^n} p(y^n|\hat{\theta})dy^n = \int p(y^n|\hat{\theta})\chi_{|y_i|<R}dy^n$$
$$\leq \int R\chi_{|y_i|<R}dy^n \leq R(2R)^n < \infty.$$

Next, we show $\int p(y^n|\hat{\theta})dy^n > 0$. Regard $\hat{\theta}$ as a function $\hat{\theta} : \mathcal{Y}^n \to \Theta$ and denote the interior of the image of $\hat{\theta}$ by $Im^0(\hat{\theta})$. For any $\theta_0 \in Im^0(\hat{\theta})$ there must be a $y_0^n$ for which $\hat{\theta}(y_0^n) = \theta_0$. For this $y_0^n$, $\exists \tau > 0$ so that $p(y_0^n|\theta_0) > \tau$: by contradiction, if no such $\tau$ existed then $p(y_0^n|\hat{\theta}) = 0$ then since $\hat{\theta}$ is a maximum we have that $p(y_0^n|\theta) = 0$ for all $\theta$. Thus, $\hat{\theta}$ is not unique.

Since $p(\cdot|\cdot)$ is continuous in $y^n$ and $\hat{\theta}$, we have that for this $\tau > 0$, there is an $r > 0$ so that for $y^n$ and $\theta$ satisfying $|y^n - y_0^n| < r$ and $|\hat{\theta} - \theta_0| < r$ we have $p(y^n|\theta) > \frac{\tau}{2} > 0$. But now, by continuity of $\hat{\theta}$, there exists $r' \leq r$ so that $|y^n - y_0^n| < r'$ implies that $|\hat{\theta} - \theta_0| < r$. Writing $B(r') = \{|y^n - y_0| < r'\}$, we have

$$\int_{y^n} p(y^n|\hat{\theta}) dy^n \;\; \geq \;\; \int_{B(r')} p(y^n|\hat{\theta}) dy^n$$

$$\geq \;\; \frac{\tau}{2} \int_{B(r')} dy^n$$

$$= \;\; \frac{\tau}{2} \times \text{Volume of a ball in } \mathbb{R}^n \text{of radius } r'$$

$$= \;\; \frac{\tau}{2} \frac{\pi^{\frac{n}{2}}}{\Gamma(\frac{n}{2}+1)} (r')^n > 0.$$

Hence, $q_{\mathsf{opt,F}}$ is well-defined.

The proof for $q_{\mathsf{opt,B}}$ is similar. There is an $R$ so that $p(y|\theta)$, $w(\hat{\theta})$, and $|y^n|$ are bounded by $R$. So, we have that

$$\int w(\hat{\theta}) p(y^n|\hat{\theta}) \mathrm{d}y^n \;\; = \;\; \int_{\{|y^n| < R\}} w(\hat{\theta}) p(y^n|\hat{\theta}) \mathrm{d}y^n$$

$$\leq \;\; R^2 \int_{\{|y^n| < R\}} \mathrm{d}y^n \leq R^2 (2R)^n < \infty.$$

To show $\int_{y^n} w(\tilde{\theta}) p(y^n|\tilde{\theta}) dy^n > 0$, choose $\theta_0 \in Im^0(\tilde{\theta})$. As before, there is a $y_0^n$ so that $\tilde{\theta}(y_0^n) = \theta_0$ and a $\tau > 0$ so that $w(\theta_0) p(y_0^n|\theta_0) > \tau > 0$. (By way of contradiction, if no such $\tau$ exists, $w(\theta_0) p(y_0^n|\theta) = 0$ for all $\theta$ so the maximum is not unique.) In fact, we can find $\tau > 0$ so that $w(\theta_0) > \tau$ and $p(y_0^n|\theta_0) > \tau$.

Since $p(\cdot|\cdot)$ is continuous in its arguments, say $y^n$ and $\theta$, there exists $r > 0$ so that $|y^n - y_0^n| < r$ and $|\hat{\theta} - \theta_0| < r$ implies $p(y^n|\tilde{\theta}) > \tau/2 > 0$. Since the prior density is continuous we can also assume that $r$ and $\tau$ can be chosen small enough that $w(\tilde{\theta}(y_0^n)) > \tau/2$. But now, by continuity of $\tilde{\theta}$, there exists $r' \leq r$, such that $|y^n - y_0^n| < r' \implies |\tilde{\theta}(y^n) - \theta_0| < r$. So,

$$\int_{y^n} w(\tilde{\theta}) p(y^n|\tilde{\theta}) dy^n \;\; \geq \;\; \int_{B(r')} w(\tilde{\theta}) p(y^n|\tilde{\theta}) dy^n$$

$$\geq \;\; \frac{\tau}{2} \int_{B(r')} w(\tilde{\theta}) dy^n$$

$$\geq \;\; \frac{\tau^2}{4} \int_{B(r')} \mathrm{d}y^n$$

$$= \;\; \frac{\tau^2}{4} \text{ Volume of a ball in } R^n \text{ of radius } r'$$

$$= \;\; \frac{\tau^2}{4} \frac{\pi^{\frac{n}{2}}}{\Gamma(\frac{n}{2}+1)} (r')^n > 0.$$

Hence, $q_{\mathsf{opt,B}}$ is well-defined.

# Appendix B: Calculus proofs for Shtarkov predictors

This section is to record proofs that are important for the sake of completeness.

## B.1. Normal Cases

First, we give the details for the derivations in Subsec. 2.2.1. The details of the derivation of (10) i.e., unknown mean, known variance, are as follows. The log ML using $\hat{\mu} = \bar{y} = \bar{y}_n$ is

$$
\begin{aligned}
\ln p(y^{n+1}|\hat{\mu}_{n+1}, \sigma^2) &= -\frac{n+1}{2}\ln(\sigma^2 2\pi) \\
&\quad -\frac{1}{2\sigma^2}\left[\sum_{i=1}^{n} y_i^2 + y_{n+1}^2 - (n+1)\times\left(\frac{n\bar{y}_n + y_{n+1}}{n+1}\right)^2\right]
\end{aligned}
\tag{83}
$$

Setting the derivative with respect to $y_{n+1}$ equal to zero and solving gives

$$
\hat{y}_{n+1} = \frac{\frac{n\bar{y}_n}{n+1}}{1 - \frac{1}{n+1}} = \bar{y}_n
\tag{84}
$$

That is, the frequentist Shtarkov predictor for the normal mean family of experts at time $n+1$ using $y^n$ is just the sample mean. To verify (84) maximises (8), we note the second derivative of (83) w.r.t. $y_{n+1}$ is negative:

$$
\begin{aligned}
\frac{d^2}{dy_{n+1}^2}\ln p(y^{n+1}|\hat{\mu}_{n+1}, \sigma^2) &= -\frac{1}{2\sigma^2}\left(2 - \frac{2}{n+1}\right) \\
&= \frac{1}{\sigma^2}\frac{-n}{n+1} < 0.
\end{aligned}
\tag{85}
$$

For the normal case with mean and variance unknown, we verify that $\hat{y}_{n+1}$ maximizes (16). So, we check if the second derivative of (16) w.r.t. $y_{n+1}$ is negative at $\hat{y}_{n+1}$. Note that

$$
\begin{aligned}
&\frac{d^2}{dy_{n+1}^2}\ln p(y^{n+1}|\hat{\mu}_{n+1}, \hat{\sigma^2}_{n+1}) \\
&= -\frac{\{\sum_{i=1}^{n} y_i^2 + y_{n+1}^2 - \frac{(n\bar{y}_n + y_{n+1})^2}{n+1}\} - (y_{n+1} - \bar{y}_n)\left\{2y_{n+1} - \frac{2(n\bar{y}_n + y_{n+1})}{n+1}\right\}}{[\sum_{i=1}^{n} y_i^2 + y_{n+1}^2 - \frac{1}{n+1}(n\bar{y}_n + y_{n+1})^2]^2}
\end{aligned}
\tag{86}
$$

The denominator of (86) is strictly positive. So, we will just focus on the numerator. Let us rename it $Num$. So, the numerator is

$$
\begin{aligned}
Num \;=\; & -\Bigg[\sum_{i=1}^{n} y_i^2 + y_{n+1}^2 - \frac{(n\bar{y}_n + y_{n+1})^2}{n+1}\} \\
& -(y_{n+1} - \bar{y}_n)\bigg\{2y_{n+1} - \frac{2(n\bar{y}_n + y_{n+1})}{n+1}\bigg\}\Bigg].
\end{aligned}
\tag{87}
$$

Substituting $y_{n+1}$ with $\hat{y}_{n+1} = \bar{y}_n$ in (87), we have,

$$
\begin{aligned}
Num \;=\; & -\Bigg[\sum_{i=1}^{n} y_i^2 + \bar{y}_n^2 - \frac{(n\bar{y}_n + \bar{y}_n)^2}{n+1}\Bigg] \\
\;=\; & -\sum_{i=1}^{n}(y_i - \bar{y}_n)^2 < 0.
\end{aligned}
\tag{88}
$$

Hence, (86) $< 0$. So, $\hat{y}_{n+1}$ maximizes (16).

Next we turn to the normal Bayes Shtarkov cases. For unknown mean but fixed variance, expression (18) gives

$$
\begin{aligned}
& \frac{d}{dy_{n+1}} \ln p(y^{n+1}|\hat{\mu}_{n+1}, \sigma^2, \mu_0, \sigma_0^2) \\
=\; & \frac{1}{\sigma^4 \tau_{n+1}} \sum_{i=1}^{n} y_i - \frac{n}{\sigma^4 \tau_{n+1}^2}\bigg(\frac{\mu_0}{\sigma_0^2} + \frac{n\bar{y}_n + y_{n+1}}{\sigma^2}\bigg) - \frac{y_{n+1}}{\sigma^2} + \frac{y_{n+1}}{\sigma^4 \tau_{n+1}} \\
& + \frac{1}{\sigma^2 \tau_{n+1}}\bigg(\frac{\mu_0}{\sigma_0^2} + \frac{n\bar{y}_n + y_{n+1}}{\sigma^2}\bigg) - \frac{1}{\sigma^4 \tau_{n+1}^2}\bigg(\frac{\mu_0}{\sigma_0^2} + \frac{n\bar{y}_n + y_{n+1}}{\sigma^2}\bigg) \\
& - \frac{1}{\sigma^2 \sigma_0^2 \tau_{n+1}^2}\bigg(\frac{\mu_0}{\sigma_0^2} + \frac{n\bar{y}_n + y_{n+1}}{\sigma^2}\bigg) + \frac{1}{\sigma^2 \sigma_0^2 \tau_{n+1}}.
\end{aligned}
\tag{89}
$$

Now, solving $\frac{d}{dy_{n+1}} \ln p(y^{n+1}|\hat{\mu}_{n+1}, \sigma^2, \mu_0, \sigma_0^2) = 0$ to find $\hat{y}_{n+1}$ we obtain $\hat{y}_{n+1} = \hat{\mu}_{MPLE}$.

It can be verified this is a maximum: differentiating again w.r.t $y_{n+1}$ gives

$$\frac{d^2}{dy_{n+1}^2} \ln p(y^{n+1}|\hat{\mu}_{n+1}, \sigma^2, \sigma_0^2, \mu_0)$$

$$= -\frac{n}{\sigma^4 \tau_{n+1}^2} \frac{1}{\sigma^2} - \frac{1}{\sigma^2} + \frac{1}{\sigma^4 \tau_{n+1}} + \frac{1}{\sigma^4 \tau_{n+1}} - \frac{1}{\sigma^6 \tau_{n+1}^2} - \frac{1}{\sigma^4 \sigma_0^2 \tau_{n+1}^2}$$

$$= \frac{1}{\sigma^2} \left[ -\frac{n\sigma^4}{\tau_{n+1}^2} - 1 + \frac{1}{\sigma^2 \tau_{n+1}} - \frac{1}{\sigma^4 \tau_{n+1}^2} - \frac{1}{\sigma^2 \sigma_0^2 \tau_{n+1}} \right]$$

$$= \frac{1}{\sigma^2} \left[ -\frac{(n+1)\sigma^4}{\tau_{n+1}^2} - \frac{1}{\sigma^2 \sigma_0^2 \tau_{n+1}} - 1 + \frac{1}{\sigma^2 \tau_{n+1}} \right]$$

$$= \frac{1}{\sigma^2} \left[ -\frac{1}{\sigma^2 \tau_{n+1}^2} \left( \frac{n+1}{\sigma^2} + \frac{1}{\sigma_0^2} \right) - 1 + \frac{2}{\sigma^2 \tau_{n+1}} \right]$$

$$= \frac{1}{\sigma^2} \left[ -\frac{1}{\sigma^2 \tau_{n+1}^2} \tau_{n+1} - 1 + \frac{2}{\sigma^2 \tau_{n+1}} \right]$$

$$= \frac{1}{\sigma^2} \left[ \frac{1}{\sigma^2 \tau_{n+1}^2} - 1 \right]. \tag{90}$$

Now,

$$\frac{1}{\sigma^2 \tau_{n+1}^2} = \frac{\frac{1}{\sigma^2}}{\frac{n+1}{\sigma^2} + \frac{1}{\sigma_0^2}} = \frac{\frac{1}{\sigma^2}}{\left( \frac{n}{\sigma^2} + \frac{1}{\sigma_0^2} \right) + \frac{1}{\sigma^2}} < 1. \tag{91}$$

Thus at $\hat{y}_{n+1} = \hat{\mu}_{MPLE}$,

$$\frac{d^2}{dy_{n+1}^2} \ln p(y^{n+1}|\hat{\mu}_{n+1}, \sigma^2, \sigma_0^2, \mu_0) < 0. \tag{92}$$

For the Bayes Shtarkov predictor for the normal case with known mean but unknown variance we recall (22) for $n+1$:

$$p(y^{n+1}|\hat{\sigma}_{n+1}^2, \mu, \alpha, \beta) \quad \propto \quad \left[ \frac{\alpha + \frac{n+1}{2} + 1}{\beta + \frac{1}{2} \sum_{i=1}^{n} (y_i - \mu)^2 + \frac{1}{2}(y_{n+1} - \mu)^2} \right]^{\alpha + \frac{n+1}{2} + 1}$$

$$\times e^{-(\alpha + \frac{n+1}{2} + 1)}. \tag{93}$$

Taking logarithm on both sides of (93), we have

$$\ln p(y^{n+1}|\hat{\sigma}_{n+1}^2, \mu, \alpha, \beta) \propto \left( \alpha + \frac{n+1}{2} + 1 \right) \ln \left( \alpha + \frac{n+1}{2} + 1 \right)$$

$$- \left( \alpha + \frac{n+1}{2} + 1 \right) \ln \left[ \beta + \frac{1}{2} \sum_{i=1}^{n} (y_i - \mu)^2 + \frac{1}{2}(y_{n+1} - \mu)^2 \right]. \tag{94}$$

Differentiating both sides of (94) w.r.t $y_{n+1}$ and equating to 0 gives:

$$
\begin{aligned}
\frac{d}{dy_{n+1}} \ln p(y^{n+1}|\hat{\sigma}^2_{n+1}, \mu, \alpha, \beta) &= 0 \\
\implies -\left(\alpha + \frac{n+1}{2} + 1\right) \frac{\frac{1}{2}2(y_{n+1} - \mu)}{\beta + \frac{1}{2}\sum_{i=1}^{n}(y_i - \mu)^2 + \frac{1}{2}(y_{n+1} - \mu)^2} &= 0 \\
\implies \hat{y}_{n+1} &= \mu.
\end{aligned}
\tag{95}
$$

Differentiating $\ln p(y^{n+1}|\hat{\sigma}^2_{n+1}, \mu, \alpha, \beta)$ twice gives

$$
\begin{aligned}
&\frac{d^2}{dy_{n+1}^2} \ln p(y^{n+1}|\hat{\sigma}^2_{n+1}, \mu, \alpha, \beta) \\
={}& \frac{-(\alpha + \frac{n}{2} + 1)}{[\beta + \frac{1}{2}\sum_{i=1}^{n}(y_i - \mu)^2 + \frac{1}{2}(y_{n+1} - \mu)^2]^2} \\
&\times \left[1\left\{\beta + \frac{1}{2}\sum_{i=1}^{n}(y_i - \mu)^2 + \frac{1}{2}(y_{n+1} - \mu)^2\right\} - (y_{n+1} - \mu)\frac{1}{2}2(y_{n+1} - \mu)\right].
\end{aligned}
\tag{96}
$$

At $\hat{y}_{n+1} = \mu$,

$$
\begin{aligned}
&\frac{d^2}{dy_{n+1}^2} \ln p(y^{n+1}|\hat{\sigma}^2_{n+1}, \mu, \alpha, \beta) \\
={}& \frac{-(\alpha + \frac{n}{2} + 1)}{[\beta + \frac{1}{2}\sum_{i=1}^{n}(y_i - \mu)^2]^2}[\beta + \frac{1}{2}\sum_{i=1}^{n}(y_i - \mu)^2] \\
={}& \frac{-(\alpha + \frac{n}{2} + 1)}{\beta + \frac{1}{2}\sum_{i=1}^{n}(y_i - \mu)^2} < 0.
\end{aligned}
\tag{97}
$$

To conclude this subsection, we derive the Bayes Shtarkov predictor for the normal with both $\mu$ and $\sigma$ unknown. Using the identity

$$
\begin{aligned}
&\sum_{i=1}^{n}(y_i - \mu)^2 + \left(\frac{\mu - \mu_0}{\sigma_0}\right)^2 \\
={}& \left(n + \frac{1}{\sigma_0^2}\right)\left[\mu - \frac{n\bar{y} + \frac{\mu_0}{\sigma_0^2}}{n + \frac{1}{\sigma_0^2}}\right]^2 - \frac{(n\bar{y} + \frac{\mu_0}{\sigma_0^2})^2}{n + \frac{1}{\sigma_0^2}} + \left(\sum_{i=1}^{n}y_i^2 + \frac{\mu_0^2}{\sigma_0^2}\right).
\end{aligned}
$$

in (25) gives

$$p(y^n|\mu,\sigma^2) \times p(\mu|\mu_0,\sigma_0^2) \times p(\sigma^2|\alpha,\beta)$$

$$\propto \left(\frac{1}{\sigma^2}\right)^{\frac{n}{2}+\alpha+1}\left(\frac{1}{\sigma^2}\right)^{\frac{1}{2}}e^{-\frac{\beta}{\sigma^2}}$$

$$\times e^{-\frac{1}{2\sigma^2}\left[\left(n+\frac{1}{\sigma_0^2}\right)\left\{\mu-\frac{n\bar{y}+\frac{\mu_0}{\sigma_0^2}}{n+\frac{1}{\sigma_0^2}}\right\}^2 - \frac{(n\bar{y}+\frac{\mu_0}{\sigma_0^2})^2}{n+\frac{1}{\sigma_0^2}} + \left(\sum_{i=1}^n y_i^2 + \frac{\mu_0^2}{\sigma_0^2}\right)\right]}$$

$$= \left(\frac{1}{\sigma^2}\right)^{\frac{n}{2}+\alpha+1}\left(\frac{1}{\sigma^2}\right)^{\frac{1}{2}}e^{-\frac{1}{\sigma^2}\left[\beta+\frac{1}{2}\left(\sum_{i=1}^n y_i^2 + \frac{\mu_0^2}{\sigma_0^2}\right) - \frac{1}{2}\frac{(n\bar{y}+\frac{\mu_0}{\sigma_0^2})^2}{n+\frac{1}{\sigma_0^2}}\right]}$$

$$\times e^{-\frac{1}{2\sigma^2}\frac{n\sigma_0^2+1}{\sigma_0^2}\left(\mu-\frac{n\bar{y}+\frac{\mu_0}{\sigma_0^2}}{n+\frac{1}{\sigma_0^2}}\right)^2}, \tag{98}$$

and hence

$$\mu|\sigma^2,\sigma_0^2,\mu_0 \sim \mathcal{N}\left(\frac{n\bar{y}+\frac{\mu_0}{\sigma_0^2}}{n+\frac{1}{\sigma_0^2}}, \frac{\sigma^2\sigma_0^2}{n\sigma_0^2+1}\right) \tag{99}$$

$$\sigma^2|\alpha,\beta \sim \mathcal{IG}\left(\alpha+\frac{n}{2}, \beta+\frac{1}{2}\left\{\sum_{i=1}^n y_i^2 + \frac{\mu_0^2}{\sigma_0^2} - \frac{(n\bar{y}+\frac{\mu_0}{\sigma_0^2})^2}{n+\frac{1}{\sigma_0^2}}\right\}\right). \tag{100}$$

Thus, we have

$$\hat{\mu}_{MPLE} = \frac{n\bar{y}+\frac{\mu_0}{\sigma_0^2}}{n+\frac{1}{\sigma_0^2}} \tag{101}$$

$$\hat{\sigma}^2_{MPLE} = \frac{\beta+\frac{1}{2}\left\{\sum_{i=1}^n y_i^2 + \frac{\mu_0^2}{\sigma_0^2} - \frac{(n\bar{y}+\frac{\mu_0}{\sigma_0^2})^2}{n+\frac{1}{\sigma_0^2}}\right\}}{\alpha+\frac{n}{2}+1}. \tag{102}$$

Replacing $\mu$ by $\hat{\mu}_{MPLE}$ and $\sigma^2$ by $\hat{\sigma}^2_{MPLE}$ for $n+1$ copies of $y$ in (98) and using

(9) for $\bar{y}_{n+1}$ gives

$$p(y^{n+1}|\hat{\mu}_{n+1,MPLE}, \hat{\sigma^2}_{n+1,MPLE}) \times p(\hat{\mu}_{n+1,MPLE}|\mu_0, \sigma_0^2)$$
$$\times p(\hat{\sigma^2}_{n+1,MPLE}|\alpha, \beta)$$

$$\propto \left[ \frac{\alpha + \frac{n+1}{2} + 1}{\beta + \frac{1}{2}\left\{\sum_{i=1}^{n} y_i^2 + y_{n+1}^2 + \frac{\mu_0^2}{\sigma_0^2} - \frac{\left\{(n+1)\frac{1}{n+1}(n\bar{y}_n + y_{n+1}) + \frac{\mu_0}{\sigma_0^2}\right\}^2}{n+1+\frac{1}{\sigma_0^2}}\right\}} \right]^{\alpha + \frac{n+2}{2} + 1}$$

$$\times e^{-\frac{(\alpha+\frac{n+1}{2}+1)\left[\beta+\frac{1}{2}\left\{\sum_{i=1}^{n} y_i^2 + y_{n+1}^2 + \frac{\mu_0^2}{\sigma_0^2} - \frac{\left\{(n+1)\frac{1}{n+1}(n\bar{y}_n + y_{n+1}) + \frac{\mu_0}{\sigma_0^2}\right\}^2}{n+1+\frac{1}{\sigma_0^2}}\right\}\right]}{\beta+\frac{1}{2}\left\{\Sigma_{i=1}^n y_i^2 + y_{n+1}^2 + \frac{\mu_0^2}{\sigma_0^2} - \frac{\left\{(n+1)\frac{1}{n+1}(n\bar{y}_n + y_{n+1}) + \frac{\mu_0}{\sigma_0^2}\right\}^2}{n+1+\frac{1}{\sigma_0^2}}\right\}}}$$

$$\times e^{-\frac{\frac{(n+1)\bar{y}_{n+1} + \frac{\mu_0}{\sigma_0^2}}{n+1+\frac{1}{\sigma_0^2}} - \frac{(n+1)\bar{y}_{n+1} + \frac{\mu_0}{\sigma_0^2}}{n+1+\frac{1}{\sigma_0^2}}}{2\hat{\sigma^2}_{n+1,MPLE}}}.$$

$$= \left[ \frac{\alpha + \frac{n+1}{2} + 1}{\beta + \frac{1}{2}\left\{\sum_{i=1}^{n} y_i^2 + y_{n+1}^2 + \frac{\mu_0^2}{\sigma_0^2} - \frac{\left\{(n\bar{y}_n + y_{n+1}) + \frac{\mu_0}{\sigma_0^2}\right\}^2}{n+1+\frac{1}{\sigma_0^2}}\right\}} \right]^{\alpha + \frac{n+2}{2} + 1}$$

$$\times e^{-(\alpha + \frac{n+1}{2} + 1)}. \tag{103}$$

Taking logarithms on both sides of (103), we get

$$\ln[p(y^{n+1}|\hat{\mu}_{n+1,MPLE}, \hat{\sigma^2}_{n+1,MPLE}) \times p(\hat{\mu}_{n+1,MPLE}|\mu_0, \sigma_0^2)$$
$$\times p(\hat{\sigma^2}_{n+1,MPLE}|\alpha, \beta)]$$
$$\propto \left(\alpha + \frac{n+2}{2} + 1\right)\ln\left(\alpha + \frac{n+1}{2} + 1\right) - \left(\alpha + \frac{n+2}{2} + 1\right)$$
$$\times \ln\left[\beta + \frac{1}{2}\left\{\sum_{i=1}^{n} y_i^2 + y_{n+1}^2 + \frac{\mu_0^2}{\sigma_0^2} - \frac{\left\{(n\bar{y}_n + y_{n+1}) + \frac{\mu_0}{\sigma_0^2}\right\}^2}{n+1+\frac{1}{\sigma_0^2}}\right\}\right]$$
$$- \left(\alpha + \frac{n+1}{2} + 1\right). \tag{104}$$

Differentiating both sides of (104) w.r.t $y_{n+1}$ and equating to 0 gives

$$\frac{-(\alpha + \frac{n+2}{2} + 1) \times \frac{1}{2}\left[2y_{n+1} - \frac{2\left(n\bar{y}_n + y_{n+1} + \frac{\mu_0}{\sigma_0^2}\right)}{n+1+\frac{1}{\sigma_0^2}}\right]}{\beta + \frac{1}{2}\left\{\sum_{i=1}^n y_i^2 + y_{n+1}^2 + \frac{\mu_0^2}{\sigma_0^2} - \frac{\left\{(n\bar{y}_n + y_{n+1}) + \frac{\mu_0}{\sigma_0^2}\right\}^2}{n+1+\frac{1}{\sigma_0^2}}\right\}} = 0$$

$$\implies y_{n+1} - \frac{2\left(n\bar{y}_n + y_{n+1} + \frac{\mu_0}{\sigma_0^2}\right)}{n+1+\frac{1}{\sigma_0^2}} = 0$$

$$\implies \hat{y}_{n+1} = \frac{n\bar{y}_n + \frac{\mu_0}{\sigma_0^2}}{n+\frac{1}{\sigma_0^2}}$$

as the point predictor as claimed.

To verify it is a maximum, we differentiate

$$\frac{d}{dy_{n+1}}\ln[p(y^{n+1}|\hat{\mu}_{n+1,MPLE}, \hat{\sigma^2}_{n+1,MPLE}) \times p(\hat{\mu}_{n+1,MPLE}|\mu_0, \sigma_0^2)$$

$$\times p(\hat{\sigma^2}_{n+1,MPLE}|\alpha, \beta)]$$

w.r.t. $y_{n+1}$ to get

$$\frac{d^2}{dy_{n+1}^2}\ln[p(y^{n+1}|\hat{\mu}_{n+1,MPLE}, \hat{\sigma^2}_{n+1,MPLE}) \times p(\hat{\mu}_{n+1,MPLE}|\mu_0, \sigma_0^2)$$

$$\times p(\hat{\sigma^2}_{n+1,MPLE}|\alpha, \beta)]$$

$$= \frac{-(\alpha + \frac{n}{2} + 2)}{\left[\beta + \frac{1}{2}\left\{\sum_{i=1}^n y_i^2 + y_{n+1}^2 + \frac{\mu_0}{\sigma_0^2} - \frac{(n\bar{y}_n + y_{n+1} + \frac{\mu_0}{\sigma_0^2})^2}{n+1+\frac{1}{\sigma_0^2}}\right\}\right]^2}\left[\left(1 - \frac{1}{n+1+\frac{1}{\sigma_0^2}}\right)^2\right.$$

$$\times \left\{\beta + \frac{1}{2}\left(\sum_{i=1}^n y_i^2 + y_{n+1}^2 + \frac{\mu_0}{\sigma_0^2} - \frac{(n\bar{y}_n + y_{n+1} + \frac{\mu_0}{\sigma_0^2})^2}{n+1+\frac{1}{\sigma_0^2}}\right)\right\}$$

$$\left. - \left(y_{n+1} - \frac{n\bar{y}_n + y_{n+1} + \frac{\mu_0}{\sigma_0^2}}{n+1+\frac{1}{\sigma_0^2}}\right)\left\{\frac{1}{2}\left(2y_{n+1} - 2\frac{(n\bar{y}_n + y_{n+1} + \frac{\mu_0}{\sigma_0^2})}{n+1+\frac{1}{\sigma_0^2}}\right)\right\}\right].$$

Renaming, we set

$$A = \frac{-(\alpha + \frac{n}{2} + 2)}{\left[\beta + \frac{1}{2}\left\{\sum_{i=1}^n y_i^2 + y_{n+1}^2 + \frac{\mu_0}{\sigma_0^2} - \frac{(n\bar{y}_n + y_{n+1} + \frac{\mu_0}{\sigma_0^2})^2}{n+1+\frac{1}{\sigma_0^2}}\right\}\right]^2}$$

$$B = \left(1 - \frac{1}{n+1+\frac{1}{\sigma_0^2}}\right)\left[\beta + \frac{1}{2}\left\{\sum_{i=1}^n y_i^2 + y_{n+1}^2 + \frac{\mu_0}{\sigma_0^2} - \frac{(n\bar{y}_n + y_{n+1} + \frac{\mu_0}{\sigma_0^2})^2}{n+1+\frac{1}{\sigma_0^2}}\right\}\right]$$

$$C = \left(y_{n+1} - \frac{n\bar{y}_n + y_{n+1} + \frac{\mu_0}{\sigma_0^2}}{n+1+\frac{1}{\sigma_0^2}}\right)\left[\frac{1}{2}\left\{2y_{n+1} - 2\frac{(n\bar{y}_n + y_{n+1} + \frac{\mu_0}{\sigma_0^2})}{n+1+\frac{1}{\sigma_0^2}}\right\}\right].$$

Now,

$$\frac{d^2}{dy_{n+1}^2}\ln[p(y^{n+1}|\hat{\mu}_{n+1,MPLE},\hat{\sigma}^2_{n+1,MPLE}) \times p(\hat{\mu}_{n+1,MPLE}|\mu_0,\sigma_0^2)$$

$$\times p(\hat{\sigma}^2_{n+1,MPLE}|\alpha,\beta)]$$

$$= \text{Term A}(\text{Term B} - \text{Term C}). \tag{105}$$

If we put $\hat{y}_{n+1}$ in each of Term A, Term B, and Term C we get the following expressions. First, the easiest one is

$$\text{Term C} = \left(\hat{y}_{n+1} - \frac{n\bar{y}_n + y_{n+1} + \frac{\mu_0}{\sigma_0^2}}{n+1+\frac{1}{\sigma_0^2}}\right)\left[\frac{1}{2}\left\{2\hat{y}_{n+1} - 2\frac{(n\bar{y}_n+\hat{y}_{n+1}+\frac{\mu_0}{\sigma_0^2})}{n+1+\frac{1}{\sigma_0^2}}\right\}\right]$$

$$= \left(\hat{y}_{n+1} - \frac{n\bar{y}_n + \hat{y}_{n+1} + \frac{\mu_0}{\sigma_0^2}}{n+1+\frac{1}{\sigma_0^2}}\right)^2$$

$$= \frac{n\hat{y}_{n+1} + \hat{y}_{n+1} + \frac{y_{n+1}}{\sigma_0^2} - n\bar{y}_n - \hat{y}_{n+1} - \frac{\mu_0}{\sigma_0^2}}{n+1+\frac{1}{\sigma_0^2}}$$

$$= \frac{\hat{y}_{n+1}(n+\frac{1}{\sigma_0^2}) - (n\bar{y}_n + \frac{\mu_0}{\sigma_0^2})}{n+1+\frac{1}{\sigma_0^2}}$$

$$= \frac{\frac{n\bar{y}_n+\frac{\mu_0}{\sigma_0^2}}{n+\frac{1}{\sigma_0^2}}(n+\frac{1}{\sigma_0^2}) - (n\bar{y}_n + \frac{\mu_0}{\sigma_0^2})}{n+1+\frac{1}{\sigma_0^2}} = 0. \tag{106}$$

Hence, (105) becomes

$$\frac{d^2}{dy_{n+1}^2}\ln[p(y^{n+1}|\hat{\mu}_{n+1,MPLE},\hat{\sigma}^2_{n+1,MPLE}) \times p(\hat{\mu}_{n+1,MPLE}|\mu_0,\sigma_0^2)$$

$$\times p(\hat{\sigma}^2_{n+1,MPLE}|\alpha,\beta)]$$

$$= \text{Term A Term B}.$$

More explicitly, the product of $A$ and $B$ is

$$\frac{-(\alpha+\frac{n}{2}+2)}{\left[\beta+\frac{1}{2}\left\{\sum_{i=1}^n y_i^2 + y_{n+1}^2 + \frac{\mu_0^2}{\sigma_0^2} - \frac{(n\bar{y}_n+y_{n+1}+\frac{\mu_0}{\sigma_0^2})^2}{n+1+\frac{1}{\sigma_0^2}}\right\}\right]^2}\frac{n+\frac{1}{\sigma_0^2}}{n+1+\sigma_0^2}$$

$$\times\left[\beta+\frac{1}{2}\left\{\sum_{i=1}^n y_i^2 + y_{n+1}^2 + \frac{\mu_0}{\sigma_0^2} - \frac{(n\bar{y}_n+y_{n+1}+\frac{\mu_0}{\sigma_0^2})^2}{n+1+\frac{1}{\sigma_0^2}}\right\}\right]$$

$$= \frac{-(\alpha+\frac{n}{2}+2)\frac{n+\frac{1}{\sigma_0^2}}{n+1+\sigma_0^2}}{\beta+\frac{1}{2}\left\{\sum_{i=1}^n y_i^2 + y_{n+1}^2 + \frac{\mu_0}{\sigma_0^2} - \frac{(n\bar{y}_n+y_{n+1}+\frac{\mu_0}{\sigma_0^2})^2}{n+1+\frac{1}{\sigma_0^2}}\right\}}.$$

So, at $\hat{y}_{n+1}$,

$$\frac{d^2}{dy_{n+1}^2} \ln[p(y^{n+1}|\hat{\mu}_{n+1,MPLE}, \hat{\sigma}^2_{n+1,MPLE}) \times p(\hat{\mu}_{n+1,MPLE}|\mu_0, \sigma_0^2)$$
$$\times p(\hat{\sigma}^2_{n+1,MPLE}|\alpha, \beta)]$$

$$= \frac{-(\alpha + \frac{n}{2} + 2)\frac{n + \frac{1}{\sigma_0^2}}{n+1+\sigma_0^2}}{\beta + \frac{1}{2}\left\{\sum_{i=1}^n y_i^2 + \hat{y}_{n+1}^2 + \frac{\mu_0^2}{\sigma_0^2} - \frac{(n\bar{y}_n + \hat{y}_{n+1} + \frac{\mu_0}{\sigma_0^2})^2}{n+1+\frac{1}{\sigma_0^2}}\right\}}$$

$$= \frac{-(\alpha + \frac{n}{2} + 2)\frac{n + \frac{1}{\sigma_0^2}}{n+1+\sigma_0^2}}{\beta + \frac{1}{2}\left\{\sum_{i=1}^n y_i^2 + (\frac{n\bar{y}_n + \frac{\mu_0}{\sigma_0^2}}{n + \frac{1}{\sigma_0^2}})^2 + \frac{\mu_0^2}{\sigma_0^2} - \frac{(n\bar{y}_n + \frac{n\bar{y}_n + \frac{\mu_0}{\sigma_0^2}}{n+\frac{1}{\sigma_0^2}} + \frac{\mu_0}{\sigma_0^2})^2}{n+1+\frac{1}{\sigma_0^2}}\right\}}$$

$$= \frac{-(\alpha + \frac{n}{2} + 2)\frac{n + \frac{1}{\sigma_0^2}}{n+1+\sigma_0^2}}{\beta + \frac{1}{2}\left\{\sum_{i=1}^n y_i^2 - \frac{(n\bar{y}_n + \frac{\mu_0}{\sigma_0^2})^2}{n + \frac{1}{\sigma_0^2}} + \frac{\mu_0}{\sigma_0^2}\right\}}. \tag{107}$$

By definition, if $y_{n+1}$ is a maximum, it will maximize the joint likelihood function

$$p(y^{n+1}|\hat{\mu}_{n+1,MPLE}, \hat{\sigma}^2_{n+1,MPLE}) \times p(\hat{\mu}_{n+1,MPLE}|\mu_0, \sigma_0^2) \times p(\hat{\sigma}^2_{n+1,MPLE}|\alpha, \beta).$$

For simplicity, we only look at the case $\mu_0 = 0$ and $\sigma_0 = 1$. Then, (107) becomes

$$\frac{d^2}{dy_{n+1}^2} \ln[p(y^{n+1}|\hat{\mu}_{n+1,MPLE}, \hat{\sigma}^2_{n+1,MPLE}) \times p(\hat{\mu}_{n+1,MPLE})$$
$$\times p(\hat{\sigma}^2_{n+1,MPLE}|\alpha, \beta)]$$
$$= -\left(\frac{n+1}{n+2}\right)\frac{\alpha + \frac{n}{2} + 2}{\beta + \frac{1}{2}(\sum_{i=1}^n y_i^2 - \frac{n^2}{n+1}\bar{y}_n^2)}. \tag{108}$$

Since

$$\sum_{i=1}^n y_i^2 - \frac{n^2}{n+1}\bar{y}_n^2 = \sum_{i=1}^n y_i^2 - n\bar{y}_n^2 + n\bar{y}_n^2 - \frac{n^2}{n+1}\bar{y}_n^2$$
$$= (n-1)Var(Y) + \frac{n}{n+1}\bar{y}_n^2 > 0,$$

we see that (108) < 0. Other values of $\mu_0$ and $\sigma_0$ are similar.

### B.2. Binomial Cases

The omitted derivation is for (32). For $y_1, y_2, \cdots, y_n, y_{n+1}$, (31) is:

$$
\begin{aligned}
p(y^{n+1}; \hat{\theta}_{MLE}) &= \left\{ \prod_{i=1}^{n+1} \binom{N}{y_i} \right\} \left( \frac{\bar{y}_{n+1}}{N} \right)^{\sum_{i=1}^{n+1} y_i} \left( \frac{N - \bar{y}_{n+1}}{N} \right)^{N(n+1) - \sum_{i=1}^{n+1} y_i} \\
&\propto \binom{N}{y_{n+1}} \left\{ \frac{n\bar{y}_n + y_{n+1}}{N(n+1)} \right\}^{n\bar{y}_n + y_{n+1}} \\
&\times \left\{ \frac{N(n+1) - (n\bar{y}_n + y_{n+1})}{N(n+1)} \right\}^{N(n+1) - (n\bar{y}_n + y_{n+1})} \\
&\propto \binom{N}{y_{n+1}} (n\bar{y}_n + y_{n+1})^{n\bar{y}_n + y_{n+1}} \left\{ \frac{1}{N(n+1)} \right\}^{n\bar{y}_n} \\
&\times \left\{ \frac{1}{N(n+1)} \right\}^{y_{n+1}} \\
&\times \{ N(n+1) - (n\bar{y}_n + y_{n+1}) \}^{N(n+1) - (n\bar{y}_n + y_{n+1})} \\
&\times \left\{ \frac{1}{N(n+1)} \right\}^{-n\bar{y}_n} \left\{ \frac{1}{N(n+1)} \right\}^{-y_{n+1}} \left\{ \frac{1}{N(n+1)} \right\}^{N(n+1)} \\
&\propto \binom{N}{y_{n+1}} (n\bar{y}_n + y_{n+1})^{n\bar{y}_n + y_{n+1}} \\
&\times \{ N(n+1) - (n\bar{y}_n + y_{n+1}) \}^{N(n+1) - (n\bar{y}_n + y_{n+1})}
\end{aligned}
\tag{109}
$$

Taking logarithms on both sides of (109), we get (32).

For the beta-binomial case, the derivation for (34) is the following. Multiply (30) and (33) to get

$$
\begin{aligned}
p(y^n | \theta) \times w(\theta | \alpha, \beta) &= \left\{ \prod_{i=1}^{n} \binom{N}{y_i} \right\} \theta^{\sum_{i=1}^{n} y_i} (1-\theta)^{Nn - \sum_{i=1}^{n} y_i} \\
&\times \frac{1}{Beta(\alpha, \beta)} \theta^{\alpha-1} (1-\theta)^{\beta-1} \tag{110} \\
&\propto \theta^{\alpha + \sum_{i=1}^{n} y_i - 1} (1-\theta)^{\beta + Nn - \sum_{i=1}^{n} y_i - 1} \\
&= \theta^{\alpha + n\bar{y}_n - 1} (1-\theta)^{\beta + Nn - n\bar{y}_n - 1}. \tag{111}
\end{aligned}
$$

Hence, $\theta | y^n, \alpha, \beta \sim Beta(\alpha + n\bar{y}_n, \beta + Nn - n\bar{y}_n)$ and

$$
\hat{\theta}_{MPLE} = \frac{\alpha + n\bar{y}_n - 1}{\alpha + \beta + Nn - 2}.
\tag{112}
$$

Substituting (112) into (110), we have

$$
\begin{aligned}
&p(y^n|\hat{\theta}_{MPLE})w(\hat{\theta}_{MPLE}, \alpha, \beta) \\
&= \frac{1}{Beta(\alpha,\beta)}\left\{\prod_{i=1}^{n}\binom{N}{y_i}\right\}\left[\frac{\alpha + n\bar{y}_n - 1}{\alpha + \beta + Nn - 2}\right]^{\alpha + n\bar{y}_n - 1} \\
&\quad \times \left[\frac{\beta + Nn - n\bar{y}_n - 1}{\alpha + \beta + Nn - 2}\right]^{\beta + Nn - n\bar{y}_n - 1}.
\end{aligned}
\tag{113}
$$

So, for $y_1, y_2, \cdots, y_n, y_{n+1}$, we can rewrite (113) as:

$$
\begin{aligned}
&p(y^{n+1}|\hat{\theta}_{MPLE})w(\hat{\theta}_{MPLE}, \alpha, \beta) \\
&\propto \binom{N}{y_{n+1}}(\alpha + n\bar{y}_n + y_{n+1} - 1)^{\alpha + n\bar{y}_n + y_{n+1} - 1} \\
&\quad \times (\beta + N(n+1) - n\bar{y}_n - y_{n+1} - 1)^{\beta + N(n+1) - n\bar{y}_n - y_{n+1} - 1}.
\end{aligned}
\tag{114}
$$

Taking logarithms on both sides of (114) we get (34).

### *B.3. Gamma Cases*

Taking the logarithm on both sides of (45), we get

$$
\ln p_\alpha(y^{n+1}|\hat{\theta}_{MLE}) \quad \propto \quad (\alpha - 1)\ln y_{n+1} - (n+1)\alpha \ln\left(\sum_{i=1}^{n} y_i + y_{n+1}\right).
\tag{115}
$$

Differentiating (115), w.r.t. $y_{n+1}$ and setting it to 0 gives

$$
\hat{y}_{n+1} \quad = \quad \frac{n(\alpha - 1)\bar{y}_n}{n\alpha + 1}.
\tag{116}
$$

To be sure this $\hat{y}_{n+1}$ is a maximum we look at three ranges of $\alpha$. If $\alpha = 1$, we return to the exponential in Subsec. 2.2.3. For $\alpha > 1$, we verify the second derivative of (115) at $\hat{y}_{n+1}$ is negative. We have

$$
\frac{d^2}{dy_{n+1}^2}\ln p_\alpha(y^{n+1}|\hat{\theta}_{MLE}) = \frac{y_{n+1}^2(n+1)\alpha + (1-\alpha)(n\bar{y}_n + y_{n+1})^2}{(n\bar{y}_n + y_{n+1})^2 y_{n+1}^2}.
\tag{117}
$$

The denominator in (117) is strictly positive. So, it is enough to look only at the numerator. Rename the numerator

$$
Q(y_{n+1}) = Q_{\bar{y}_n}(y_{n+1}) \quad = \quad y_{n+1}^2(1 + n\alpha) + (1-\alpha)(2y_{n+1}n\bar{y}_n + n^2\bar{y}_n^2).
\tag{118}
$$

Now, replacing $y_{n+1}$ with $\hat{y}_{n+1}$, we get,

$$
\begin{aligned}
Q(\hat{y}_{n+1}) \quad &= \quad \frac{n^2(\alpha-1)^2\bar{y}_n^2}{(n\alpha+1)^2}(1+n\alpha) + (1-\alpha)\frac{2n(\alpha-1)\bar{y}_n}{n\alpha+1}n\bar{y}_n + (1-\alpha)n^2\bar{y}_n^2 \\
&= \quad -\frac{n^2\bar{y}_n^2}{n\alpha+1}(\alpha-1)\alpha(n+1)
\end{aligned}
\tag{119}
$$

, For $\alpha > 1$, (119) is negative. Hence, (45) is maximised at $y_{n+1} = \hat{y}_{n+1}$.

For $\alpha \in (0, 1)$, we can see directly that (45) is maximized at $\hat{y}_{n+1} = 0$. Simply rewrite (45) as

$$p(y^{n+1}|\hat{\theta}_{MLE}, \alpha) \quad \propto \quad \left(\frac{1}{y_{n+1}}\right)^{1-\alpha} \left(\frac{1}{n\bar{y}_n + y_{n+1}}\right)^{(n+1)\alpha}. \tag{120}$$

Now, both terms are maximised at $y_{n+1} = 0$.

For the Bayesian Gamma Shtarkov family of experts we want to show (50). Start by substituting (49) in (48) to get

$$
\begin{aligned}
&p(y^n|\hat{\theta}_{MPLE}, \alpha) \times w(\hat{\theta}_{MPLE}|\alpha_0, \beta_0) \\
&= \frac{\beta_0^{\alpha_0}}{\Gamma(\alpha_0)} \left(\prod_{i=1}^{n} y_i^{\alpha-1}\right) \frac{1}{\Gamma(\alpha)^n} \left(\frac{n\alpha + \alpha_0 - 1}{n\bar{y}_n + \beta_0}\right)^{n\alpha + \alpha_0 - 1} \\
&\times e^{-\frac{n\alpha + \alpha_0 - 1}{n\bar{y}_n + \beta_0}(n\bar{y}_n + \beta_0)} \\
&\propto \left(\prod_{i=1}^{n} y_i^{\alpha-1}\right) \left(\frac{1}{n\bar{y}_n + \beta_0}\right)^{n\alpha + \alpha_0 - 1}. 
\end{aligned}
\tag{121}
$$

For $n + 1$ samples $y_1, y_2, \cdots, y_n, y_{n+1}$ (121) can be rewritten as

$$
\begin{aligned}
&p(y^{n+1}|\hat{\theta}_{MPLE}, \alpha) \times w(\hat{\theta}_{MPLE}|\alpha_0, \beta_0) \\
&\propto \left(\prod_{i=1}^{n} y_i^{\alpha-1}\right) y_{n+1}^{\alpha-1} \left(\frac{1}{\beta_0 + (n+1)\bar{y}_{n+1}}\right)^{(n+1)\alpha + \alpha_0 - 1} \\
&\propto y_{n+1}^{\alpha-1} \left(\frac{1}{\beta_0 + n\bar{y}_n + y_{n+1}}\right)^{(n+1)\alpha + \alpha_0 - 1}.
\end{aligned}
\tag{122}
$$

Taking logarithm on both sides of (122), we get

$$
\begin{aligned}
&\ln[p(y^{n+1}|\hat{\theta}_{MPLE}, \alpha)w(\hat{\theta}_{MPLE}|\alpha_0, \beta_0)] \\
&= (\alpha - 1)\ln y_{n+1} - \{(n+1)\alpha + \alpha_0 - 1\} \\
&\times \ln\left(\beta_0 + n\bar{y}_n + y_{n+1}\right)
\end{aligned}
\tag{123}
$$

Differentiating both sides of (123) w.r.t. $y_{n+1}$, we have

$$
\begin{aligned}
&\frac{d}{dy_{n+1}}[\ln p(y^{n+1}|\hat{\theta}_{MPLE}, \alpha) \times w(\hat{\theta}_{MPLE}|\alpha_0, \beta_0)] \\
&= \frac{\alpha - 1}{y_{n+1}} - \frac{(n+1)\alpha + \alpha_0 - 1}{\beta_0 + n\bar{y}_n + y_{n+1}}.
\end{aligned}
\tag{124}
$$

Equating (124) to 0 and solving for $y_{n+1}$ gives us (50).

To see that (50) is a maximum, start with $\alpha < 1$. Then,

$$p(y^{n+1}|\hat{\theta}_{MPLE}, \alpha) \times w(\hat{\theta}_{MPLE}|\alpha_0, \beta_0)$$

$$\propto \quad y_{n+1}^{\alpha-1} \left( \frac{1}{\beta_0 + n\bar{y}_n + y_{n+1}} \right)^{(n+1)\alpha + \alpha_0 - 1}.$$

$$= \quad \left( \frac{1}{y_{n+1}} \right)^{1-\alpha} \left( \frac{1}{\beta_0 + n\bar{y}_n + y_{n+1}} \right)^{(n+1)\alpha + \alpha_0 - 1}. \tag{125}$$

We can see directly that (125) is maximized at $\hat{y}_{n+1} = 0$.

For $\alpha = 1$,

$$p(y^{n+1}|\hat{\theta}_{MPLE}, \alpha = 1) \times w(\hat{\theta}_{MPLE}|\alpha_0, \beta_0)$$

$$\propto \quad \left( \frac{1}{y_{n+1}} \right)^{1-1} \left( \frac{1}{\beta_0 + n\bar{y}_n + y_{n+1}} \right)^{(n+1)+\alpha_0 - 1}. \tag{126}$$

which is also maximum at $\hat{y}_{n+1} = 0$.

For $\alpha > 1$, differentiate (124) w.r.t. $y_{n+1}$ to find

$$\frac{d^2}{d^2 y_{n+1}} \ln[p(y^{n+1}|\hat{\theta}_{MPLE}, \alpha) w(\hat{\theta}_{MPLE}|\alpha_0, \beta_0)]$$

$$= \quad \frac{-(\alpha-1)(\beta_0 + n\bar{y}_n + y_{n+1})^2 + [(n+1)\alpha + \alpha_0 - 1]y_{n+1}^2}{y_{n+1}^2(\beta_0 + n\bar{y}_n + y_{n+1})^2}$$

$$= \quad \frac{Numerator}{Denominator}. \tag{127}$$

The *Denominator* is positive and the *Numerator* at $\hat{y}_{n+1} = \frac{(\alpha-1)(\beta_0 + n\bar{y}_n)}{n\alpha + \alpha_0}$ is

$$-(\alpha-1)\left[ \beta_0 + n\bar{y}_n + \frac{(\alpha-1)(\beta_0 + n\bar{y}_n)}{n\alpha + \alpha_0} \right]^2$$

$$+ [(n+1)\alpha + \alpha_0 - 1]\frac{(\alpha-1)^2}{(n\alpha + \alpha_0)^2}(\beta_0 + n\bar{y}_n)^2$$

$$= \quad -(\alpha-1)(n\alpha + \alpha_0)(n\alpha + \alpha_0 + \alpha - 1)\frac{(\beta_0 + n\bar{y}_n)^2}{(n\alpha + \alpha_0)^2} < 0. \tag{128}$$

## Appendix C: Bayes Predictors

In this subsection, we give the results needed for the method of moments arguments used in the first two subsections of Sec. 3.

### C.1. Method of Moments for IID Bias in GPP's

Here we give the calculations used in Subsec. 3.1.2. We start with the first moment condition. To show (63), recall from (56) that

$$Y^n|a^n, \sigma^2 \sim \mathcal{N}(a^n, \sigma^2(I+K)_{n \times n}) \tag{129}$$

and so

$$Y'^n = (I + K)^{\frac{1}{2}} Y^n \sim \mathcal{N}(a^n, \sigma^2 I_{n \times n}). \tag{130}$$

Also from (57), we have

$$a^n \sim \mathcal{N}(\gamma 1^n, \sigma^2 \delta^2 I_{n \times n}). \tag{131}$$

If we define

$$S'_2 \;\; = \;\; \frac{1}{n-1} \sum_{i=1}^{n} (Y'_i - \bar{Y'})^2,$$

then

$$
\begin{aligned}
E(S'_2) \;\; &= \;\; E\left[ \frac{1}{n-1} \sum_{i=1}^{n} (Y'_i - \bar{Y'})^2 \right] \\
&= \;\; E\left[ \frac{1}{n-1} \left( \sum_{i=1}^{n} Y'^2_i - n\bar{Y'}^2 \right) \right].
\end{aligned} \tag{132}
$$

Now, from

$$E(Y'_i) = EE(Y'_i | a_i, \sigma^2) = \gamma, \tag{133}$$

we get

$$
\begin{aligned}
E(\bar{Y'}) = EE(\bar{Y'} | a^n, \sigma^2) \;\; &= \;\; E\left[ \frac{1}{n} \sum_{i=1}^{n} E(Y_i | a_i, \sigma^2) \right] \\
&= \;\; E\left[ \frac{1}{n} \sum_{i=1}^{n} a_i \right] = \gamma.
\end{aligned}
$$

We also have the identity

$$E(Y'^2_i) = Var(Y'_i) + E^2(Y'_i). \tag{134}$$

Now

$$
\begin{aligned}
Var(Y'_i) \;\; &= \;\; EVar(Y'_i | a_i, \sigma^2) + VarE(Y'_i | a_i, \sigma^2) \\
&= \;\; E(\sigma^2) + Var(a_i | \sigma^2) = \sigma^2 + \sigma^2 \delta^2.
\end{aligned} \tag{135}
$$

From (133), (134) and, (135), we have

$$E(Y'^2_i) \;\; = \;\; \sigma^2 + \sigma^2 \delta^2 + \gamma^2. \tag{136}$$

and we recall the identity

$$E(\bar{Y}^2) \;\; = \;\; Var(\bar{Y'}) + E^2(\bar{Y'}). \tag{137}$$

So, it follows that

$$
\begin{aligned}
Var(\bar{Y}') &= Var E(\bar{Y}'|a^n, \sigma^2) + E Var(\bar{Y}'|a^n, \sigma^2) \\
&= Var\left(\frac{1}{n}\sum_{i=1}^{n} a_i | a^n, \sigma^2\right) + E Var\left(\frac{1}{n}\sum_{i=1}^{n} Y_i' | a^n, \sigma^2\right) \\
&= \frac{1}{n^2}\sum_{i=1}^{n} Var(a_i) + E\left(\frac{1}{n^2}\sum_{i=1}^{n} Var(Y_i') | a^n, \sigma^2\right) \\
&= \frac{\sigma^2 \delta^2}{n} + E\left(\frac{1}{n^2} n \sigma^2\right) \\
&= \frac{\sigma^2 \delta^2}{n} + \frac{\sigma^2}{n}.
\end{aligned}
\tag{138}
$$

Hence from (134), (137) and (138), we have

$$
E(\bar{Y}'^2) = \frac{\sigma^2 \delta^2}{n} + \frac{\sigma^2}{n} + \gamma^2.
\tag{139}
$$

Then from (132), (136), and (139) we have

$$
\begin{aligned}
E(S_2') &= \frac{1}{n-1}\left[\sum_{i=1}^{n}(\sigma^2 + \sigma^2\delta^2 + \gamma^2) - n\left(\frac{\sigma^2\delta^2}{n} + \frac{\sigma^2}{n} + \gamma^2\right)\right] \\
&= \sigma^2(1 + \delta^2).
\end{aligned}
$$

Hence,

$$
E\left(\frac{S_2'}{1+\delta^2}\right) = \sigma^2.
\tag{140}
$$

So, setting $\hat{\sigma}^2 = \frac{S_2'}{1+\delta^2}$ we have unbiasedness.

For a second moment condition to use in the method of moments argument in Subsec. 3.1.2, we find $Var(\hat{\sigma}^2)$. We know that if $X \sim \mathcal{N}(\mu, \sigma^2)$, then

$$
E(X^4) = \mu^4 + 6\mu^2\sigma^2 + 3\sigma^4
\tag{141}
$$

Using this with $Y'$, we start with

$$
E(\bar{Y}'^4) = E E(\bar{Y}'^4 | a^n, \sigma^2).
$$

From (130), we can say that

$$
\bar{Y}'|a^n, \sigma^2 \sim \mathcal{N}(\bar{a^n}, \frac{\sigma^2}{n})
$$

$$
E(\bar{Y}'^4 | a^n, \sigma^2) = \bar{a}^4 + 6\bar{a}^2\frac{\sigma^2}{n} + 3\frac{\sigma^4}{n^2}
\tag{142}
$$

From (131), we have,

$$\bar{a} \sim \mathcal{N}(\gamma, \frac{\sigma^2 \delta^2}{n}).$$

We have,

$$\begin{aligned}
E(\bar{a}^2) &= Var(\bar{a}) + E^2(\bar{a}) \\
&= \frac{\sigma^2 \delta^2}{n} + \gamma^2.
\end{aligned}$$

Also from (141), we can say that,

$$E(\bar{a}^4) = \gamma^4 + 6\frac{\gamma^2 \sigma^2 \delta^2}{n} + \frac{3}{n^2}\sigma^4 \delta^4. \tag{143}$$

Hence from (142) and (143), we get

$$\begin{aligned}
E(\bar{Y'}^4) &= E(\bar{Y'}^4 | a^n, \sigma^2) \\
&= E(\bar{a}^4) + 6\frac{\sigma^2}{n}E(\bar{a}^2) + \frac{3}{n^2}\sigma^4. \\
&= \gamma^4 + \frac{6}{n}\gamma^2 \sigma^2 \delta^2 + \frac{3}{n^2}\sigma^4 \delta^4 + \frac{6}{n^2}\sigma^4 \delta^2 + \frac{6}{n}\sigma^2 \gamma^2 + \frac{3}{n^2}\sigma^4. \tag{144}
\end{aligned}$$

Now,

$$\begin{aligned}
E(S_2'^2) &= E\left[\left\{\frac{1}{n-1}\sum_{i=1}^{n}(Y_i' - \bar{Y'})^2\right\}^2\right] \\
&= \frac{1}{(n-1)^2}\left[\sum_{i=1}^{n}E(Y_i'^2)\sum_{j=1}^{n}E(Y_j'^2) - nE(\bar{Y'}^2)\sum_{i=1}^{n}E(Y_i'^2)\right. \\
&\quad \left. -nE(\bar{Y'}^2)\sum_{j=1}^{n}E(Y_j'^2) + n^2 E(\bar{Y'}^4)\right] \tag{145}
\end{aligned}$$

Then from (136), (139), and (144), we have

$$\begin{aligned}
E(S_2'^2) &= \frac{1}{(n-1)^2}\left[n^2(\sigma^2 + \sigma^2 \delta^2 + \gamma^2)^2 - 2n^2(\sigma^2 + \sigma^2 \delta^2 + \gamma^2)\right. \\
&\quad \left\{\gamma^2 + \frac{1}{n}(\sigma^2 + \sigma^2 \delta^2)\right\} + n^2\left(\gamma^4 + \frac{6}{n}\gamma^2 \sigma^2 \delta^2 + \frac{3}{n^2}\sigma^4 \delta^4 + \frac{6}{n^2}\sigma^4 \delta^2\right. \\
&\quad \left.\left. +\frac{6}{n}\sigma^2 \gamma^2 + \frac{3}{n^2}\sigma^4\right)\right] \\
&= \frac{1+\delta^2}{(n-1)^2}[\sigma^4(n^2 - 2n + 3)(1 + \delta^2) + 4n\sigma^2 \gamma^2]. \tag{146}
\end{aligned}$$

Hence,

$$Var\left(\frac{S_2'}{1+\delta^2}\right) = \frac{1}{(1+\delta^2)^2}[E(S_2'^2) - E^2(S_2')].$$

From ([132](#)) and ([146](#)), we have

$$Var\left(\frac{S_2'}{1+\delta^2}\right) = Var(\hat{\sigma}^2) = \frac{1}{(n-1)^2}[\sigma^4(n^2-2n+3) + \frac{4n}{1+\delta^2}\sigma^2\gamma^2] - \sigma^4$$

$$= \frac{2\sigma^2}{(n-1)^2}\left[\sigma^2 + \frac{2n\gamma^2}{1+\delta^2}\right]. \quad (147)$$

### C.2. Proof of Theorem *3.2*

This proof has been adapted from [13]. The terms that do not contain $\gamma$ in the proof have not been changed.

We start by noting that the joint prior distribution is given by

$$
\begin{aligned}
w(a^n,\sigma^2) &= \mathcal{N}(\gamma 1^n, \sigma^2\delta^2 I_{n\times n}) \; \mathcal{IG}(\alpha,\beta) \\
&= \frac{e^{-\frac{1}{2\sigma^2}(a^n-\gamma^n)'(\delta^2 I_{n\times n})^{-1}(a^n-\gamma^n)}}{(2\pi)^{\frac{n}{2}}(\sigma^2\delta^2)^{\frac{n}{2}}}\frac{\beta^\alpha}{\Gamma(\alpha)}\left(\frac{1}{\sigma^2}\right)^{\alpha+1}e^{-\frac{\beta}{\sigma^2}}.
\end{aligned}
$$

Set,

$$
\begin{aligned}
\mu^{\gamma n} &= [(I+K)^{-1}_{n\times n} + (\delta^2 I_{n\times n})^{-1}]^{-1}[(I+K)^{-1}_{n\times n}y^n + (\delta^2 I_{n\times n})^{-1}\gamma^n] \\
&= V_{n\times n}[(I+K)^{-1}_{n\times n}y^n + \gamma(\delta^2 I_{n\times n})^{-1}1^n] \quad (148) \\
\beta_n^{*\gamma} &= \beta + \frac{1}{2}y'^n(I+K)^{-1}_{n\times n}y^n + \frac{1}{2}\gamma^{n'}[\delta^2 I_{n\times n}]^{-1}\gamma^n - \frac{1}{2}\mu^{\gamma'n}V^{-1}_{n\times n}\mu^{\gamma n} \\
&= \beta + \frac{1}{2}y'^n[(I+K)^{-1}_{n\times n} - (I+K)^{-1}_{n\times n}V_{n\times n}(I+K)^{-1}_{n\times n}]y^n \\
&\quad -\frac{1}{\delta^2}y'^n(I+K)^{-1}_{n\times n}V_{n\times n}\gamma^n + \frac{n}{2}\frac{\gamma^{n'}\gamma^n}{\delta^2} - \frac{1}{2}\frac{1}{\delta^4}\gamma'^n V_{n\times n}\gamma^n. \quad (149)
\end{aligned}
$$

$$(150)$$

We have, $w(a^n,\sigma^2|y^n) = \mathcal{L}_1(y^n|a,\sigma^2) \times w(a^n,\sigma^2)$ Then,

$$
\begin{aligned}
w(a^n,\sigma^2|y^n) &= \frac{\beta^\alpha}{(2\pi)^n|I+K|^{\frac{1}{2}}(\delta^2)^{\frac{n}{2}}\Gamma(\alpha)}\left(\frac{1}{\sigma^2}\right)^{\alpha^*+1} \\
&\quad \times e^{-\frac{1}{\sigma^2}[\frac{1}{2}(a^n-\mu^{\gamma n})'V^{-1}_{n\times n}(a^n-\mu^{\gamma n})]}e^{-\frac{1}{\sigma^2}\beta^{*\gamma}} \\
\\
m(y^n) &= \frac{|V_{n\times n}|^{\frac{1}{2}}}{(2\pi\delta^2)^{\frac{n}{2}}|I+K|^{\frac{1}{2}}}\frac{\beta^\alpha}{\Gamma(\alpha)}\frac{\Gamma(\alpha^*-\frac{n}{2})}{\beta_n^{*\gamma^{\alpha^*-\frac{n}{2}}}}.
\end{aligned}
$$

Then,

$$\frac{m(y^{n+1})}{m(y^n)} = c\frac{(\beta_{n+1}^{*\gamma})^{-(\alpha+\frac{n+1}{2})}}{(\beta_n^{*\gamma})^{-(\alpha+\frac{n}{2})}} \quad (151)$$

where, c has been defined in the Festschrift paper. Generalizing (149) for $(n+1)$ observations, we get,

$$
\begin{aligned}
\beta_{n+1}^{\gamma*} &= \beta + \frac{1}{2}y'^{n+1}[(I+K)_{n+1\times n+1}^{-1} \\
&\quad -(I+K)_{n+1\times n+1}^{-1}V_{n+1\times n+1}(I+K)_{n+1\times n+1}^{-1}]y^{n+1} \\
&\quad -\frac{1}{\delta^2}y'^{n+1}(I+K)_{n+1\times n+1}^{-1}V_{n+1\times n+1}\gamma^{n+1} + \frac{n+1}{2}\frac{\gamma'^{n+1}\gamma^{n+1}}{\delta^2} \\
&\quad -\frac{1}{2}\frac{1}{\delta^4}\gamma'^{n+1}V_{n+1\times n+1}\gamma^{n+1}.
\end{aligned}
\tag{152}
$$

Redefine

$$
\begin{aligned}
\Gamma_2^{\gamma,n+1} &= \frac{1}{\delta^2}(I+K)_{n+1\times n+1}^{-1}V_{n+1\times n+1}\gamma^{n+1} \\
\Delta^\gamma &= \frac{n+1}{2}\frac{\gamma^2}{\delta^2} - \frac{1}{2}\frac{\gamma^2}{\delta^4}1'^{n+1}V_{n+1\times n+1}1^n
\end{aligned}
$$

$$
\beta_{n+1}^{\gamma*} = \beta + \frac{1}{2}y'^{n+1}\Gamma_{1,n+1\times n+1}y^{n+1} - y'^{n+1}\Gamma_2^{\gamma,n+1} + \Delta^\gamma.
\tag{153}
$$

, where $\Gamma_{1,n+1\times n+1}$ has been defined in the Festschrift paper.

Now, we partition $y^{n+1}$, $\Gamma_{1,n+1\times n+1}$, and $\Gamma_2^{n+1}$. Write

$$
\begin{aligned}
y'^{n+1}\Gamma_{1,n+1\times n+1}y^{n+1} &= \begin{pmatrix} y^n & y_{n+1} \end{pmatrix} \left[ \begin{array}{c|c} \Gamma_{1,n\times n} & g_1^{*n} \\ \hline g_1'^{*n} & g_1^* \end{array} \right] \begin{pmatrix} y^n \\ y_{n+1} \end{pmatrix} \\
&= y'^n\Gamma_{1,n\times n}y^n + 2y'^n g_1^{n*}y_{n+1} + y_{n+1}^2 g_1^*
\end{aligned}
\tag{154}
$$

and

$$
y'^{n+1}\Gamma_2^{n+1,\gamma} = \begin{pmatrix} y^n & y_{n+1} \end{pmatrix} \begin{pmatrix} \Gamma_2^{n,\gamma} \\ g_2^* \end{pmatrix} = y'^n\Gamma_2^{n,\gamma} + y_{n+1}g_2^*.
\tag{155}
$$

Using (154) and (155) in (153), we get

$$
\begin{aligned}
\beta_{n+1}^{*,\gamma} &= \beta + \frac{1}{2}y'^n\Gamma_{1,n\times n}y^n - y'^n\Gamma_2^{n,\gamma} + \Delta^\gamma + \frac{1}{2}g_1^*y_{n+1}^2 - y_{n+1}(g_2^* - y'^n g_1^{n*}) \\
&= \beta + \frac{g_1^*}{2}(y_{n+1} - A_1^*)^2 + A_2^*.
\end{aligned}
\tag{156}
$$

where,

$$
\begin{aligned}
A_1^* &= \frac{g_2^* - y'^n g_1^{*n}}{g_1^*} \\
A_2^* &= \frac{1}{2}y'^n\Gamma_{1,n\times n}y^n - y'^n\Gamma_2^{n,\gamma} + \Delta^\gamma - \frac{1}{2g_1^*}(g_2^* - y'^n g_1^{*n})^2.
\end{aligned}
\tag{157}
$$

Now, since $m(y^n)$ is the marginal density of $y^n$ and, $m(y^{n+1})$ is the marginal density of $y^{n+1}$,

$$\int_{\mathbb{R}} \frac{m(y^{n+1})}{m(y^n)} \mathrm{dy_{n+1}} = 1.$$

From (151) we have that

$$\int_{\mathbb{R}} c \times \frac{\beta_{n+1}^{*,\gamma -\left(\alpha+\frac{n+1}{2}\right)}}{\beta_n^{*,\gamma -\left(\alpha+\frac{n}{2}\right)}} \mathrm{dy_{n+1}} = 1. \tag{158}$$

So solving for $c$ gives

$$c = \frac{\beta_n^{*,\gamma -\left(\alpha+\frac{n}{2}\right)}}{\int_{\mathbb{R}} \beta_{n+1}^{*,\gamma -\left(\alpha+\frac{n+1}{2}\right)} \mathrm{dy_{n+1}}}.$$

Using (158) in (151), we have

$$\frac{m(y^{n+1})}{m(y^n)} = \frac{\beta_{n+1}^{*,\gamma -\left(\alpha+\frac{n+1}{2}\right)}}{\int_{\mathbb{R}} \beta_{n+1}^{*,\gamma -\left(\alpha+\frac{n+1}{2}\right)} \mathrm{dy_{n+1}}}. \tag{159}$$

Rename, $\beta^{**,\gamma} = \beta + A_2^*$.

$$\beta_{n+1}^{*,\gamma -(\alpha+\frac{n+1}{2})} = \beta^{**,\gamma -(\alpha+\frac{n}{2})}\beta^{**,\gamma -\frac{1}{2}}\left[1 + \frac{g_1^*}{2\beta^{**,\gamma}}(y_{n+1} - A_1^*)^2\right]^{-\left(\alpha+\frac{n+1}{2}\right)} \tag{160}$$

By definition, the t-density is given by

$$St_v(\tau, \Sigma)(g) = \frac{\Gamma(\frac{v+d}{2})}{\Gamma(\frac{v}{2})\pi^{\frac{d}{2}}|v\Sigma|^{\frac{1}{2}}}\left(1 + \frac{(g-\tau)'\Sigma^{-1}(g-\tau)}{v}\right)^{-\frac{v+d}{2}}. \tag{161}$$

So if we let

$$v = 2\alpha, d = 1, \Sigma = \frac{\beta^{**\gamma}}{\frac{2\alpha+n}{2}}\frac{1}{g_1^*}, g = y_{n+1}, \tau = A_1^*. \tag{162}$$

and use (162) in (161), we get

$$St_{2\alpha+n}\left(A_1^*, \frac{\beta^{**,\gamma}}{\frac{2\alpha+n}{2}}\right)(y_{n+1}) = \frac{\Gamma(\frac{2\alpha+n+1}{2})}{\Gamma(\frac{2\alpha+n}{2})}g_1^{*\frac{1}{2}}\frac{1}{(2\pi)^{\frac{1}{2}}}\beta^{**,\gamma -\frac{1}{2}}$$

$$\times \left[1 + \frac{g_1^*}{2\beta^{**,\gamma}}(y_{n+1} - A_1^*)^2\right]^{-\frac{2\alpha+n+1}{2}}.$$

Hence,

$$\beta^{**,\gamma-\frac{1}{2}}\left[1+\frac{g_1^*}{2\beta^{**,\gamma}}(y_{n+1}-A_1^*)^2\right]^{-\left(\alpha+\frac{n+1}{2}\right)}$$
$$=\frac{\Gamma(\frac{2\alpha+n+1}{2})}{\Gamma(\frac{2\alpha+n}{2})}\frac{(2\pi)^{\frac{1}{2}}}{g_1^{*\frac{1}{2}}}\times St_{2\alpha+n}\left(A_1,\frac{\beta^{**}}{2\alpha+n}\right)(y_{n+1}). \tag{163}$$

Using (163) in (160), and (160) in (159), we have

$$\frac{m(y^{n+1})}{m(y^n)} = \frac{\beta^{**,\gamma-(\alpha+\frac{n}{2})}}{\int_{\mathbb{R}}\beta_{n+1}^{*,\gamma}{}^{-(\alpha+\frac{n+1}{2})}\mathrm{dy}_{n+1}}\frac{\Gamma(\frac{2\alpha+n}{2})}{\Gamma(\frac{2\alpha+n+1}{2})}\frac{(2\pi)^{\frac{1}{2}}}{(g_1^*)^{\frac{1}{2}}}$$
$$\times St_{2\alpha+n}\left(A_1,\frac{\beta^{**,\gamma}}{2\alpha+n}\right)(y_{n+1}). \tag{164}$$

Since $\frac{m(y^{n+1})}{m(y^n)}=m(y_{n+1}|y^n)$ is a density, $\int_{\mathbb{R}}\frac{m(y^{n+1})}{m(y^n)}\mathrm{dy}_{n+1}=1$. Integrating the right hand side of (164) w.r.t $y_{n+1}$ gives that

$$\frac{\beta^{**,\gamma-(\alpha+\frac{n}{2})}}{\int_{\mathbb{R}}\beta_{n+1}^{*,\gamma}{}^{-(\alpha+\frac{n+1}{2})}\mathrm{dy}_{n+1}}\frac{\Gamma(\frac{2\alpha+n}{2})}{\Gamma(\frac{2\alpha+n+1}{2})}\frac{(2\pi)^{\frac{1}{2}}}{(\gamma_1)^{\frac{1}{2}}}\int_{\mathbb{R}}St_{2\alpha+n}\left(A_1,\frac{\beta^{**,\gamma}}{2\alpha+n}\right)(y_{n+1})\mathrm{dy}_{n+1} \tag{165}$$

equals 1, since $y_{n+1}$ is only in the argument of the $t$ distribution. The integral of the $t$ distribution being one means (165) gives

$$\frac{\beta^{**,\gamma-(\alpha+\frac{n}{2})}}{\int_{\mathbb{R}}\beta_{n+1}^{*,\gamma}{}^{-(\alpha+\frac{n+1}{2})}\mathrm{dy}_{n+1}}\frac{\Gamma(\frac{2\alpha+n}{2})}{\Gamma(\frac{2\alpha+n+1}{2})}\frac{(2\pi)^{\frac{1}{2}}}{(g_1^*)^{\frac{1}{2}}}=1. \tag{166}$$

Finally using (166) in (164), we get

$$m(y_{n+1}|y^n) = St_{2\alpha+n}\left(A_1,\frac{\beta^{**,\gamma}}{2\alpha+n}\right)(y_{n+1}). \quad \square$$

### C.3. *Method of moments for Non-identical Biases in GPP's*

From (67), we have,

$$a^n \sim \mathcal{N}(\gamma^n,\sigma^2\delta^2 I_{n\times n}).$$

Hence,

$$\bar{a} \sim \mathcal{N}(\bar{\gamma}=\frac{1}{n}\sum_{i=1}^n\gamma_i,\frac{\sigma^2\delta^2}{n}).$$

From (130),

$$E(Y_i')=\gamma_i. \tag{167}$$

and

$$E(\bar{Y'}) = \tilde{\gamma}. \qquad (168)$$

Hence from (167) and (135), we have

$$E(Y_i'^2) = \sigma^2 \delta^2 + \sigma^2 + \gamma_i^2. \qquad (169)$$

From (168) and (138), we have

$$E(\bar{Y'}^2) = \frac{\sigma^2}{n}(1 + \delta^2) + \tilde{\gamma}^2. \qquad (170)$$

From (132), (169), and (170), we have

$$
\begin{aligned}
E(S_2') &= \frac{1}{n-1}\left[\sum_{i=1}^{n}(\sigma^2\delta^2 + \sigma^2 + \gamma_i^2) - n\left\{\frac{\sigma^2}{n}(1+\delta^2) + \tilde{\gamma}^2\right\}\right] \\
&= \sigma^2(1+\delta^2) + \frac{1}{n-1}\sum_{i=1}^{n}(\gamma_i - \bar{\gamma})^2. \qquad (171)
\end{aligned}
$$

Define $S_2^\gamma = \frac{1}{n-1}\sum_{i=1}^{n}(\gamma_i - \bar{\gamma})^2$. Then, (171) becomes

$$E(S_2') = \sigma^2(1 + \delta^2) + S_2^\gamma. \qquad (172)$$

Then

$$E\left(\frac{S_2' - S_2^\gamma}{1 + \delta^2}\right) = \sigma^2.$$

Define

$$\hat{\sigma}^2 = \frac{S_2' - S_2^\gamma}{1 + \delta^2}. \qquad (173)$$

$$Var\left(\frac{S_2' - S_2^\gamma}{1 + \delta^2}\right) = \frac{Var(S_2')}{(1 + \delta^2)^2}. \qquad (174)$$

We can show that

$$E(\bar{Y'}^4) = \tilde{\gamma}^4 + \frac{6}{n}\tilde{\gamma}^2\sigma^2\delta^2 + \frac{3}{n^2}\sigma^4\delta^4 + \frac{3}{n^2}\sigma^4 + \frac{6}{n^2}\sigma^4\delta^2 + \frac{6}{n}\sigma^2\tilde{\gamma}^2. \quad (175)$$

From (145), (169), (139), and (175), we have

$$
\begin{aligned}
E(S_2'^2) &= \frac{1}{(n-1)^2}\bigg[\sum_{i=1}^n(\sigma^2\delta^2+\sigma^2+\gamma_i^2)\sum_{j=1}^n(\sigma^2\delta^2+\sigma^2+\gamma_j^2)\\
&\quad -n\bigg\{\frac{\sigma^2}{n}(1+\delta^2)+\bar{\gamma}^2\bigg\}\sum_{i=1}^n(\sigma^2\delta^2+\sigma^2+\gamma_i^2)\\
&\quad -n\bigg\{\frac{\sigma^2}{n}(1+\delta^2)+\bar{\gamma}^2\bigg\}\sum_{j=1}^n(\sigma^2\delta^2+\sigma^2+\gamma_j^2)\\
&\quad +n^2\bigg(\bar{\gamma}^4+\frac{6}{n}\bar{\gamma}^4+\frac{3}{n^2}\sigma^4\delta^4+\frac{3}{n^2}\sigma^4+\frac{6}{n^2}\sigma^4\delta^2+\frac{6}{n}\sigma^2\bar{\gamma}^2\bigg)\bigg]\\
&= \frac{1}{(n-1)^2}[\sigma^4(n^2-2n+3)(1+\delta^2)^2+2\sigma^2(n-1)(1+\delta^2)\sum_{i=1}^n\gamma_i^2\\
&\quad -2n(n-3)\sigma^2\bar{\gamma}^2(1+\delta^2)+(\sum_{i=1}^n\gamma_i^2-n\bar{\gamma}^2)^2]\\
&= \frac{1}{(n-1)^2}[\sigma^4(n^2-2n+3)(1+\delta^2)^2+2\sigma^2(n-1)(1+\delta^2)\sum_{i=1}^n\gamma_i^2\\
&\quad -2n(n-3)\sigma^2\bar{\gamma}^2(1+\delta^2)]+S_2^{\gamma 2}. \quad\quad (176)
\end{aligned}
$$

From (171) and (176) (174) becomes

$$
\begin{aligned}
Var\bigg(\frac{S_2'}{1+\delta^2}\bigg) &= \frac{1}{(1+\delta^2)^2}\bigg[\frac{1}{(n-1)^2}\{\sigma^4(n^2-2n+3)(1+\delta^2)^2\\
&\quad +2\sigma^2(n-1)(1+\delta^2)\sum_{i=1}^n\gamma_i^2-2n(n-3)\sigma^2\bar{\gamma}^2(1+\delta^2)\}\\
&\quad +(n-1)^2 S_2^{\gamma 2}-\{\sigma^2(1+\delta^2)+S_2^\gamma\}^2\bigg]\\
&= \sigma^2\bigg[\frac{2\sigma^2}{(n-1)^2}-\frac{2}{1+\delta^2}S_2^\gamma\\
&\quad +\frac{2}{n-1}\frac{1}{1+\delta^2}\bigg(\sum_{i=1}^n\gamma_i^2-\frac{n(n-3)}{n-1}\bar{\gamma}^2\bigg)\bigg]\\
&= \frac{2\sigma^2}{(n-1)^2}\bigg[\sigma^2+\frac{2n\bar{\gamma}^2}{1+\delta^2}\bigg]. \quad\quad (177)
\end{aligned}
$$

Since (65) and (66) continue to hold in the INID case, we can use (173) and (177) to get $\hat{\alpha}$ and $\hat{\beta}$.

# Appendix D: Further Computational Details

## D.1. Details of Implementation

In this subsection we give the details of implementation of the 12 methods grouped into their four classes as indicated in Table 1. Again, the reader can skip to the results in Subsec. 6.2 if desired. For all of our computed results here, we used a burn-in of 10% of the sample size for each data set. We used the burn-in set to get our first predictor. When we used a representative subset from streaming $K$-means, we set $K = 200$ and simply took the first 200 distinct points as the initial representative set.

### D.1.1. Shtarkov

For the Shtarkov method, we chose our 'experts' to be normal distributions. In all cases where the mean was unknown the Shtarkov predictor was either the mean itself or an affine function of the mean defined by the hyperparameters in the prior. In these cases, the predictor defaulted to the mean for large $n$ This was simple to implement in one-pass or to modify by using a representative set.

### D.1.2. Bayes

We used five Bayesian methods. Even though the conformal method was based on the posterior as a nonconformity we did not regard this as a Bayes method because it did not use the posterior directly to get a prediction. Moreover, there are other technique that use a density for a nonconformity that we did not use simply because code was not available. This was a judgment call and readers may prefer to regard this conformal predictor as Bayesian.

We grouped the five Bayesian methods into three that were based on GPP's and two that were based on DPP's.

1. **GPP methods**

    We used three GPP based predictors. For the first, the GPP with no random bias, the predictor is given in equation (54) where the terms $K_{11}$ and $K_{12}$ need to be specified. For the second, the predictor in GPPRB (IID Additive case) is specified in Theorem 3.1 where the form of the predictor has a function $A_1$ in it. This function depends on other functions $g_1$, $g_2$, $g_1^n$ which can be explicitly written as the functions of the hyperparameters $\gamma$, $\delta$, the variance matrix $K_{n+1 \times n+1}$ and the data. for the third, the predictor in GPPRB (INID Additive bias case) is given by $A_1^*$ where the details about the predictor can be found in Theorem 3.2. For this method the variance matrix $K_{n \times n}$ must be chosen.

In all cases, we chose the variance matrices to correspond to an $AR(1)$ correlation matrix which has the form

$$K_{n \times n} = \begin{pmatrix} 1 & \rho & \cdots & \rho^{n-1} \\ \rho & 1 & \cdots & \rho^{n-2} \\ \vdots & \vdots & \ddots & \vdots \\ \rho^{n-1} & \rho^{n-2} & \cdots & 1 \end{pmatrix}$$

and we set $\rho = 0.8$. We have explained at the end of Subsubsec. 3.1.2 why we choose a small value of $\delta$. Here and elsewhere we set $\delta = .1$. For the GPP(INID) case, we chose,

$$\hat{\gamma}_{n+1} = \frac{1}{n} \sum_{i=1}^{n} \gamma_i.$$

2. **DPP methods**

The DPP predictor is given in (70). The base measure $F_0$ was chosen to be the discrete uniform on $[\min\{y_1, \ldots, y_n\}, \max\{y_1, \ldots, y_n\}]$.

When we used a representative subset to form $DPP_{rep}$ we used $K = 200$ as described at the end of Subsec. 6.1.

### D.1.3. Hash function based methods

We used four hash function based predictors. Two were the mean and median of the EEDF generated by the Count-Min sketch. These are given in equations (78) and (79).

For the representative set forms of these two predictors, that we included only for the sake of comparison, as hash based methods are already one-pass, we followed the same procedure as for Shtarkov and Bayes.

### D.1.4. Conformal

We simply used the technique of [8] who provided code. Their code was not one-pass so we only used their method with a representative subset. To be explicit, the conformity measure $C(y^{n+1})$ was the posterior from a normal density equipped with conjugate priors:

$$p(y_{n+1}|y^n) = \frac{\Gamma(\frac{2a_\sigma+1}{2})}{\sqrt{2a_\sigma}\pi\Gamma(\frac{2a_\sigma}{2})} \left( \frac{1}{\sqrt{\sigma_t^2}} \left( 1 + \frac{1}{2a_\sigma} \frac{(y_i - \mu_\theta)^2}{\sigma_t^2} \right)^{-\frac{2a_\sigma+1}{2}} \right)$$

where

$$\tau_\theta^2 = \left( \frac{1}{\tau^2} + n \right)^{-1}, \mu_\sigma = \left( \frac{\mu}{\tau^2} + 1_n^T y^n \right) \tau_\theta^2, a_\sigma = a + n,$$

$$b_\sigma = b + y^{nT} y^n + \frac{\mu^2}{\tau^2} - (\tau_\theta^2)^{-1} \mu_\theta^2, \sigma_t^2 = \frac{b_\sigma}{a_\sigma}(1 + \tau_\theta^2).$$

For this method, the predictor is given by the mean of the prediction bounds from (80) with $\alpha = .15$, the default value in fabContinuousPrediction.

### D.2. Further Examples

Here we provide graphs for four more data sets. The results from these graphs were used in Subsec. 6.3 and in particular in Table 2.

The four data sets are:

1. The Colombia data set can be found at
   https://data.world/hdx/f402d5ef-4a74-4036-8829-f04d6f38c8e9
   which provides a description. It contains daily values of precipitation (mm) in Columbia over a period of four years ending in 2013. We used the first 5000 rows of the value column of the data set.

2. The accelerometer data set can be found at [9], which provides a description. We used the first 10,000 rows of the data set and used the column "$y$" for our computing. For the HBP computations with this data set we used $d_K = 25$, $W_K = 50$, and $K = 200$.

3. The Real Estate data set can be found at
   https://www.kaggle.com/datasets/derrekdevon/real-estate-sales-2001-2020.
   This data set contains information on real estate from all across the world. We used the first 10,000 rows and the column "Sales Ratio"  which is the ratio of the Assessed value and the Sale Amount.

4. The Parking data set can be found at
   https://www.kaggle.com/datasets/mhmdkardosha/parking-birmingham.
   This data set contains information on parking capacity and parking occupancy of several car parks in Birmingham. We used the first 10000 rows and the column labeled "Occupancy".

The graphs of the sensitivity curves for all 12 methods follow.
For completeness we give a few details of our implementations.

- Except for the Colombia data set we used the same specifications for the other data sets as mentioned in the accelerometer data set. For the Colombia dataset we used $d_K = 10, K = 100$ for the hash function methods.
- For Accelerometer, Fig. 6, $\sigma_{RV} = .18$ with $\tau$ spaced at interval length of 0.02. Here, the HBP median (one pass) method and the GPPRB method show the best performances.
- For Real Estate, Fig. 7, $\sigma_{RV} = 630$ with $\tau$ at equal intervals of 90. The GPPRB(INID), GPPnoRB and the HBP median (rep.) methods had the best performances.
- For Parking, Fig. 8, $\sigma_{RV} = 198$ with $\tau$ located at equal intervals of 33. The HBP median (one pass) had the best performance.
- For Colombia, Fig. 9, $\sigma_{RV} = 900$ and $\tau$ is chosen at interval lengths of 100. The HBP median (one pass) and GPPRB methods had the best performance.

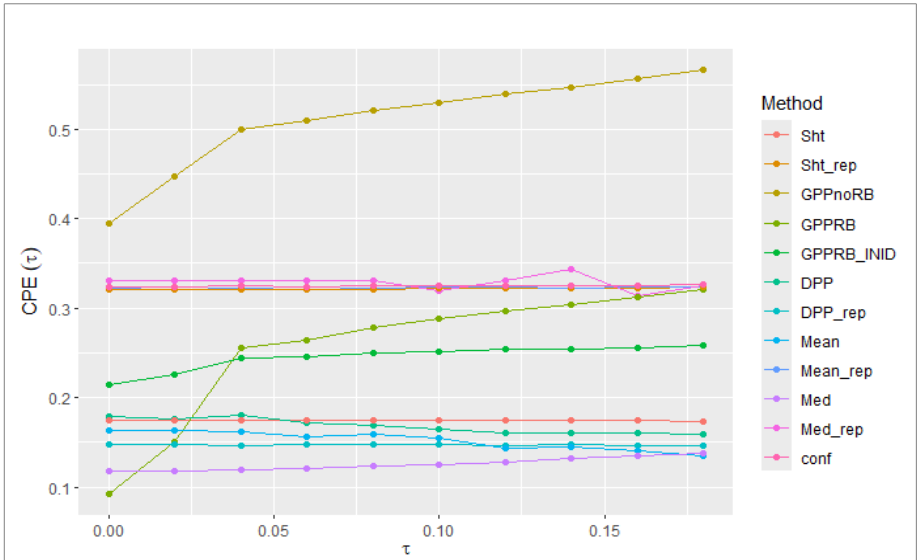

Fig 6. Accelerometer *data*.

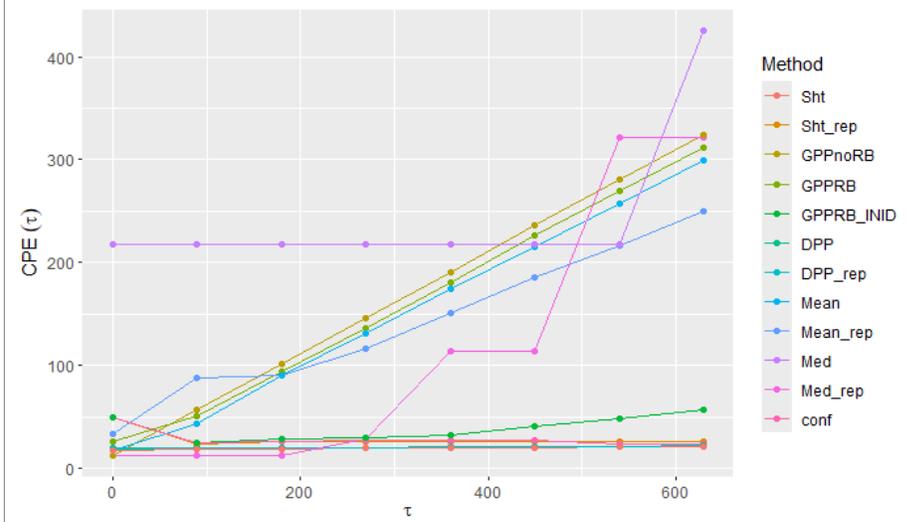

Fig 7. Real Estate *data*.

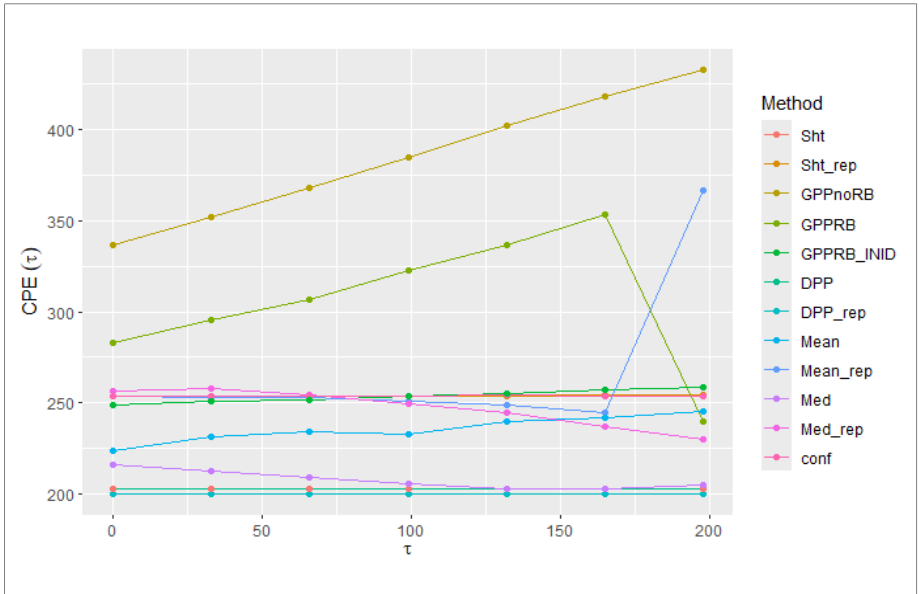

FIG 8. **Parking** *data.*

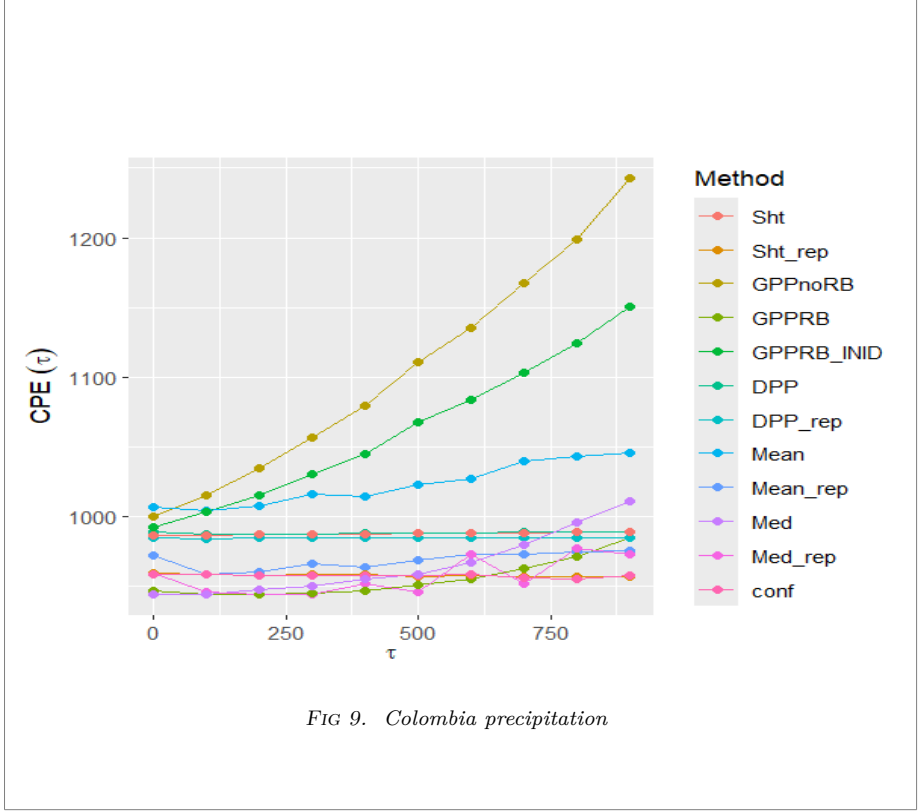

FIG 9. *Colombia precipitation*